\documentclass[letterpaper,11pt]{article}
\usepackage{graphics,graphicx,epsfig,wrapfig}
\usepackage{longtable}  
\usepackage{amsmath,verbatim,amssymb,color,lscape}
\usepackage{kotex}
\usepackage{natbib}
\usepackage{lastpage}
\usepackage{multirow} 
\usepackage{authblk}
\usepackage{bm}
\usepackage[ruled,vlined]{algorithm2e}
\usepackage[normalem]{ulem}
\usepackage[export]{adjustbox}

\newcommand{\hide}[1]{}

\makeatletter

\usepackage[outerbars,color]{changebar}
\ifx\pdfoutput\undefined
\else\ifnum\pdfoutput>0
  \usepackage{pdfcolmk}
\fi\fi
\cbcolor{black}

\usepackage[margin=1.0in]{geometry}
\usepackage{tikz}
\usetikzlibrary{bayesnet}
\usetikzlibrary{fit,positioning}

\newfont{\rmm}{cmr10 at 11pt}
\rmm

\title{How social networks influence human behavior: An integrated latent space approach for differential social influence}

\author[1,2]{Jina Park}
\author[1,2]{Ick Hoon Jin}
\author[3]{Minjeong Jeon}
\affil[1]{Department of Applied Statistics, Yonsei University. South Korea.}
\affil[2]{Department of Statistics and Data Science, Yonsei University. South Korea.}
\affil[3]{School of Education and Information Studies, University of California, Los Angeles. USA.}
\date{}

\begin{document}
\maketitle

\begin{abstract}
How social networks influence human behavior has been an interesting topic in applied research. Existing methods often utilized scale-level behavioral data to estimate the influence of a social network on human behavior. This study proposes a novel approach to studying social influence that utilizes item-level behavioral measures. Under the latent space modeling framework, we integrate the two interaction maps for respondents' social network data and item-level behavior measures. The interaction map visualizes the association between the latent homophily of the respondents and their behaviors measured at the item level in a low-dimensional latent space, revealing the potential, differential social influence effects across specific behaviors measured at the item level. We also measure overall social influence as the impact of the interaction map configuration contributed by the social network data on the behavior data. The performance and properties of the proposed approach are evaluated via simulation studies. We apply the proposed model to an empirical dataset to demonstrate how the students' friendship network influences their participation in school activities. 
\end{abstract}

\noindent {\bf Keywords:} Differential Social influence; Network analysis; Item response model; latent space model; latent space item response model; Peer Network
\newpage

\section{Introduction}

A frequently asked question in education and social science research is how contexts influence the behavioral outcomes of individuals. The influence process can take various forms. For example, students' emotional or social support networks in classrooms with their teachers, mentors, or friends are likely to play an important role in shaping the students' behaviors. The impact of individuals' social networks on their behavior is typically referred to as \textit{social influence} in the literature \citep{Sijtsema2010, Sun2011, Simpkins2013, Shakarian2015}.

Social influence has been widely studied in applied research using various analytic methods for cross-sectional and longitudinal data. For instance, \citet{Urberg1997} and \citet{Mercken2010} investigated how a network of close friends influenced adolescents' smoking behavior with longitudinal analysis. \citet{Cheng2014} analyzed cross-sectional data on high school adolescents to examine how friends' physical activities influenced the adolescents' physical activity levels. \citet{Maria2016} examined how students' academic performance was influenced by their shared interest network by applying cross-sectional network analysis.

While numerous social influence models are available in the literature, most approaches process behavior of interest at the scale level and treat them as a person (or actor) level variable for analysis. These models include network autocorrelation models, autologistic actor attribute models, linear-in-means models, and stochastic actor-oriented models \citep{Ord1975, Doreian1989, Robins2001b,Leenders2002, Daraganova2013, Dittrich2019,  Sweet:2020}. These approaches typically focus on understanding the overall impact of social networks on behavioral data measured at the aggregated or scale level. When behavioral data are related to the human mind and psychology that are not directly observable, they are often measured based on questionnaires consisting of multiple items (indicating specific symptoms or behaviors). However, using aggregated behavioral measures ignores the fact that social influence can manifest in different patterns and structures across specific/individual behavioral items. Motivated by this limitation of existing methods, we propose an approach that allows us to inspect potentially differential social influence effects across multi-item behavioral measures in a longitudinal data setting where an individual's social network is measured (at time $t$) before their item-level behavior of interest (at time $t+1$). Importantly, our approach offers a tool for visually examining the effects of differential social influence in a low-dimensional latent space, which we also call an \textit{interaction map} following \cite{jeon:2020}. 

We will model two data types in a common modeling framework based on latent space, using the latent space model (LSM) for social network data \citep{Hoff:2002} and the latent space item response model (LSIRM; \citealt{jeon:2020}) for behavior item response data. LSM, a widely used approach for social network data, measures the distances between the latent positions of individuals to model the probabilities of ties among them. The between-people distances indicate individuals' interactions in the network of interest, whereas shorter distances indicate stronger between-people connections. LSIRM, a recent development in the psychometrics literature, is a specialized extension of the LSM for item response analysis. LSIRM views binary item response data as bipartite network data representing relationships between respondents and items. It measures the distances between respondents and items as a penalty term to model the respondents' probabilities of giving correct (or positive) responses, where the respondent-item distances represent their relations given the respective main effects. Shorter respondent-item distances, for example, indicate that respondents have stronger relations with the items, meaning that respondents are likely to provide positive responses to those items. 

A common feature of the LSM and the LSIRM is that both models locate individuals in a latent space based on their relationships with peers (for the social network) and behavior items (for the item response network). To measure social influence, we integrate LSM and LSIRM such that the latent positions of respondents from LSIRM are determined from LSM, reflecting their social relationships or homophily at time $t$. The item latent positions are then determined from LSIRM based on whether and how the respondents' homophily measured at time $t$ is related to the respondents' item-level behaviors measured at time $t+1$. In this way, the estimated latent space helps us to overview the structure and patterns of differential social influence across specific item-level behavioral measures. In addition, the size of the overall social influence can also be evaluated with the weight of the distance term in the LSM. That is, in our integrated framework, overall social influence is defined as the impact of social homophily on the respondents' behavior, where the respondents' latent positions reflect social homophily, and respondents' behavior is their responses to the test items (i.e., item-level behavioral measures). 
Differential social influence, on the other hand, reveals how social influence manifests across different, item-level behavioral measures, in the latent space. 

It is worth noting that researchers have defined and measured social influence in diverse ways in the literature. For example, the network autocorrelation model (\citealp{Ord1975, Doreian1989, Leenders2002}, \citealp{Frank2004, Zheng2010, Scott2012, FUJIMOTO2013, Carr2015, sewell2017network}, \citealp{Dittrich2019}) and the linear-in-means model \citep{Manski1993, GoldsmithPinkham2013} use the observed social network to estimate social influence, while the autologistic actor attribute model \citep{Robins2001b, Daraganova2013, Parker2021} uses network statistics to measure social influence. Our definition and measurement of social influence are somewhat different from the existing frameworks discussed above. We believe that our definition of social influence is sensible, and bringing this new perspective can be seen as a useful contribution. Additional elaborations on our definition of social influence are provided in Section \ref{sec:influence}. Further, it is important to stress that  our proposed approach enables us to inspect potentially differential social influence effects across specific, item-level behavioral measures, which is a unique perspective and contribution to the literature. 

The rest of the paper is organized as follows. In Section 2, we start by providing some background on the proposed approach. We briefly describe the latent space modeling framework for social network data (LSM) and item response data (LSIRM). We then briefly discuss some existing social influence models in the literature. In Section 3, we present the formulation of the proposed Bayesian model of social influence. We also provide details of the Bayesian estimation procedure. In Section 4, a real data example is provided with a detailed analysis of the proposed approach. In Section 5, we conduct simulation studies to evaluate the performance of our proposed approach. Finally, we conclude our paper with a summary and discussion in Section 6. 

\section{Background}

\subsection{Latent Space Model}

The latent space model (LSM; \citealt{Hoff:2002, Handcock:2007, Krivitsky:2009, Raftery:2012, Rastelli:2015, Sewell:2015, Gollini:2016}) is one of the well-developed statistical models for analyzing network data. In LSM, the probability of an edge between nodes $k$ and $l$ depends on the distance between their latent positions in a $D$-dimensional Euclidean latent space. In general, the smaller the distance between nodes in the latent space, the greater the probability that they are connected. 

To formulate the model, let ${\bf Y}_{n \times n}$ be a peer network among the respondents, where $y_{kl}$ is a connection between the respondent $k$ and $l$. Note that ${\bf Y}$ is an undirected network and there are no self-edges in the network ${\bf Y}$. Let ${\bf Z}$ be a $N \times D$ latent position matrix where ${\bf z}_k = (z_{k1}, \cdots, z_{kD})$ is the $D$-dimensional vector indicating the position of node $k$ in the $D$-dimensional latent space. Let $\alpha$ be the intercept term of LSM and $\gamma$ be the weight parameter of the Euclidean distance term. Let $\boldsymbol\Theta^n = \{ {\bf Z}, \alpha, \gamma \}$ be the collection of model parameters of the LSM, where superscript $n$ represents that the parameter set for the LSM. Then, LSM can then be written as
\begin{equation}\label{eq:lsm}
    P\Big({\bf Y} \mid \boldsymbol\Theta^n\Big) = \prod_{k \neq l} P\Big(y_{kl} \mid {\bf z}_k, {\bf z}_l, \alpha, \gamma\Big) = \prod_{k \neq l}\frac{\exp\left(\alpha - \gamma||{\bf z}_k - {\bf z}_l||\right)^{y_{kl}}}{1+\exp\left(\alpha - \gamma||{\bf z}_k - {\bf z}_l||\right)},
\end{equation}
where $||{\bf z}_k - {\bf z}_l||$ is the Euclidean distance between node $k$ and $l$ in a low-dimensional Euclidean space. We selected a two-dimensional space in this paper, which has convenience for visualization and is a conventional choice in the literature. The weight parameter $\gamma \geq 0$ for the distance term represents the magnitude of social interactions present in the network data of interest. The weight $\gamma$ is assumed to be positive so that a larger distance indicates weaker connections between people.  $\gamma = 0$ indicates no meaningful presence of social interactions in the peer network data. The LSM with $\gamma = 0$ is equivalent to the Erd\"os-Renyi model \citep{ErRe60}, in which each edge is independent of the connections of other pairs and the connection probability for all nodes is the same.

To estimate the parameters of the LSM model, a Bayesian approach is frequently used \citep{Hoff:2002, Handcock:2007}. The prior distributions for the model parameters can be given as
\begin{equation}\label{eq:pr_si_lsm}
    \alpha \sim \mbox{N}\Big(0, \sigma_{\alpha}^2\Big), \quad 
    {\bf z}_k \sim \mbox{N}\Big({\bf 0}, {\bf I}_d\Big), \quad \mbox{and} \quad 
    \mbox{log}(\gamma) \sim \mbox{N}\Big(\mu_{\gamma}, \sigma_{\gamma}^2\Big).
\end{equation}
Here, we assume that ${\bf z}_k$ follows an independent spherical multivariate normal distribution. To ensure the identifiability of the weight parameter $\gamma$, we fix the scale of ${\bf z}_k$ to one \citep{Handcock:2007, jeon:2020}. 

\begin{figure}[htbp]
    \centering
    \begin{tabular}{cc}
    (a) & (b)   \\ 
    \includegraphics[width=0.4\textwidth]{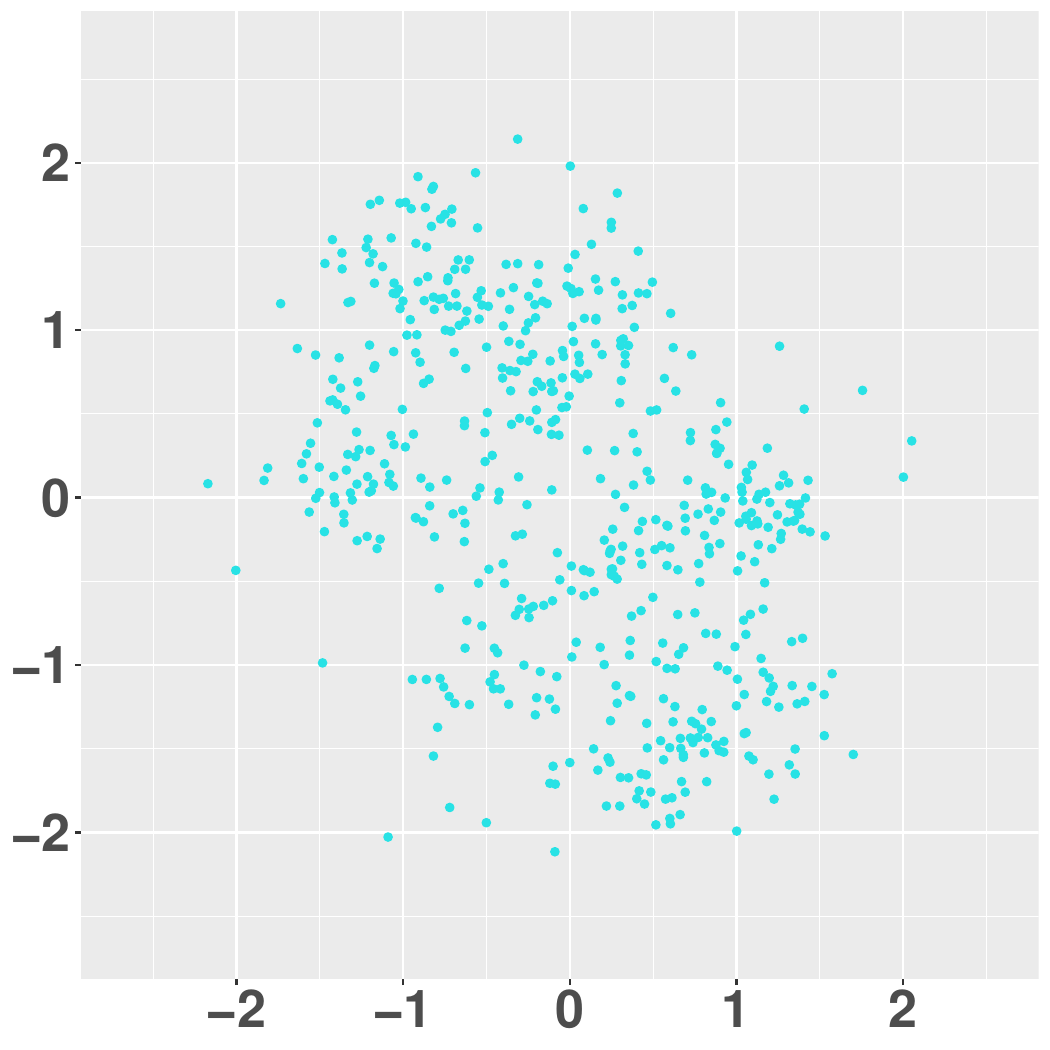} 
    &\includegraphics[width=0.4\textwidth]{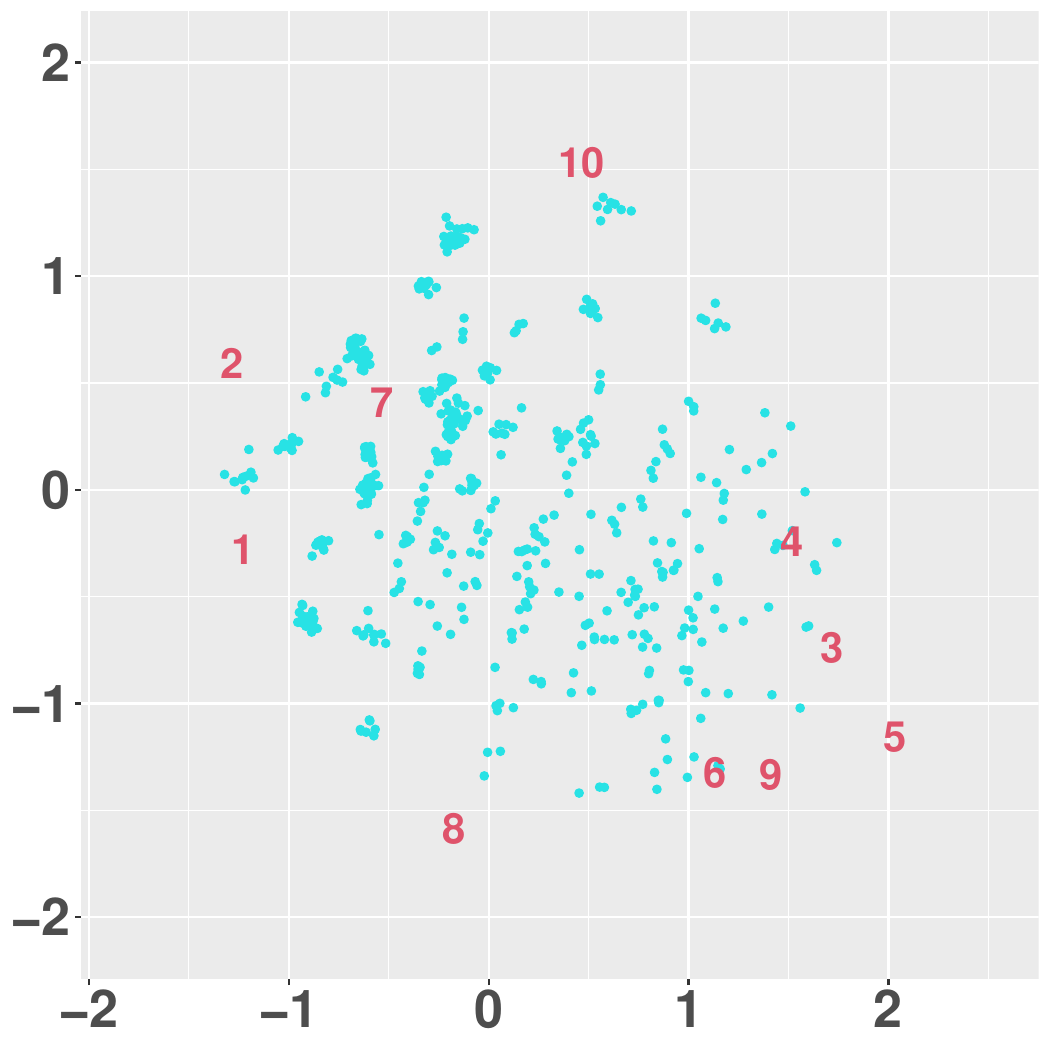} \\
    \end{tabular}
    \caption{
    (a) A latent space of close friends of 539 US middle school students estimated based on the LSM. Blue dots represent students. (b) An interaction map of students and test items is estimated based on the LSIRM. Red numbers represent ten school activity items, and colored dots represent students. 
    }
    \label{fig:lsm}
\end{figure}

Distances between respondents, i.e., their relationship with others, can be represented in a two-dimensional Euclidean space, as illustrated in Figure \ref{fig:lsm}(a). The latent space shown in Figure \ref{fig:lsm}(a) represents a close friend network of 539 US middle school students,  while small light blue dots represent the students. Shorter distances between dots indicate closer relationships or a higher likelihood of their being close friends. This latent space shows that there are roughly two groups or sub-networks of close friends in this chosen school (this is School 59 from the empirical example described in Section \ref{sec:application}). 

\subsection{Latent Space Item Response Model}

A latent space item response model \citep[LSIRM;][]{jeon:2020} is a unique extension of LSM for binary item response data. LSIRM assumes both item and respondents are embedded in a low-dimensional interaction map such that the probability of a correct (or positive) response decreases by the distance between the respondents' and the items' positions in the interaction map. The resulting interaction map represents the interactions between respondents and items after taking into account the person and item characteristics (e.g., person ability and item difficulty) as the main effects. 

To formulate the LSIRM, suppose ${\bf X}_{n \times p}$ is an item response data, where $x_{ki}$ is a binary response of respondent $k$ to item $i$. Let $\boldsymbol\beta = \{\beta_i\}$ and $\boldsymbol\theta = \{\theta_k\}$ be the item easiness parameter and person characteristic parameter, respectively. Let ${\bf Z}$ and ${\bf W}$ be a $N \times D$ and $P \times D$ latent position matrix where ${\bf z}_k = (z_{k1}, \cdots, z_{kD})$ and $(w_{i1}, \cdots, w_{iD})$ are the $D$-dimensional vector indicating the position of the respondent $k$ and that indicating the position of item $i$, respectively. Let $\boldsymbol\Theta^r = \{\boldsymbol\beta, \boldsymbol\theta, {\bf Z}, {\bf W}, \delta, \sigma^2\}$ be the collection of the LSIRM model parameters, where superscript $r$ represents the parameter set for the LSIRM. Then, LSIRM can be written as
\begin{equation}\label{eq:lsirm}
    P\Big({\bf X} \mid \boldsymbol\Theta^r \Big) = \prod_{k=1}^n \prod_{i=1}^p P\Big(x_{ki} \mid \beta_i, \theta_k, \gamma, {\bf z}_k, {\bf w}_i\Big)
    = \prod_{k=1}^n \prod_{i=1}^p \frac{\exp\left(\beta_i + \theta_k - \gamma||{\bf z}_k - {\bf w}_i||\right)^{x_{ki}}}{1+\exp\left(\beta_i + \theta_k - \gamma ||{\bf z}_k - {\bf w}_i||\right)},
\end{equation}
where $||{\bf z}_k - {\bf w}_i||$ is the Euclidean distance between the latent positions of the respondent $k$ and the item $i$ and $\gamma > 0$ is the weight of the distance term. As in the LSM, we chose a two-dimensional Euclidean space to represent respondent-item distances from the LSIRM for visualization. $\gamma > 0$ ensures that a large respondent-item distance decreases the probability of the respondent's giving a positive (or correct) response to the item. Further, the weight parameter $\gamma$ is of crucial importance as it represents the importance or impact of the interaction map (of the respondents and items) on the probability of giving positive responses, after taking into account the person and item main effects, $\theta_k$ and $\beta_i$. When $\gamma = 0$, the LSIRM reduces to a conventional item response model, specifically the Rasch model, which assumes that the main effects, $\theta_k$ and $\beta_i$ are sufficient to explain the correct probability. 

A Bayesian approach is used to estimate the LSIRM. The prior distributions of the LSIRM parameters can be given as follows: 
\begin{equation}\label{eq:pr_si_lsirm}
\begin{split}
    \beta_i &\sim {\mbox N}\Big(0, \sigma_{\beta}^2\Big), \quad 
    \theta_k \mid \sigma^2 \sim {\mbox N}\Big(0, \sigma^2\Big), \quad 
    \sigma^2 \sim \mbox{Inv-G}\Big(a_{\sigma}, b_{\sigma}\Big),\\
    {\bf w}_i &\sim \mbox{N}\Big({\bf 0}, {\bf I}_d\Big),  \quad 
    {\bf z}_i \sim \mbox{N}\Big({\bf 0}, {\bf I}_d\Big), \quad
    \quad  \mbox{and} \quad \log(\gamma) \sim \mbox{N}\Big(\mu_{\gamma}, \sigma_{\gamma}^2\Big).
\end{split}
\end{equation}

Figure \ref{eq:lsm}(b) illustrates an estimated two-dimensional interaction map from the LSIRM. The blue dots represent the 539 US middle school students, the same students used in Figure \ref{eq:lsm} (a), while the red numbers represent ten items about students' school activities. Details of the data and items are described in Section \ref{sec:application}. The shorter distances between the respondents and the items indicate that the respondents were more likely to respond positively to the items (i.e. they had a higher probability of participating in the corresponding school activities). It is important to mention that respondents in a similar region tend to show a similar profile of measured behaviors. For example, the respondents on the bottom left of the space are close to Items 3, 4, 5, 6, and 9 while far apart from items 1, 2, 7, and 10. This means that those respondents are similar in terms of their school activity patterns, such that they are more likely to participate in activities 3 (theater/drama), 4 (music), 5 (arts), 6 (clubs in other schools), and 9 (reading) but less likely to participate in the school activity of 1 (sports within schools), 2 (sports outside school), 7 (dating in schools) and 10 (video games) given their overall attribute parameters.

\subsection{Existing Social Influence Models}

Here, we briefly review existing social influence models and related approaches in the literature. Mathematical formulations and additional details of these models are provided in Supplementary Material 1. First, network autocorrelation models are widely-used to measure social influence (\citealp{Ord1975, Doreian1989, Leenders2002}, \citealp{Frank2004, Zheng2010, Scott2012, FUJIMOTO2013, Carr2015, sewell2017network}, \citealp{Dittrich2019}). These models quantify the strength of a peer effect (i.e., social influence) while controlling for individuals' characteristics and network autocorrelations. In addition, the linear-in-means model \citep[LIMM;][]{Manski1993} is available to estimate social interaction and peer effects. Specifically, the LIMM analyzes exogenous effects (i.e., the influence of peer characteristics) and endogenous effects (i.e., the influence of peer outcomes). The autologistic actor attribute model (ALAAM; \citealt{Robins2001b, Daraganova2013, Parker2021}) is another approach for social influence. ALAAM leverages exponential random graph models (ERGM; \citealt{Holland1981, Frank1986, Robins2007, David2007, Fien12}) for network configuration. 

In addition, the stochastic actor-oriented model (SAOM) \citep{Snijders2001, Snijders2010, Block2019} is another well-recognized method for analyzing social influence from longitudinal panel data. The SAOM models the joint interdependent co-evolution of network ties and individual behaviors. The SAOM assumes that the network ties change depending on the structure of the network itself as well as the behavior, while the behavioral changes depend on the current behavior as well as the network structure \citep{Snijders2017}. This represents the interdependent social influence and endogenous selection of ties in a unifying framework \citep{Snijders2017}.  

Further, latent space modeling (LSMs) approaches, discussed in Section 2.1, have also been used to study social influence. For example, \citet{Sweet:2020} proposed how the latent space model for network data can be used in the social diffusion model \citep{Valente2005}, which is a temporal version of the network autocorrelation model. This approach is related to our proposed model in the sense that both utilize LSMs for identifying social relationships between respondents from the network data of interest. However, the two approaches are different in terms of how respondents' social relationships are used for understanding and estimating their impacts on the respondents' behavioral outcomes. Existing LSM-based social influence approaches are based on the autocorrelation model for behaviors, whereas our approach is based on the latent space model for behavior data. 

Lastly, joint latent network models are another line of related models that focus on modeling respondents' social networks and attributes jointly. For example, \citet{Wang:2019} proposed a joint attribute and person latent space model (APLSM) that merges information from the social network and multivariate person covariates. \citet{Fosdick2015} proposed a joint model to test the dependencies between a network and person attributes. Similarly, joint latent network models and stochastic block models for multilayer networks or multidimensional networks \citep{Gollini:2016, Michael2017, DAngelo:2019} are related to this class of models in the sense that multiple networks are involved in the model. 

These joint models are similar to our approach in the sense that respondents' social networks and attributes (behaviors or item responses) or multiple networks (for social relations and relations with items) are involved in the models. However, the most important differences are that these models do not directly measure or quantify social influence. In contrast, our model enables us to evaluate the presence (vs. absence) and the size of social influence with a specific model parameter. In the subsequent section, we elaborate on how the proposed model is formulated to estimate social influence in an integrated latent space modeling framework.

\section{Model}
\label{sec:model}

In the proposed framework for social influence, we measure whether and how an individual's social network at time point $t$ influences their behaviors at time $t+1$, as their responses to behavior items. If social influence exists, individuals with connections in the peer network at time $t$ are likely to show a higher level of similarity in their behavior at time $t+1$. In this setting, we propose the framework integrating the LSM for social network data at time $t$ and behavior item response data at time $t+1$ to estimate social influence.

\begin{figure}[htbp]
    \centering
    \includegraphics[width=0.55\textwidth]{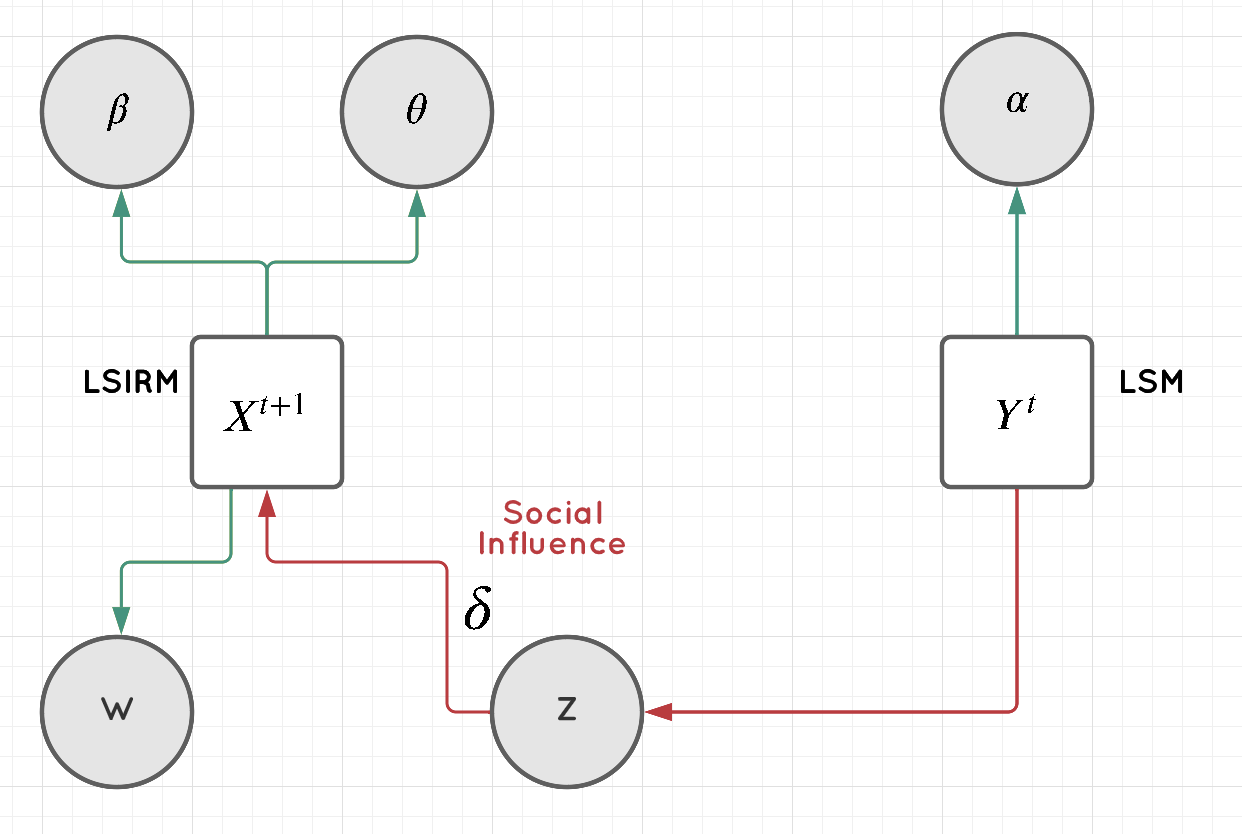}
    \caption{LSIRM for social influence. This diagram shows how social influence can be estimated with the revised LSIRM (for behavior response data at time $t+1$, ${\bf X}^{t+1}$) connected to the LSM (for social network at time $t$, ${\bf Y}^{t}$). The red line with an arrowhead shows the process of social influence. Overall social influence is estimated as the impact of the interactions between latent social homophily (${\bf Z }$) obtained from network data ${\bf Y}^{t}$ through the LSM and the multivariate (item-level) behavior response data ${\bf X}^{t+1}$.
    }
    \label{fig:illustration}
\end{figure}

We refer to the proposed model \textit{LSIRM for social influence}. The framework of the LSIRM for social influence is represented in Figure \ref{fig:illustration}. As shown in Figure \ref{fig:illustration}, two types of datasets are required to apply the proposed approach:${\bf X}_{n \times p}^{t+1}$ for an item response matrix at time $t+1$ and ${\bf Y}_{n \times n}^{t}$ for a peer network matrix at time $t$, where $n$ is the number of respondents and $p$ is the number of items (from the test used to measure the target behavior of interest). Subscripts of ${\bf X^{t+1}}$ and ${\bf Y^{t}}$ are suppressed from now on for succinctness.

\subsection{LSIRM for Social Influence}\label{sec:influence}

The adapted LSIRM for social influence can be formulated as follows:
\begin{equation} 
\label{eq:adapt_lsirm}
P\Big({\bf X^{t+1}} \mid {\bf Z}, {\boldsymbol\Theta^r}^* \Big) = 
\prod_{k=1}^n \prod_{i=1}^p \frac{\exp\Big(\beta_i + \theta_k - \delta||{\bf z}_k - {\bf w}_i||\Big)^{x_{ki}}}
    {1 + \exp\Big(\beta_i + \theta_k - \delta||{\bf z}_k - {\bf w}_i||\Big)},
\end{equation}
\noindent where ${\boldsymbol\Theta^r}^*= \{\boldsymbol\beta, \boldsymbol\theta, {\bf W}, \delta, \sigma^2\}$ is the collection of the adapted LSIRM model parameters. ${\bf Z}$ is the respondents' latent positions updated from the LSM (Equation \ref{eq:lsm}) applied to social network data at time $t$, representing the respondents' social relationships or homophily at time $t$. This is the primary difference of the adapted LSIRM from the standard LSRIM (in Equation (\ref{eq:lsirm})). All model parameters are estimated simultaneously in our Bayesian estimation framework (details are provided in Section \ref{sec:estimation}).

\paragraph*{Weight parameter ${\bf \delta}$: Overall social influence}

In the proposed framework, the weight parameter $\delta \geq 0$ represents the overall social influence as it quantifies the impact of the associations between respondents' social homophily (represented by their latent positions) and the items (that represent respondents' behaviors or activities) on their endorsement of the items. Social homophily, the latent positions of the respondents, are determined by the respondents' social networks. If social homophily has nothing to do with the respondents' activities, in other words, if there is no meaningful association between social homophily and the respondents' behaviors, the $\delta$ parameter will be close to zero. In this case, the respondents' probability of endorsing the activities is explainable by the main effect parameters of the activities and respondents. On the other hand, if social homophily is meaningfully related to the respondents' activities, its impact will be captured by the non-zero $\delta$ parameter value, such that a smaller distance between a respondent and an activity is associated with a higher probability of endorsing, i.e., showing, the activity. 

\begin{figure}[htbp]
    \centering
    \includegraphics[width=0.35\textwidth]{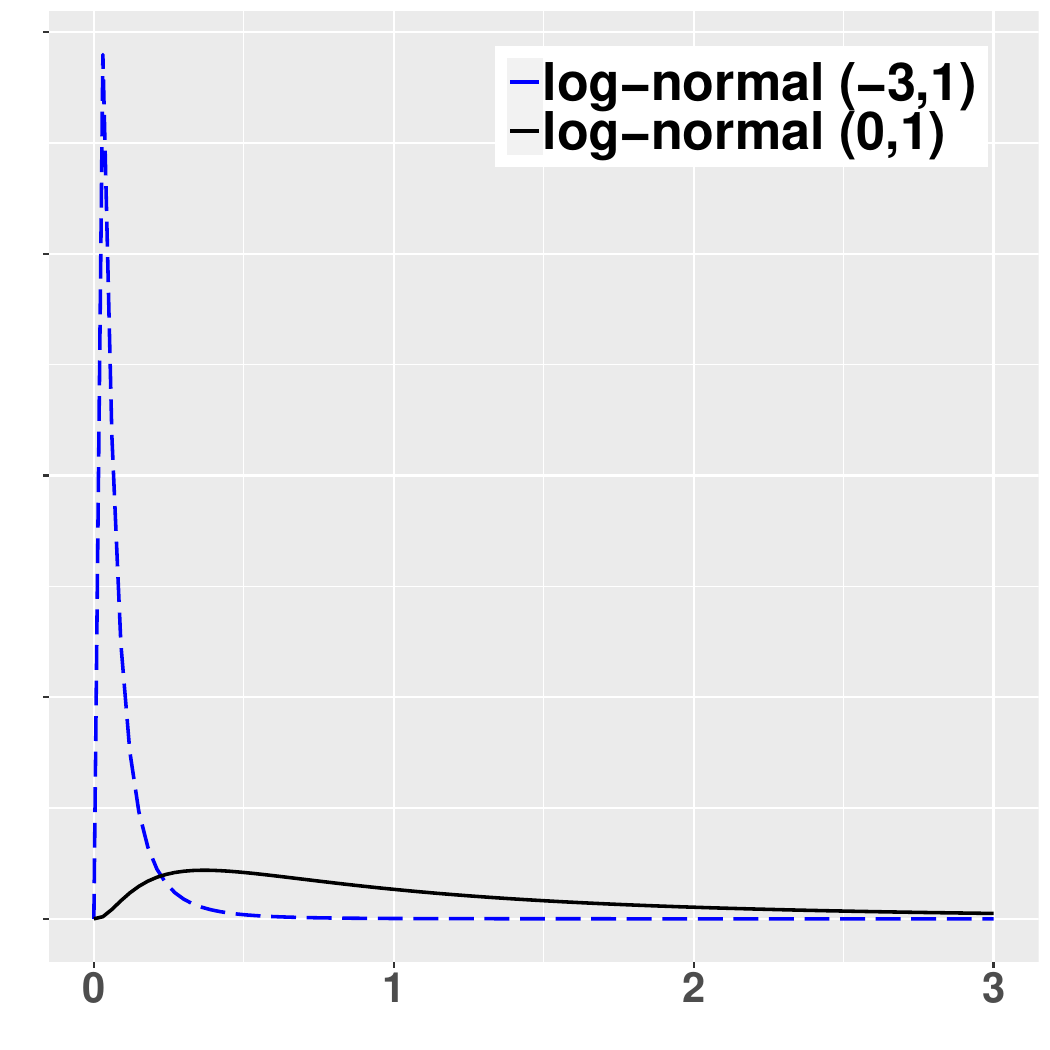}
    \caption{The two-component mixture prior distribution of $\delta$. The blue dashed and black solid lines represent $\mbox{Log-normal} (-3,\ 1 )$ and $\mbox{Log-normal} (0,\ 1 )$, respectively. 
    }
    \label{fig:prior_delta}
\end{figure}

To evaluate the presence and size of the overall social influence, we assign a two-component mixture prior to $\delta$, which is given as:
\begin{equation}\label{eq:delta_prior}
    \delta \sim (1-\pi)\ \mbox{Log-normal} \Big(-3,\ 1 \Big) + \pi \ \mbox{Log-normal} \Big(0, \ 1\Big), 
\end{equation}
where $\pi \in \{0,1\}$, where $\mbox{Log-normal} (-3,\ 1)$ has a mode of 0.02, a mean of 0.08, and a standard deviation of 0.11, and $\mbox{Log-normal} (0,\ 1)$ has a mode of 0.37, a mean of 1.65, and a standard deviation of 2.16 (Figure  \ref{fig:prior_delta} displays the two prior distributions). The first log-normal prior is concentrated near zero, while the second log-normal prior is more spread out on the positive real number line. If social homophily is meaningfully associated with the respondents' activities, we expect $\pi=1$, and otherwise $\pi=0$. We evaluate $\omega = p(\pi = 1) \in [0,1]$ assuming $\omega \sim \mbox{Beta}(1,1)$. We assume there is evidence of social influence when $\omega \geq 0.5$. If social influence is present, we also evaluate the size of social influence based on the $\delta$ estimate.
{If $\omega \leq 0.5$, it suggests that the latent space configuration has little impact on respondents' behaviors, implying that item-level behavioral measures may not show meaningfully distinctive patterns, i.e., little evidence of differential social influence effects.  In this case, one may go with an existing social influence model based on aggregated behavioral measures such as the linear-in-means model, the network autocorrelation model, and the stochastic actor-oriented model.}

\paragraph*{
{Configuration of adapted interaction map: Differential social influence effects}
}

{When we have evidence that overall social influence is present, we interpret the configuration of the interaction map estimated from the model, which can reveal differential social influence effects across item-level behavioral measures.} Specifically, the latent positions of the behavior items and their distances to the respondents help us understand how people with stronger or weaker social connections have endorsed the behavior items differently. Because the positions of the respondents, ${\bf Z}$, have been determined based on their peer network, the respondents with stronger social connections are closer to each other than those with weaker social connections in the interaction map. For example, suppose that two respondents, A and B, were close to each other in an estimated adapted interaction map. And  {suppose} three behavior items 1, 2, and 3 were close to respondents A and B. This means that the two respondents with stronger social connections had a similar likelihood of showing the three specific behavior types.  {In this sense, the configuration of the interaction map from the proposed model presents differential social influence because the social influence is in effect specifically for the three specific behaviors, not necessarily for all behaviors.} 

\begin{figure}[htbp]
    \centering
    \includegraphics[width=0.45\textwidth]{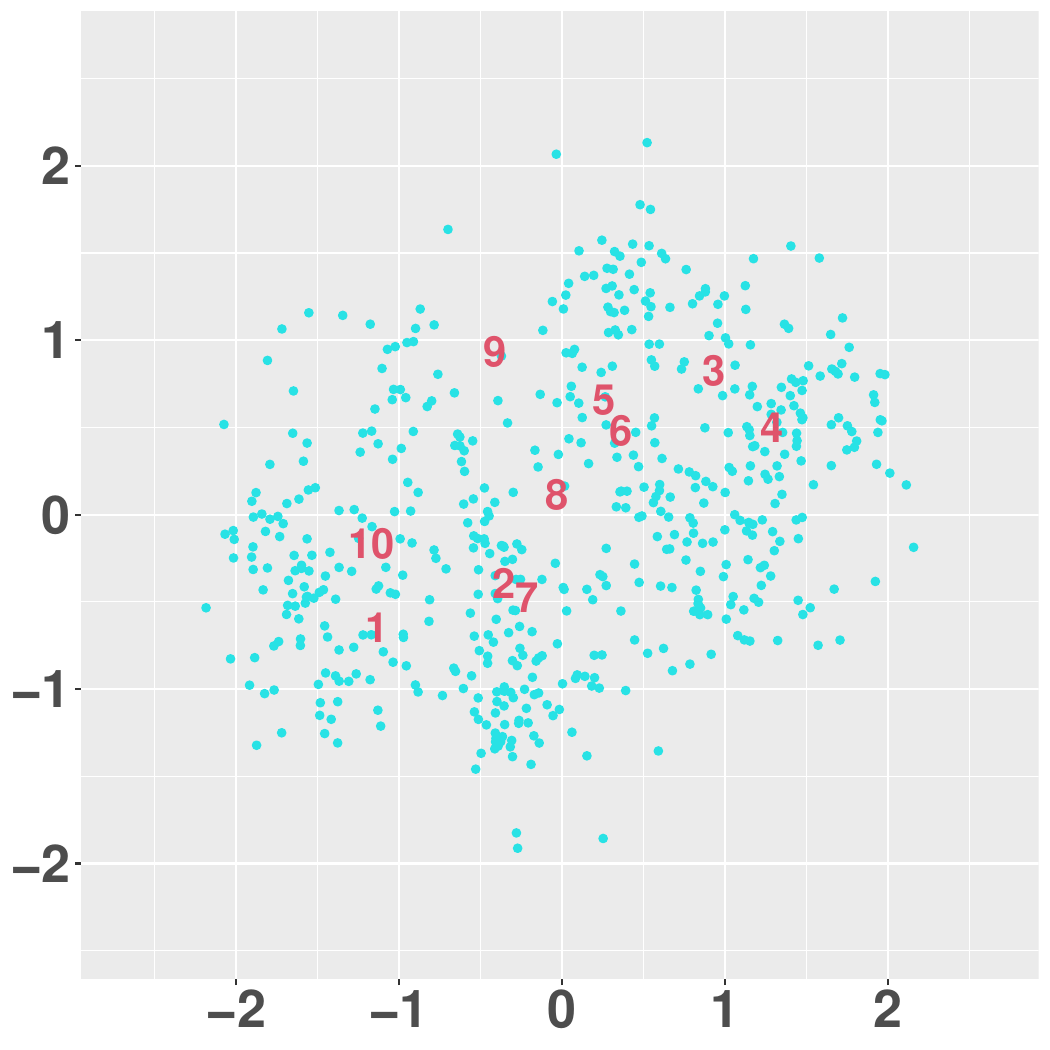}
    \caption{
    An illustration of an adapted interaction map for social influence. Red numbers represent ten school activity items, and colored dots represent 539 US middle school students (same as the data used in Figure \ref{fig:lsm}(a) and (b). 
    }
    \label{fig:lsm2}
\end{figure}

Figure \ref{fig:lsm2} illustrates an example of an interaction map estimated from LSIRM for social influence, based on the same dataset used in Figure \ref{fig:lsm}(a) and (b). The red numbers are ten school activity items, and the colored dots are US middle school students. Student locations and their distances to others represent their close friend network, and the student configuration is the same as the one shown in Figure \ref{fig:lsm}(a). The item locations are somewhat different from the patterns observed in Figure \ref{fig:lsm}(b) because the item locations in Figure \ref{fig:lsm2} reflect the associations between respondents' social homophily and their activities, whereas the item locations in Figure \ref{fig:lsm}(b) reflects the associations/interactions between individual respondents and the activities. 

\subsection{Posterior Distribution and MCMC Estimation}\label{sec:estimation}
We apply a fully Bayesian method for estimating the proposed approach using Markov chain Monte Carlo (MCMC). The parameters of the LSM and the adapted LSIRM for social influence are estimated based on two separate data sets at two different time points. Therefore, we present the posterior distribution of the two models separately here, even though the two sets of model parameters are updated in a single sampler, as shown below. The posterior distribution of the Bayesian LSM for the social network data at time $t$ can be written as follows: 
\begin{equation}\label{eq:si_lsm}
\begin{split}
    \pi\Big(\boldsymbol\Theta^n \mid {\bf Y}^t \Big) &\propto P\Big({\bf Y}^t \mid \boldsymbol\Theta^n \Big) \pi\Big(\alpha\Big) \pi\Big(\gamma\Big) \pi\Big({\bf Z}\Big) \\
    &= \prod_{k \neq l} \frac{\exp\Big(\alpha - \gamma ||{\bf z}_k - {\bf z}_l||\Big)^{y_{kl}^t}}
    {1 + \exp\Big(\alpha - \gamma||{\bf z}_k - {\bf z}_l||\Big)}\pi\Big(\alpha\Big) \pi\Big(\gamma\Big)\pi\Big({\bf Z}\Big),  
\end{split}
\end{equation}
where ${\bf Y}^t $ is $n \times n$ social network data at time $t$ for $n$ respondents, $\boldsymbol\Theta^n = \{ {\bf Z}, \alpha, \gamma \}$ is the parameter space of LSM, and $P\big({\bf Y}^t \mid \boldsymbol\Theta^n \big)$ is the LSM for the social network data ${\bf Y}^t$, which was given in Equation (\ref{eq:lsm}). The prior distributions $\pi\big(\alpha\big)$, $\pi\big(\gamma\big)$, and $\pi\big({\bf Z}\big)$ were provided in Equation (\ref{eq:pr_si_lsm}).

The posterior distribution of the adapted LSIRM for social influence can be written as follows. 
\begin{equation}\label{eq:si_lsirm}
\begin{split}
    \pi\Big( {\boldsymbol\Theta^r}^* \mid {\bf X}^{t+1}, {\bf Z}\Big) &\propto P\Big({\bf X}^{t+1} \mid {\bf Z},  {\boldsymbol\Theta^r}^*\Big) \pi\Big(\boldsymbol\beta\Big) \pi\Big(\boldsymbol\theta \mid \sigma^2\Big) \pi\Big(\sigma^2\Big) \pi\Big({\bf W}\Big) \pi\Big(\delta\Big)\\ 
    &= \prod_{k=1}^n \prod_{i=1}^p \frac{\exp\Big(\beta_i + \theta_k - \delta||{\bf z}_k - {\bf w}_i||\Big)^{x_{ki}^{t+1}}}
    {1 + \exp\Big(\beta_i + \theta_k - \delta||{\bf z}_k - {\bf w}_i||\Big)}
    \pi\Big(\boldsymbol\beta\Big) \pi\Big(\boldsymbol\theta \mid \sigma^2\Big) \pi\Big(\sigma^2\Big) \pi\Big({\bf W}\Big) \pi\Big(\delta\Big), 
\end{split}
\end{equation}
\noindent where ${\bf X}^{t+1}$ is the $n \times p$ item response data for $n$ respondents to $p$ behavior items at time $t+1$, ${\boldsymbol\Theta^r}^* =\{\boldsymbol\beta, \boldsymbol\theta, {\bf W}, \delta, \sigma^2\} $ is the parameter space for the adapted LSIRM, $P\big({\bf X}^{t+1} \mid {\bf Z}, {\boldsymbol\Theta^r}^*\big)$ is the adapted LSIRM given ${\bf Z}$, the respondent latent positions obtained from LSM. The prior distributions for $\beta_i$, $\theta_k$, $\sigma^2$, and ${\bf{w}}_i$ were given in Equation (\ref{eq:pr_si_lsirm}) and the prior distribution for $\delta$ was given in Equation (\ref{eq:delta_prior}). 

The MCMC sampler involves updating all parameters of the LSM and then updating all parameters of the adapted LSIRM given {\bf Z} from the LSM. {The details of MCMC sampling are given in Supplementary Material 1.} For all data analysis provided in the current paper, we use the following values for the priors: $\sigma_{\alpha} = \sigma_{\beta} = 2.5$,$\mu_{\gamma} = 0.0$, $\sigma_{\gamma} = 1.0$ and $a_{\sigma} = b_{\sigma} = 0.001$. For proposals, Gaussian proposal distributions are used, which are centered at the current values of the parameters of the latent positions. The variances of the proposal distributions are set to have reasonable acceptance rates (around 0.2 - 0.5). To evaluate the convergence of the MCMC algorithm, we use trace plots along with Gelman-Rubin diagnostics \citep{Gelman:1992}.

\subsection{Other Topics}\label{sec:other_topic}

\paragraph*{Identifiability}

The LSIRM for social influence has the same identifiability issue for latent positions as other latent space models as distances are invariant to translations, rotations, and reflections \citep{Hoff:2002}. We addressed the latent positions' identifiability issue by post-processing the MCMC samples with Procrustes matching \citep{gower_generalized_1975}, which is a commonly applied method for addressing the identifiability issue above in the literature \citep{Hoff:2002, jeon:2020}. We applied Procrustes matching to the MCMC samples, and all Bayesian inferences were made based on the posterior samples obtained after Procrustes matching. 

The proposed model has no other identifiability issue. For example, the $\delta$ parameter is identifiable by fixing the scale of the latent position prior: in our approach, we set ${\bf w}_i \sim N({\bf 0}, \sigma^2{\bf I})$ where $\sigma=1$. The $\delta$ parameter plays the same role as the scale parameter $\sigma$ in standard latent space models. Since $\sigma$ is fixed, the $\delta$ parameter is estimable. The weights of latent spaces have been estimated in a similar way in the literature \citep[e.g.,][]{Handcock:2007, jeon:2020}. 

\paragraph*{Latent Space Dimension}

We choose 2-dimensional latent space in this study. In the latent space modeling literature, 2-dimensions are routinely selected for the ease of visualization and interpretation, and we followed that convention. One may consider using relative model fit statistics such as BIC to determine the dimension of the latent space; however, such an approach has not yet been rigorously justified in the literature. Showing the adequacy of using BIC for dimension decisions in latent space modeling is beyond the scope of the current study. 

\paragraph*{Model Fit Evaluation} 

We evaluate the goodness-of-fit (GOF) of the proposed model using the method proposed by \citet{Hunter:2008}. This is a commonly used method for GOF evaluations for social network models. The idea is to compare the percentage of item $i$ answered positively, $p_{\cdot i}$ from the original data with the probability of positively responding item $i$, $\hat{{\bf p}}_{\cdot i}$, estimated from the model. 
Specifically, 
\begin{enumerate}
\item Calculate 
$p_{\cdot i} = \frac{1}{n} \sum_{k=1}^n x_{ki}.$
where $x_{ki}$ is the response of the respondent $k$ to the item $i$ and $n$ is the total number of respondents.
\item Calculate $\hat{{\bf p}}_{\cdot i}$ 
from each MCMC sample where 
\[\hat{p}_{\cdot i}^{(l)} = \frac{1}{n}\sum_{k=1}^n\hat{p}_{ki}^{(l)}, \qquad \hat{p}_{ki}^{(l)} = \frac{\exp\Big(\beta_i^{(l)} + \theta_k^{(l)} - \delta^{(l)}||{\bf z}_k^{(l)} - {\bf w}_i^{(l)}||\Big)}{1 + \exp\Big(\beta_i^{(l)} + \theta_k^{(l)} - \delta^{(l)}||{\bf z}_k^{(l)} - {\bf w}_i^{(l)}||\Big)}.
\]
where $\boldsymbol\beta^{(l)}$, $\boldsymbol\theta^{(l)}$, $\delta^{(l)}$, ${\bf Z}^{(l)}$, and ${\bf W}^{(l)}$ are MCMC samples at the iteration of $l$.
\end{enumerate}
We then evaluate whether $\hat{{\bf p}}_{\cdot i}$ can sufficiently capture $p_{\cdot i}$ from the original data. We also compute posterior predictive p-values, described in detail in Chapter 6 of \citet{Gelman:2013}. Posterior predictive p-values closes to 0.5 indicate a reasonable fit of the model to the data. 

\section{Real Data Application}\label{sec:application}

\subsection{Data}

For an empirical illustration of the proposed approach, we used data from the ``Changing Climates of Conflict'' study \citep{Paluck:2016, Paluck:2020}. This is a large-scale field study designed to measure the effects of anti-conflict interventions across 56 New Jersey public middle schools with 24,191 students in 2012 and 2013. To illustrate the proposed approach, we analyzed data from all schools, while selecting four schools for illustration purposes. The four schools were selected considering students' racial and economic backgrounds in the schools. For example, we chose School 36 with the highest percentage of white students (83\% of 533 students) and School 45 with the highest percentage of Hispanic and African American students (82\% of 295 students). In addition, we included School 59 with the second-lowest percentage of students receiving free or reduced lunch (11\% of 538 students) and School 44 with the highest percentage of students receiving free or reduced lunch (69\% of 228 students).\footnote{School 36 showed the lowest percentage of students receiving free or reduced lunch. Since School 36 was already selected, we chose School 44, the next one.} We provided the data analysis results for the other 52 schools in Supplementary Material 2. {The data are available on the ICPSR website {\tt https://www.icpsr.umich.edu/web/ICPSR/studies/37070}.}

Students were asked to write down their friends (up to ten) and best friends (up to two) in their schools in the fall of 2012. We constructed a binary adjacency matrix per school from both their friends and best friends within a school; we refer to them \textit{close friends} in this paper. Table \ref{tab:network_summary} summarizes some characteristics of the schools selected for data analysis, including network summary statistics and average student background information.

\begin{table}[htpb]
\centering
\begin{tabular}{rrrrrrrrr}  \\\hline 
 & Nodes &  Edges & Male & White & H/AA  & F/R  & LEP & CEN \\ \hline
School 36 & 533 & 3120 & 0.52 & 0.83 & 0.07 & 0.08 & 0.01 & 0.30\\ 
School 45 & 295 & 1005 & 0.51 & 0.04 & 0.82 & 0.65 & 0.04 & 0.25\\ 
School 44 & 228 & 977 & 0.56 & 0.08 & 0.67 & 0.69 & 0.06 & 0.32\\ 
School 59 & 538 & 3914 & 0.48 & 0.74 & 0.11 & 0.11 & 0.02 & 0.43\\ \hline
\end{tabular}
\caption{\label{tab:network_summary}
Some background characteristics of the four schools. Nodes (number of nodes (students)), edges (number of edges), male, White, H/AA (Hispanic and African American), F/R (free or reduced lunch), LEP (limited English proficiency) are the proportions in the students of the corresponding categories. CEN is the mean of eigenvector centrality, a measure of the influence of a node on a network. 
}
\end{table}

In addition, the students were asked to indicate which school activities they participated in among the following ten activities in the spring of 2013: (1) sports within schools, (2) sports outside schools, (3) theater or drama, (4) music, (5) arts, (6) clubs in other schools, (7) dating within schools, (8) homework, (9) reading, and (10) video games. {These are students' school participation/activity behaviors. That is, the construct being measured here is students' overall levels of school participation/activity. Existing scales utilize similar behaviors to measure students' school participation/activity construct, e.g., the Scale of Involvement in Extracurricular Activities for Children (SIEAC) \citep{FernandesMatias2019}. In addition, many studies have used similar school participation/activity behaviors in a wide range of applied research \citep[e.g.,][]{Barber2001, Eccles2003, Mahoney2003, Fredricks2006, Feldman2005, Larson2006, Gardner2008, Knifsend2011, McCabe2016}. 
}

\begin{figure}[htbp]
    \centering
    \begin{tabular}{cccc}
    (a) School 36 & (b) School 45 & (c) School 44 & (d) School 59 \\
    \includegraphics[width=0.225\textwidth]{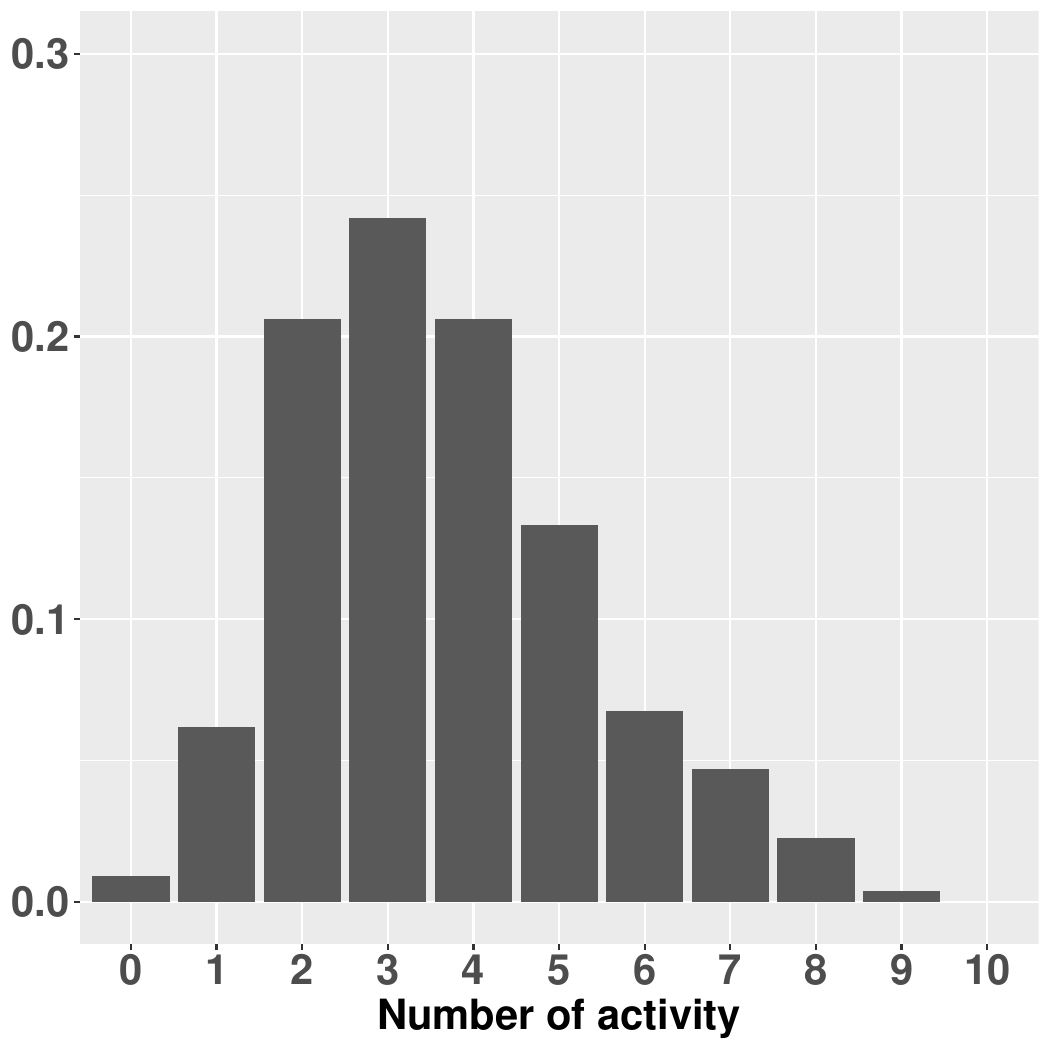} & 
    \includegraphics[width=0.225\textwidth]{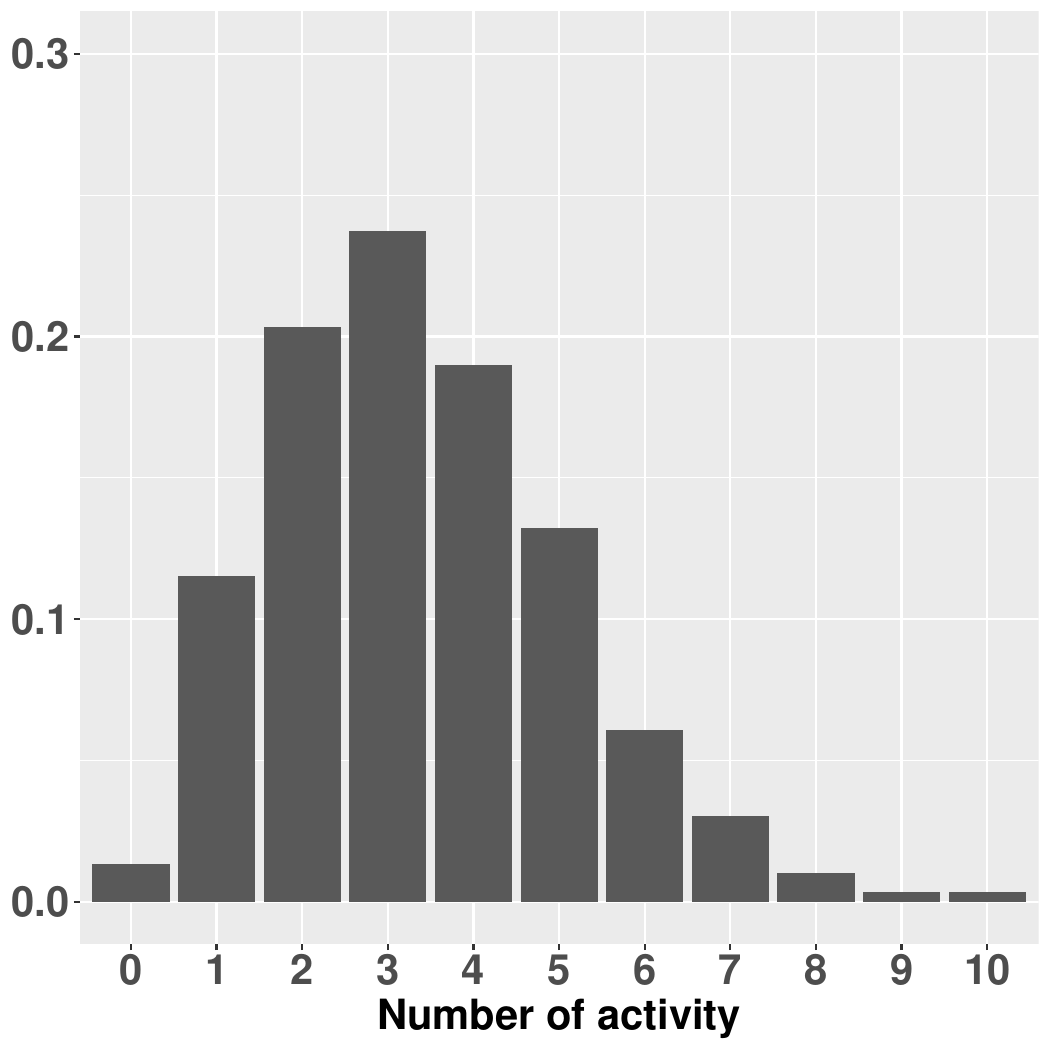} &
    \includegraphics[width=0.225\textwidth]{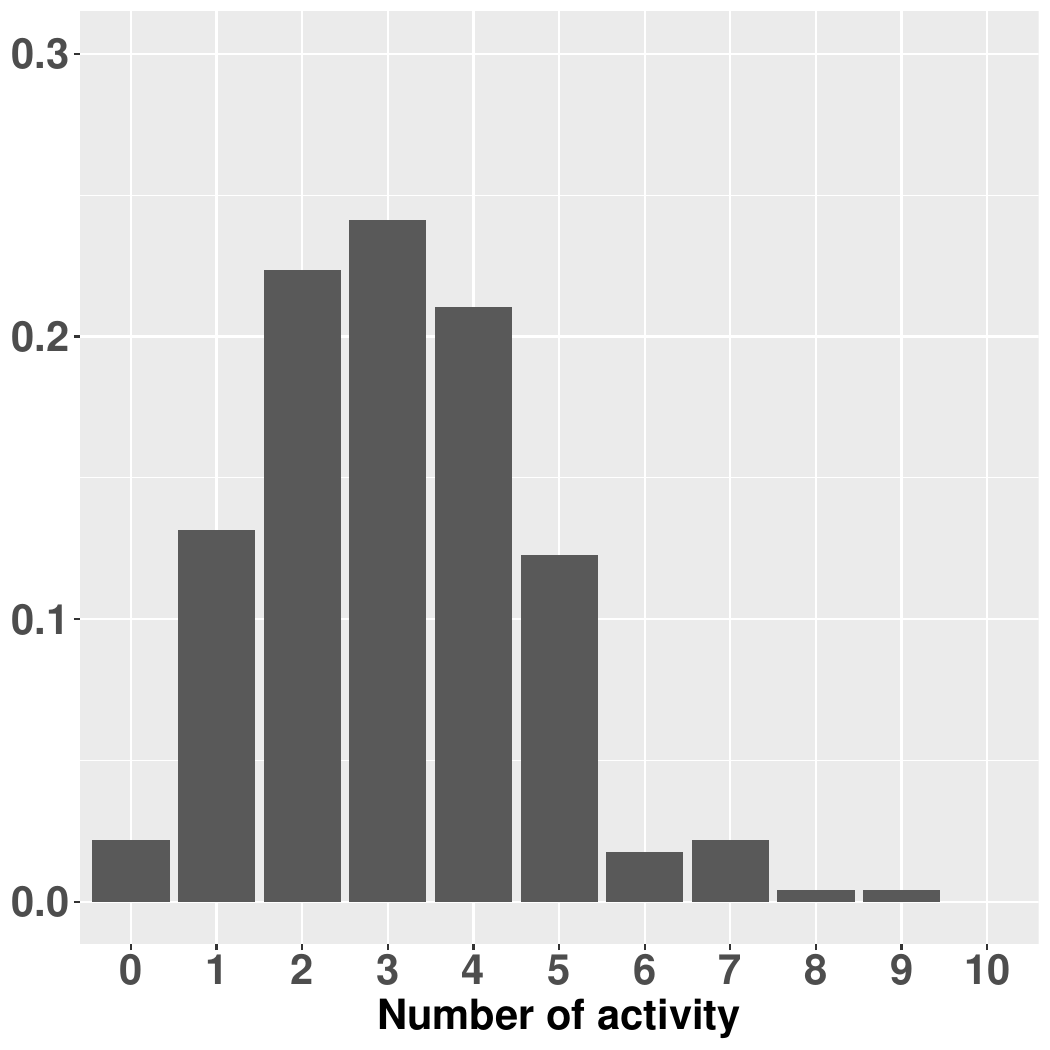} & 
    \includegraphics[width=0.225\textwidth]{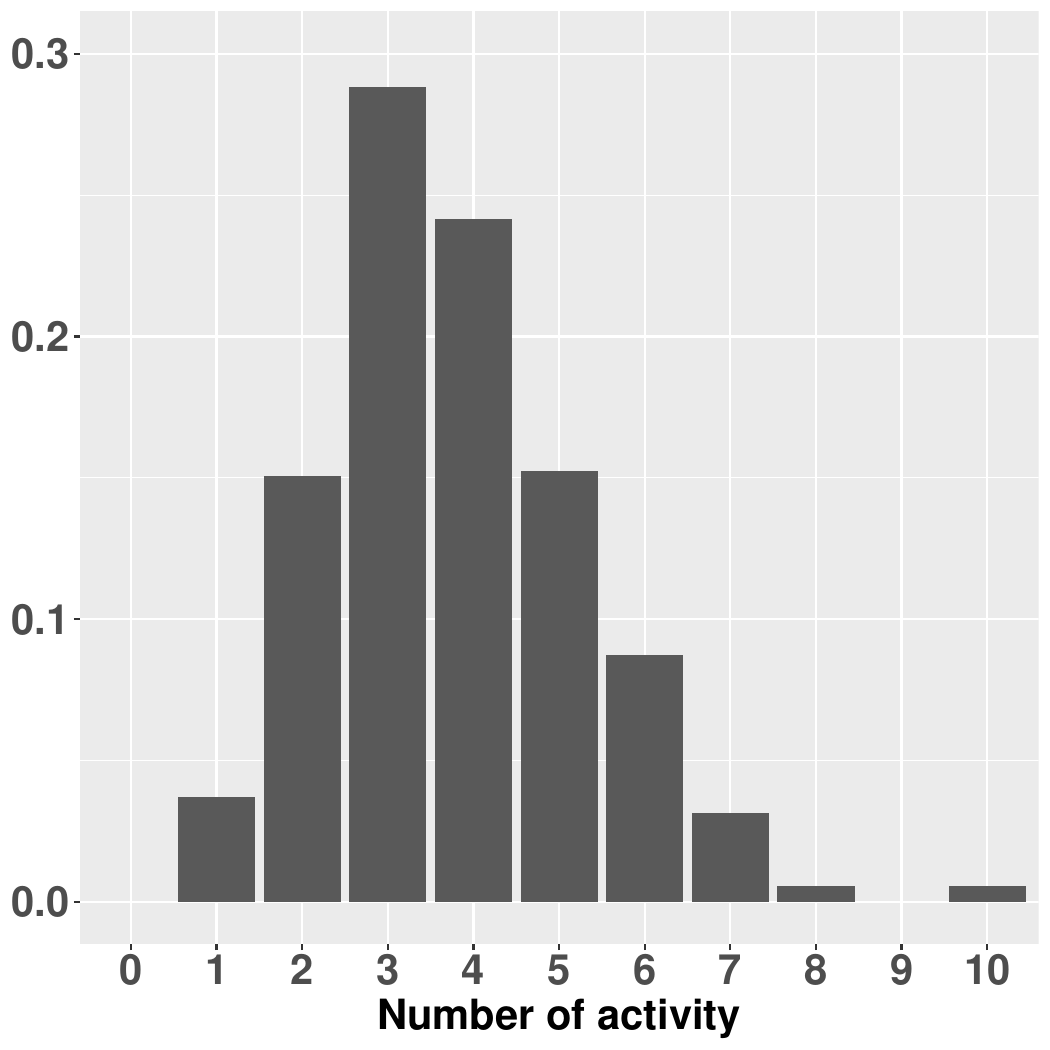} 
    \end{tabular}
    \caption{\label{fig:num_activity}
    The histograms of the total number of activities that students were participating in.}
\end{figure}

Figure \ref{fig:num_activity} presents the histogram of the total number of activities reported in the student response data per school. The distributions of the total number of school activities are fairly similar across the four schools. In all schools, two to four activities were most frequently observed. Table \ref{tab:activity_prop} shows the proportion of students who participated in each school activity per school. Overall, Activity 2 (sports outside schools) and Activity 10 (video games) were the most frequently participated activities, while Activity 3 (theater/drama) and Activity 6 (clubs outside schools) were the least frequently participated activities in the four schools. 

\begin{table}[htbp]
\centering
\begin{tabular}{rrrrrrrrrrr} \hline
Activity & 1 & 2 & 3 & 4 & 5 & 6 & 7 & 8 & 9 & 10 \\ \hline
School 36 & 24.20 & 75.42 & 10.88 & 43.90 & 20.45 & 17.64 & 41.65 & 40.53 & 23.45 & 65.85 \\ 
School 45 & 19.32 & 49.15 & 14.24 & 49.83 & 23.73 & 10.17 & 32.88 & 23.05 & 29.83 & 85.76 \\ 
School 44 & 29.39 & 59.65 & 6.14 & 36.84 & 17.98 & 4.39 & 29.39 & 19.30 & 22.37 & 83.77 \\ 
School 59 & 52.04 & 76.02 & 8.55 & 37.17 & 13.38 & 15.80 & 31.78 & 60.97 & 18.03 & 63.94 \\ \hline
\end{tabular}

\caption{\label{tab:activity_prop}
Percentage of the student participants in each activity per school. Activity 1 to 10 indicate participation in: (1) sports in school, (2) sports outside school, (3) theater/drama, (4) music, (5) arts, (6) clubs in other schools, (7) dating in school, (8) homework, (9) reading, and (10) video games. 
}
\end{table}

\subsection{Estimation} 

We apply the proposed approach to each school's data to evaluate how their close friend network influenced their school activities. The MCMC algorithm was run for 30,000 iterations for each school dataset, with the first 5,000 iterations being discarded as a burn-in process. From the remaining 25,000 iterations, 5,000 samples were collected at a time-space of 5 iterations. Jumping rules for the proposal distributions are provided in Supplementary Material 1. 

\subsection{Comparison with Existing Models} \label{sec:comp}

We consider three existing models as a comparison with the proposed approach: {the linear network autocorrelation model \citep{Ord1975}, the linear-in-means model \citep{Manski1993}, and 
the stochastic actor-oriented model (SAOM) \citep{Snijders2001}.} 

First, the network autocorrelation regression model was set up with the counts of positive responses to school activity items, representing the target behavior of interest, as dependent variables, and an adjacency matrix of a peer network as a weight for measuring the autocorrelation. Suppose that for the binary behavior response data ${\bf X}=\{x_{ki}\}$, we define $S_k = \sum_i x_{ki}$. Then we have
\[
    S_k = \rho \sum_{l \neq k} y_{kl} S_l + \epsilon_k, \qquad \epsilon_k \sim N\Big(0, \sigma^2\Big),
\]
where $\rho$ is a network autocorrelation parameter that estimates the autoregression of each $S_k$ value on its neighbors in the peer network, representing social influence. 
The network autocorrelation model was estimated using the {\tt lnam} function in the {\tt sna} package in R. 

Second, {in the linear-in-means model, $S_k$ (the counts of positive responses to the school activity items) is used as the target behavior.} We additionally assume that each student's responses linearly depend on the mean response of their friends because the respondent's covariates information is not included in our application data. Then, the model is given by:
\begin{equation*}
    S_k = \alpha + \beta_{\bar{y}}\frac{1}{M_k}\sum_{l=1}^{N}y_{kl}S_l + \epsilon_k , \qquad \epsilon_k \sim N\Big(0, \sigma^2\Big),
\end{equation*}
where $\beta_{\bar{y}}$ indicates the endogenous social influence effect and $M_k$ is the number of friends of the respondents $k$. 

{
Third, the SAOM models the co-evolution of changes in the network ties and individual behaviors, although the changes are modeled from the point of view of the individuals (i.e., actors). 
The SAOM also uses an aggregated measure (i.e., the count of positive responses to the school activity items) as the target behavioral variable.  
Specifically, to measure social influences,  an average similarity statistic \citep{Steglich2010} is utilized in the SAOM, which is given as 
\[
s_{\text{similarity}, i}\left({\bf y}, {\bf x}\right) = \frac{1}{\sum_{h}y_{ih}} \sum_j y_{ij}(\text{sim} (x_i, x_j)- \overline{\text{sim}}),\\
\]
where $\text{sim} (x_i, x_j) =  1 - \frac{|x_i - x_j|}{\Delta_x}$ with $\Delta_x = \max_{ij} |x_i - x_j|$. Here, the coefficient of the similarity statistic, $\beta_{\text{similarity}}$, indicates the propensity of actors to be similar to their peers in the behavior of interest (i.e., social influence). The SAOM was estimated using the {\tt RSiena} package in R \citep{Ruth2022}.  Note that to estimate social influence parameters using {\tt RSiena}, both network and item response datasets from all time points are necessary. In other words, the SAOM is not applicable when one set of measures is available for network and behavior (i.e., network measures at time $t$ and behavioral measures at time $t+1$), different from all other three models in comparison.   
}

{Note that social influence parameters in these models may not be directly comparable to our model  due to the differences in how social influence is defined and set up in these models.} For example, the network autocorrelation model, the linear-in-means model{, and the SAOM} take univariate data as input; therefore, aggregated measures (the number of positive responses; count data) are generated for the behavior data input for these models. In contrast, the proposed model takes the original item-level data (binary response data) as behavioral data input. 
In addition, the network autocorrelation model, the linear-in-means model{,  and the SAOM} use observed social networks to estimate social influence, whereas we use latent social homophily estimated from social network data in estimating social influence. 
These differences can lead to different results for social influence. {Most importantly, we would like to highlight that while these models in the comparison report only the overall social influence effects, we additionally provide information on differential social influence across item-level behavioral measures, through the interaction map that reveals the associations between latent social homophily and specific behaviors. None of the existing methods considered above provides information on differential social influence effects as they employ aggregated behavioral measures, not paying attention to item-level behaviors. 
}

\subsection{Results}

\paragraph*{Overall Social Influence}

Table \ref{tab:application} lists the overall social influence parameter estimates from the four models, the proposed model, the network autocorrelation model, the linear-in-means model, 
{
and the stochastic actor-oriented model (SAOM). For the proposed model, we listed the posterior means of $\delta$, and the $\omega$ values (posterior probability of having overall social influence). 
}

\begin{table}[htpb]
\centering
\footnotesize{
{
\begin{tabular}{ccccccccccc} \hline
& \multicolumn{2}{c}{\bf Proposed model} & \multicolumn{2}{c}{\bf SAOM} &
\multicolumn{3}{c}{\bf Linear-in-means model} & \multicolumn{3}{c}{\bf Autocorrelation model} \\
School & $\hat{\delta}$  & $\omega$ & Estimate & P-value & Estimate & \multicolumn{2}{c}{95\% CI} & Estimate & \multicolumn{2}{c}{95\% CI}\\ \hline
36 & 0.349 & 0.781 &  3.288 & 0.0165 &  0.180 & (-0.053, & 0.412) & 0.069 & (0.066, & 0.073) \\ 
45 & 0.055 & 0.042 & -0.016 & 0.9901 & -0.064 & (-0.329, & 0.200) & 0.096 & (0.087, & 0.105) \\ 
44 & 0.067 & 0.067 &  1.096 & 0.5258 & -0.122 & (-0.427, & 0.183) & 0.085 & (0.076, & 0.093) \\ 
59 & 0.149 & 0.268 &  1.776 & 0.2469 &  0.017 & (-0.255, & 0.289) & 0.058 & (0.055, & 0.060) \\ \hline
\end{tabular}
} }
{
\caption{ \label{tab:application}
The posterior mean of the overall influence parameters, and the probability of having social influence $\omega$ for Schools 36, 45, 44, and 59. The estimated social influence parameters, stochastic actor-oriented model, and credible interval from the linear-in-means model and network autocorrelation model.}
}
\end{table}

Based on the proposed approach, about 32\% schools (18 schools) presented meaningful social influence among the 56 schools with $\omega >0.5$ (detailed data analysis results are provided in Supplementary Material 2). Among the four select schools, School 36 showed evidence of meaningful social influence ($\omega =0.781$), while insufficient evidence for social influence was found for the other schools. The posterior mean of $\delta$ is 0.349, which is 1.418 on the odds ratio scale. 
{
The SAOM reports that School 36 shows significant overall social influence effects, similar to the proposed approach. The linear-in-means model reports that social influence is trivial in all four schools, although it is marginally significant in School 36.  The network autocorrelation model reports that all four schools have a minor but significant social influence. 
}

\begin{figure}[htbp]
    \centering
    \begin{tabular}{cc}
    (a) & (b)\\
    \includegraphics[width=0.4\textwidth]{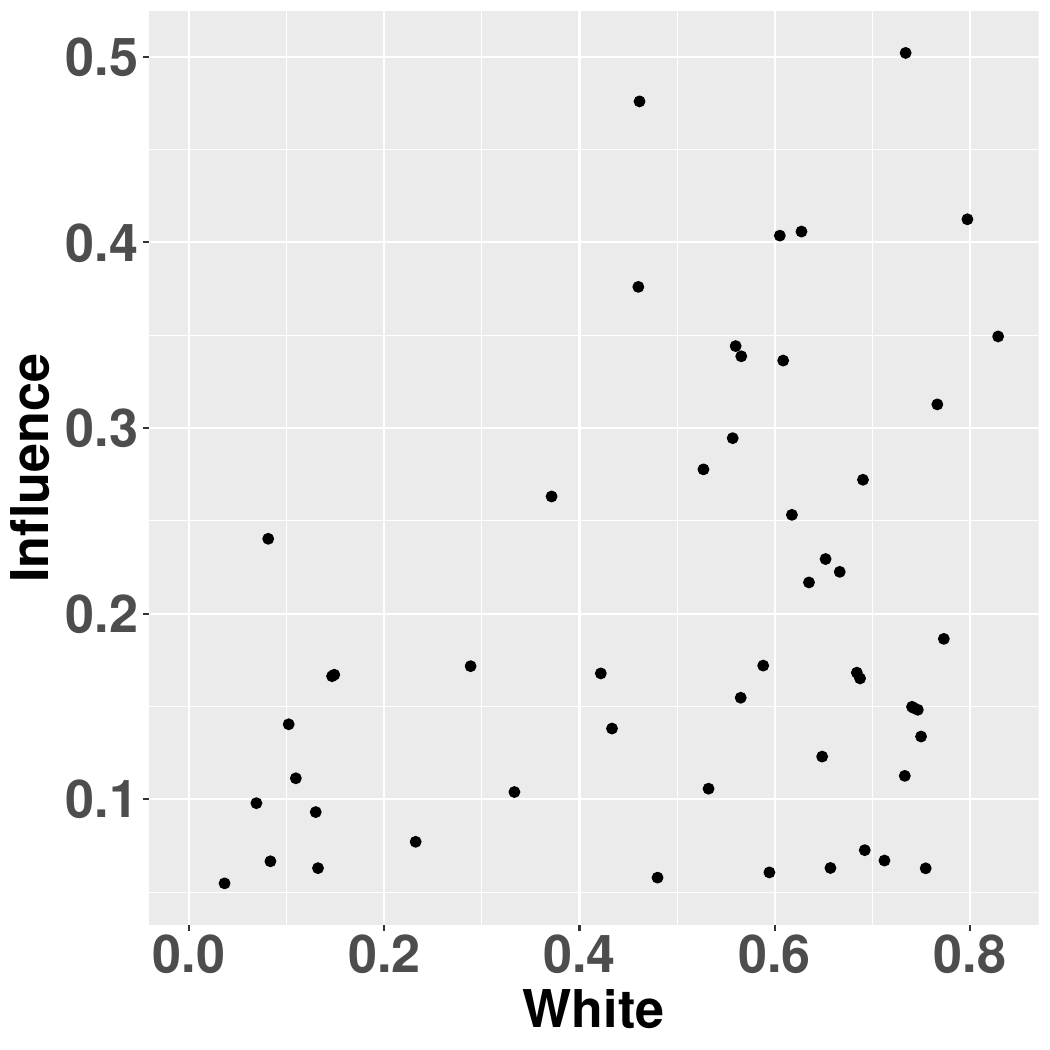} & 
    \includegraphics[width=0.4\textwidth]{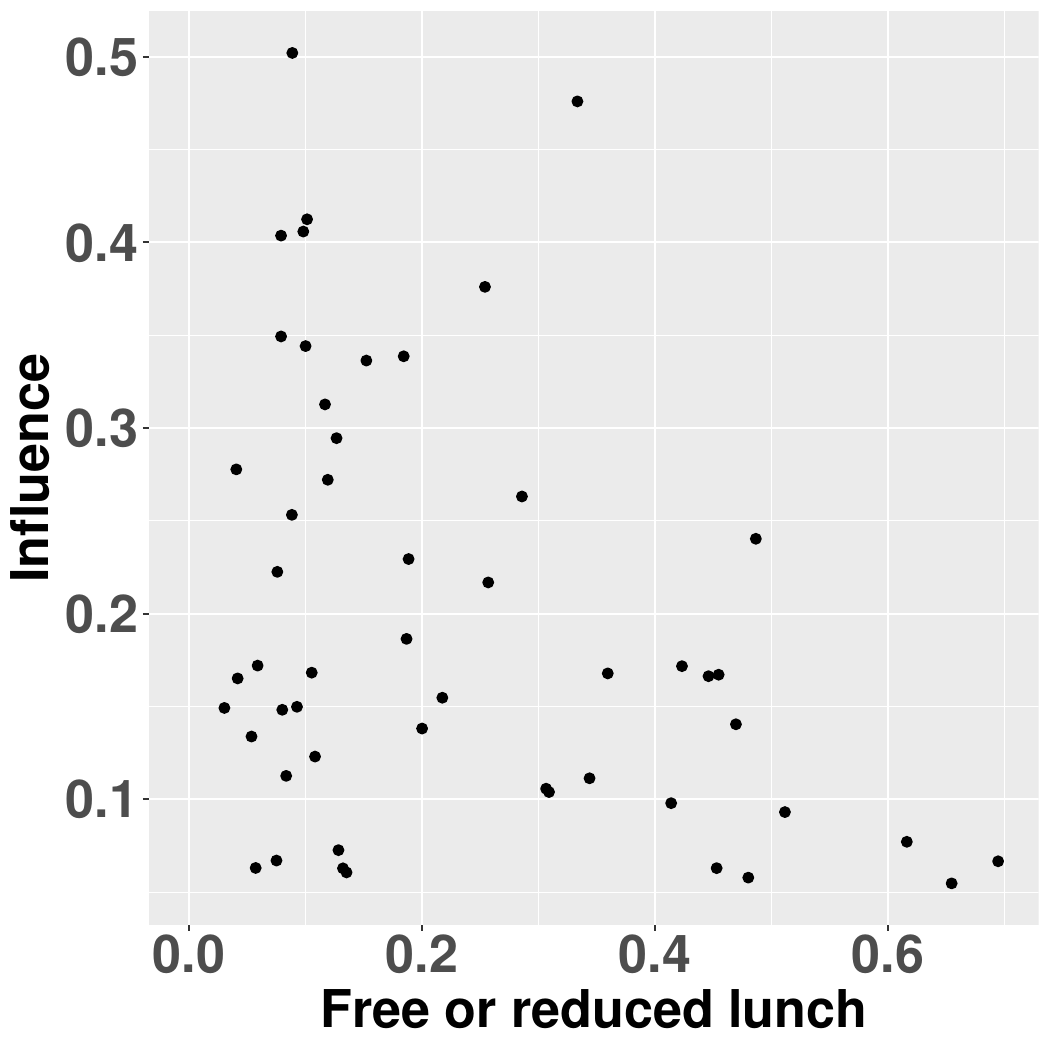} \\ 
    \end{tabular}
    \caption{\label{fig:scatter_delta}
    (a) Scatter plot between the proportion of white students and the posterior medians of the social influence parameter $\delta$ across the 56 schools.
    (b) Scatter plot between the proportion of low-income students (who received free or reduced lunch) and the posterior medians of the social influence parameter $\delta$ across the 56 schools. }
\end{figure}

We note that School 36 is the school with the highest percentage of white students among the 56 schools. Thus, we explore potential relationships between social influence estimates and schools’ background characteristics, such as the proportion of white students and low-income students. Figure \ref{fig:scatter_delta} displays the scatter plots with the two background variables. The Pearson correlation is 0.3176 and -0.3530 for Figure \ref{fig:scatter_delta}(a) and Figure \ref{fig:scatter_delta}(b), respectively. That is, we found minimal evidence that a high proportion of white students and a low proportion of low-income students might be related to a higher level of social influence in the data, meaning that the students' peer relationships influenced the types of school activities (considered in the data) that the students participated in. 

\paragraph*{{Interaction Map: Differential social Influence}}

\begin{figure}[htbp]
    \centering
    \begin{tabular}{c}
    \includegraphics[width=0.45\textwidth]{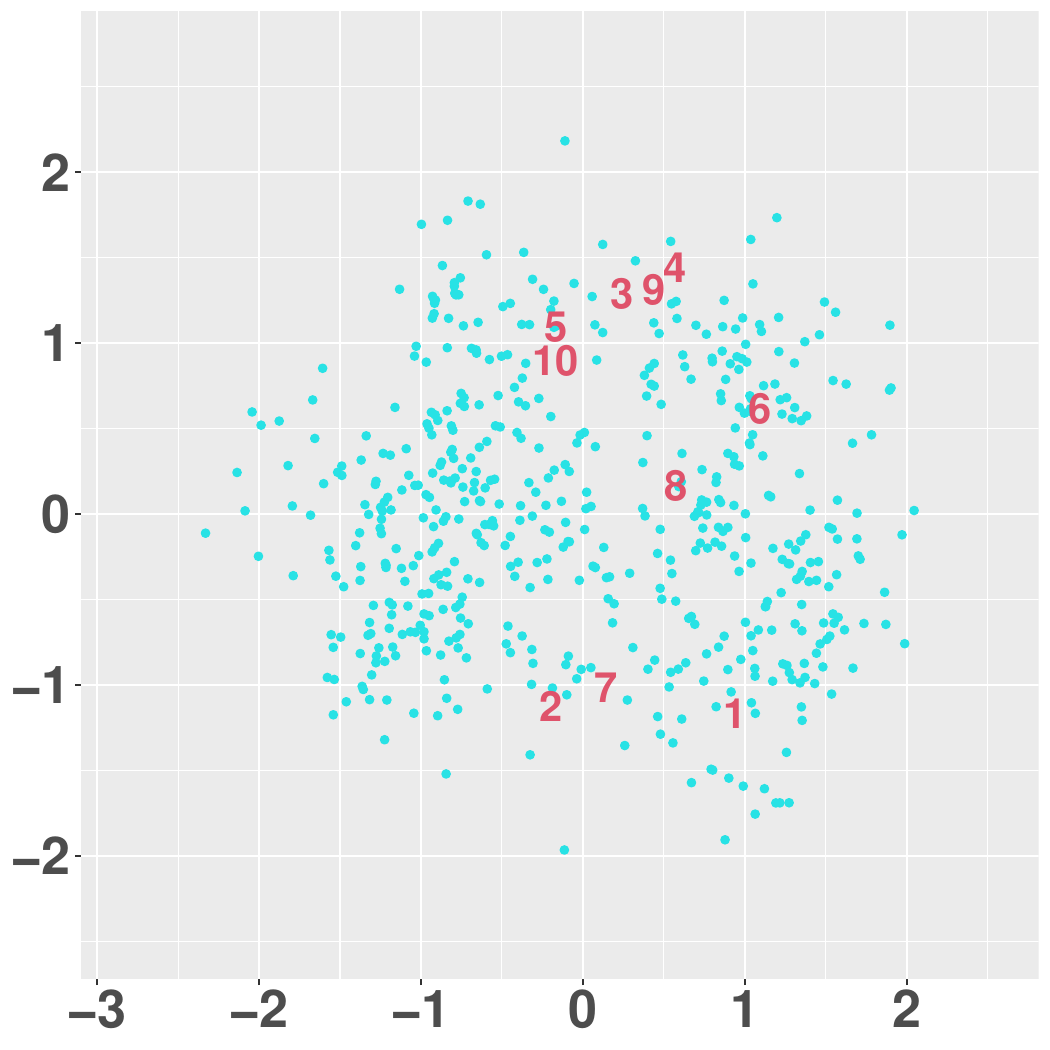} \\
    \end{tabular}
    \caption{\label{fig:interaction_lunch_race}
    Interaction map for School 36}
\end{figure}

Figure \ref{fig:interaction_lunch_race} displays the estimated interaction map of School 36 from the proposed approach, which is the only school with significant overall social influence effects among the four selected schools. The blue dots indicate the latent positions of students reflecting their close friend relationships, or social homophily, within the school. The red numbers indicate the latent positions of the ten school activities. A short distance between a student and an activity indicates a higher probability of endorsing the activity for the student. Items that are close to each other on the map are similar in the sense that they are commonly shown activities by a particular group of students who are friends with each other. 

{
The map shows roughly three clusters of activities: Activities 1 (sports in school), 2 (sports outside school), and 7 (dating) form a cluster, positioned at the bottom of the space. Activities 3 (theater/drama), 4 (music), 5 (arts), 9 (reading), and 10 (video games) make another cluster located at the top of the space make another group. Activity 6 (clubs at other schools) and Activity 8 (homework) make a third cluster, positioned between the two other item clusters. It is interesting that the first and second item clusters consist of physical (or more active) vs. non-physical activities, respectively, while the third item cluster is mixed in terms of that criterion. This suggests that those students near physical activity items increase the likelihood of participating in physical activity due to peer influence. Similarly, those students close to non-physical activity items increase the tendency of engaging in non-physical activity under the influence of friends. That is, the result suggests that who they were friends with in the fall of 2021 impacted the kind of activities (physical or less physical) the students were more likely to show in the spring of 2013 in this particular school, which is evidence of differential social influence effects. 
}

{
Note that the interaction map of social influence may not necessarily match the prevalence of activities in the data because the latent space configuration represents conditional dependencies that are not explained by the person and item intercept parameters in the model. For example, Activity 2 in Figure \ref{fig:interaction_lunch_race} is located in the far bottom area of the map, not necessarily being close to most respondents, even though the activity's prevalence rate is very high. Since students in a friendship relationship are closely located in the latent space of social influence, a friend group being close to an activity indicates that the friend group is likely to participate in the activity given their overall activity levels.
}

\paragraph*{Other Parameter Estimates}

\begin{figure}[htbp]
    \centering
    \begin{tabular}{ccc}
    (a) $\alpha$ & (b) $\gamma$ & (c) $\beta_i$\\ 
    \includegraphics[width=0.3\textwidth]{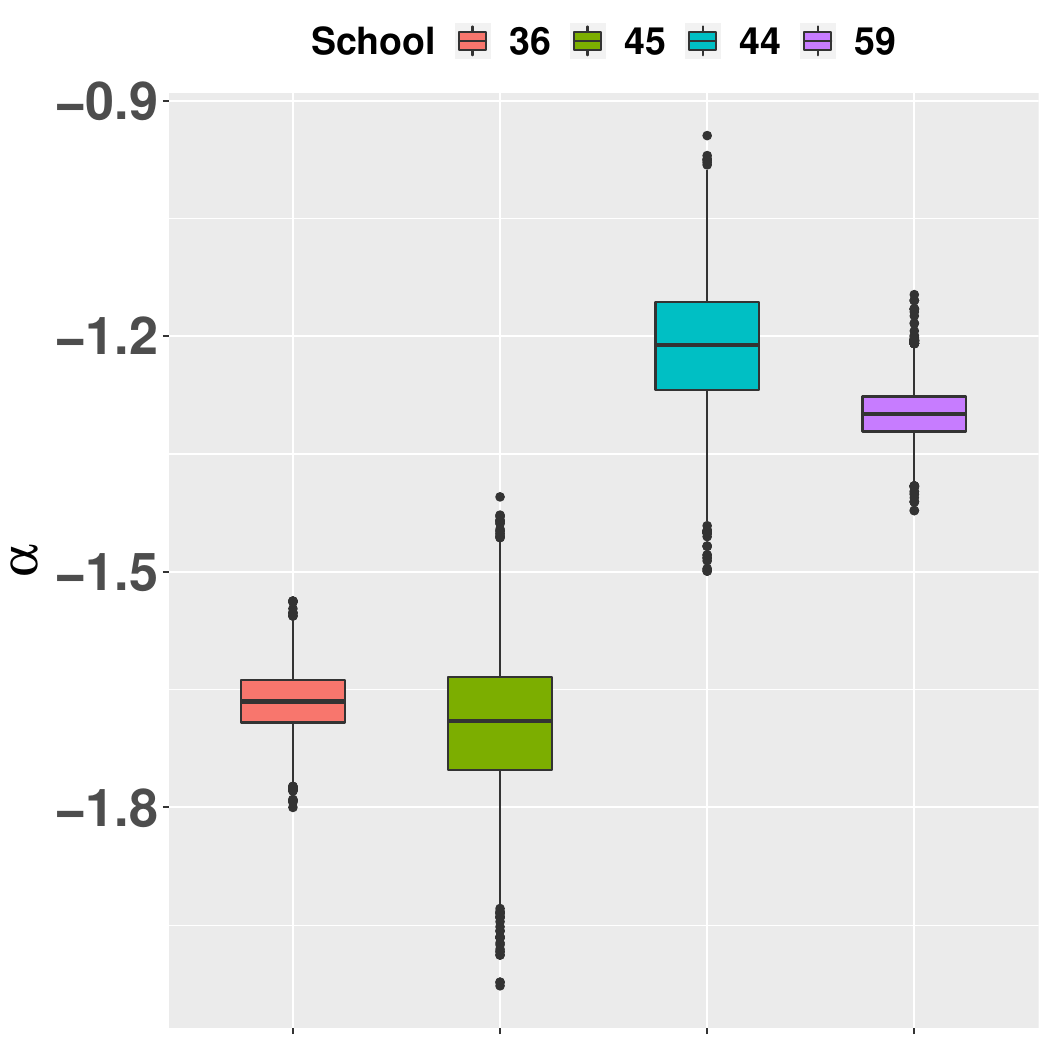} & 
    \includegraphics[width=0.3\textwidth]{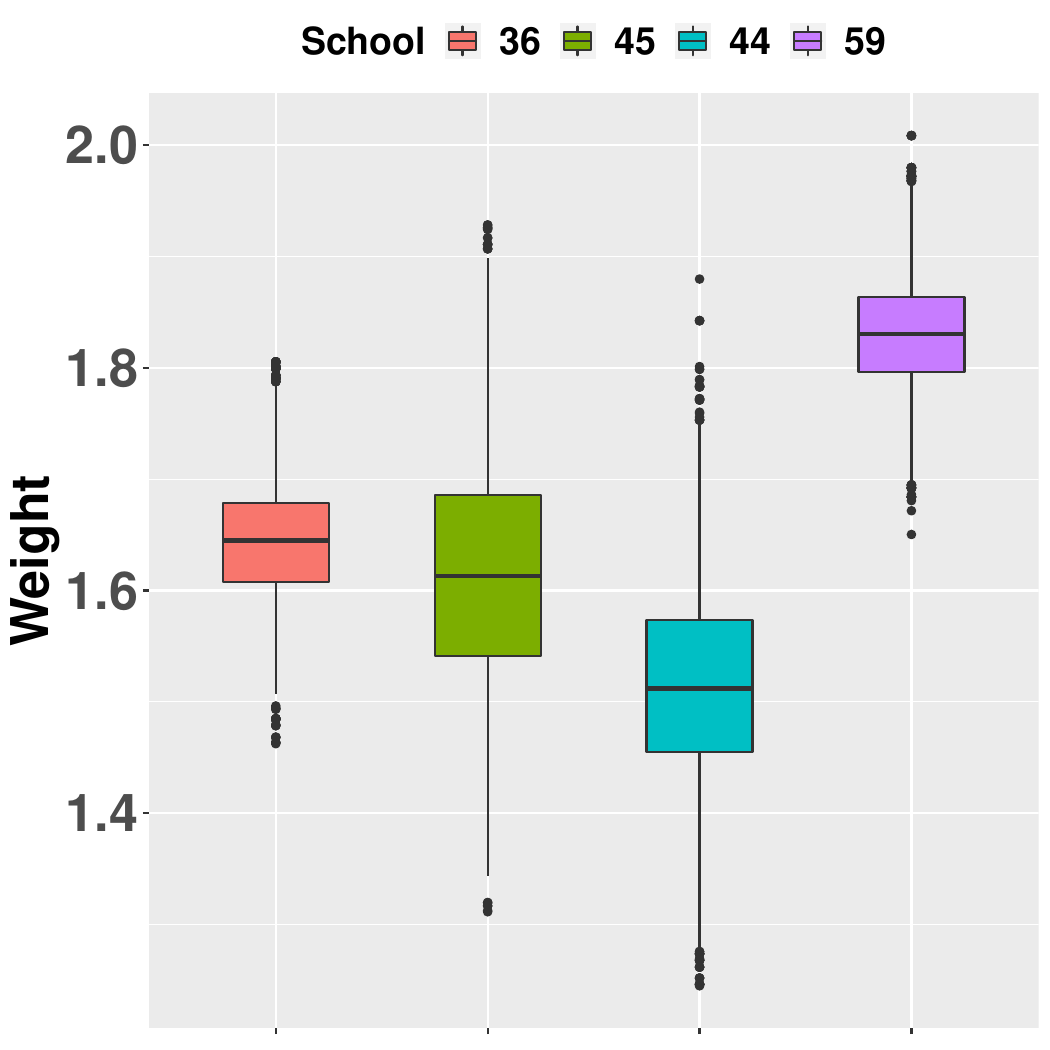} & 
    \includegraphics[width=0.3\textwidth]{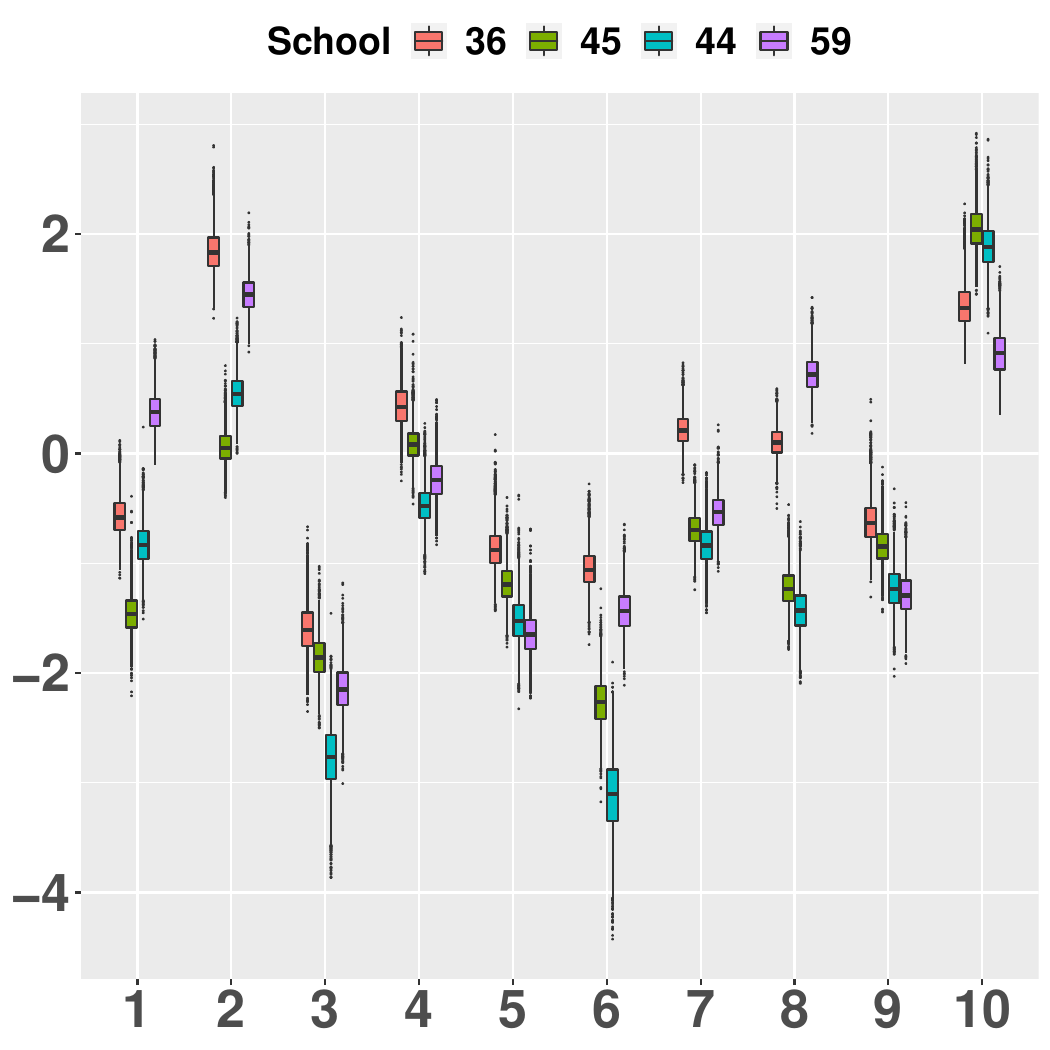} 
    \end{tabular}
    \caption{\label{fig:parameter_race_lunch}
    Boxplots of the posterior samples of the other model parameters from the  four select schools. The red, green, blue, and purple boxplots indicate School 36, 45, 44, and 59, respectively. $\alpha$ and $\gamma$ are the intercept and weight parameters of the LSM.  $\beta_i$ is the item easiness parameter of the adapted LSIRM. 
    }
\end{figure}

Figures \ref{fig:parameter_race_lunch} summarize the posterior samples of the other model parameters for the four selected schools: $\alpha$ and $\gamma$ are the intercepts and weight parameters of the LSM. $\beta_i$ is the item easiness parameters of the adapted LSIRM for social influence. It is interesting to observe that School 36, with significant social influence, tends to have larger $\beta_i$ values for most activities (except Activity 10), while $\alpha$ is somewhat low compared to other schools. This means that the students in School 36 showed a higher probability of endorsing the ten school activities, while the students in the other schools, such as Schools 44 and 59, reported more close friends than School 36. 

\paragraph*{Model Fit Evaluation} 

\begin{figure}[htbp]
    \centering
    \begin{tabular}{cc}
    (a) School 36 & (b) School 45\\
    \includegraphics[width=0.4\textwidth]{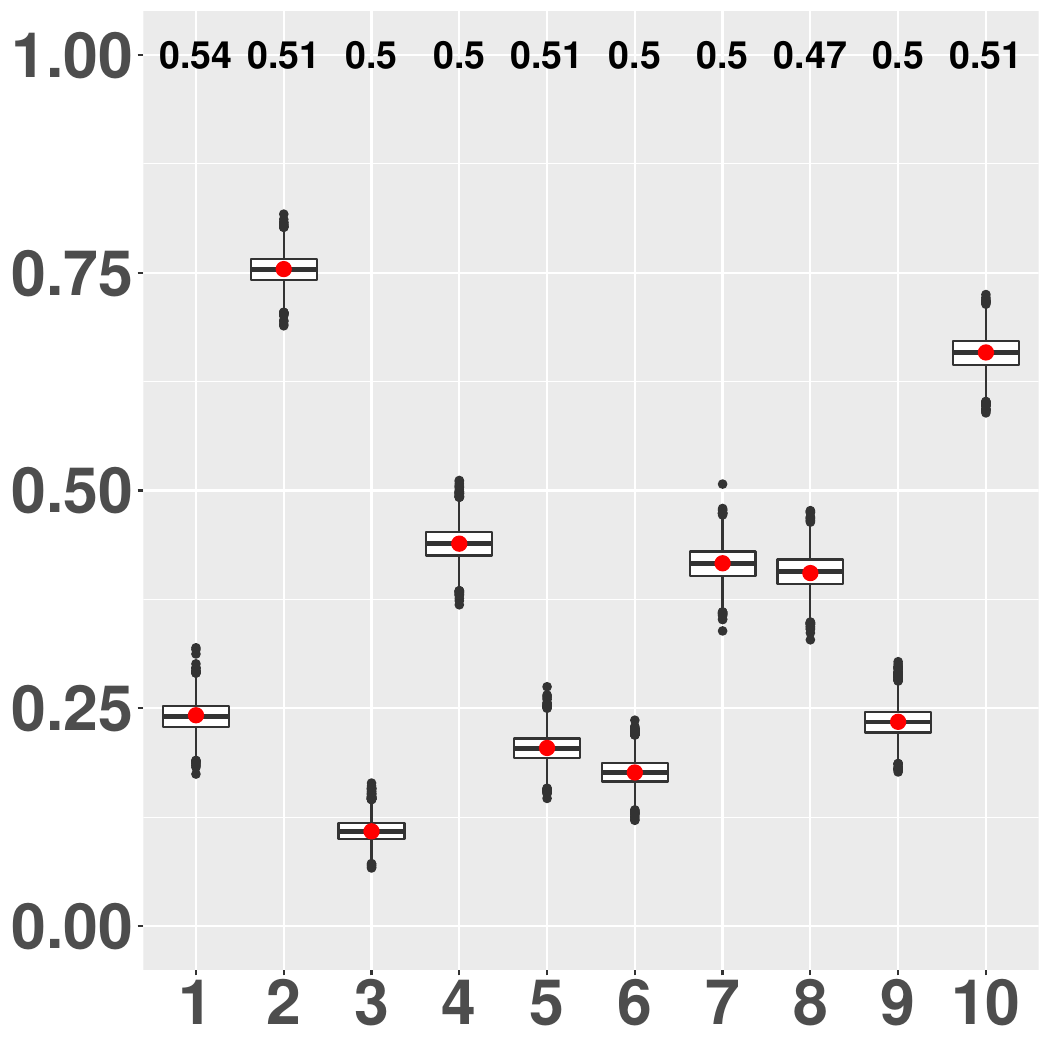} & 
    \includegraphics[width=0.4\textwidth]{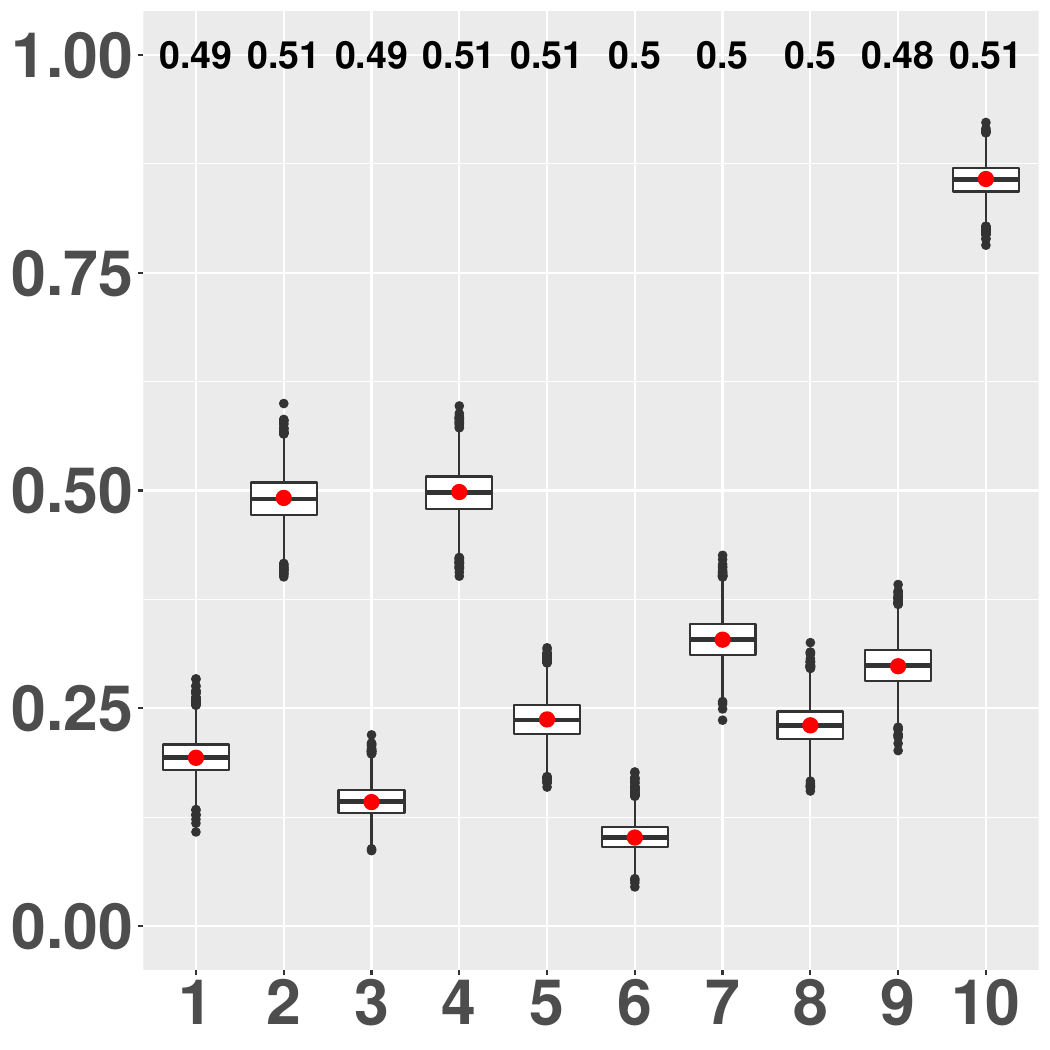} \\
    (c) School 44 & (d) School 59\\
    \includegraphics[width=0.4\textwidth]{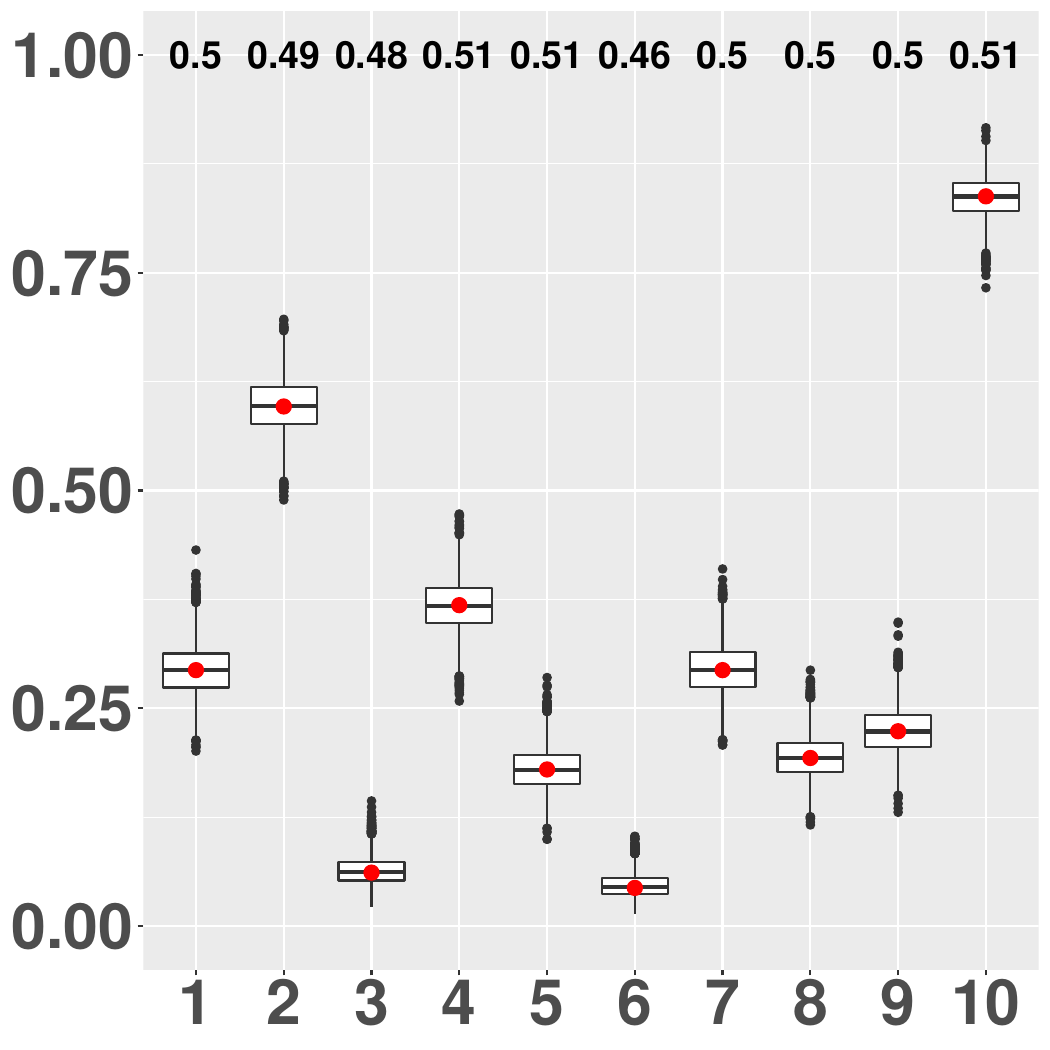} & 
    \includegraphics[width=0.4\textwidth]{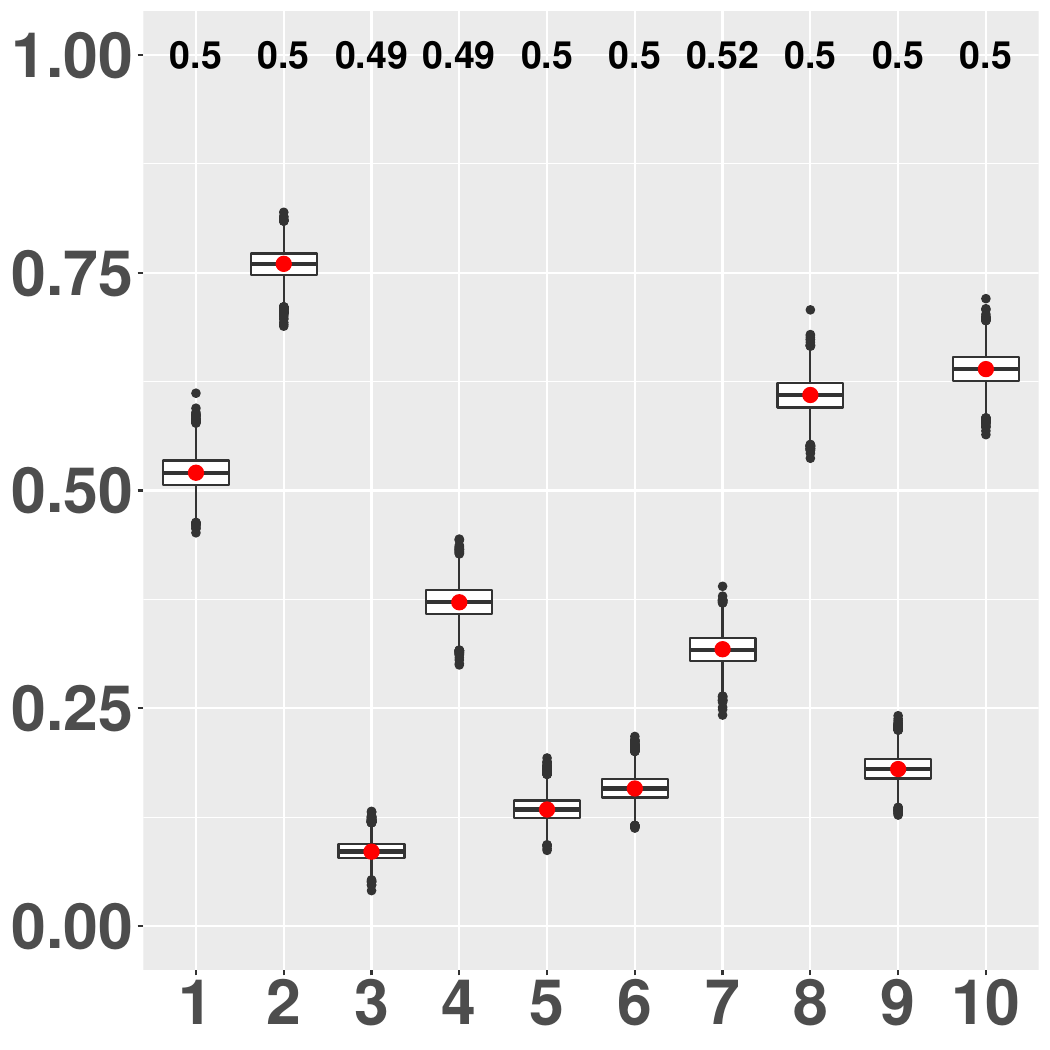} \\
    \end{tabular}
    \caption{\label{fig:gof}
    The boxplot of the predicted probability of positively responding item $i$, $\hat{p_{\cdot i}}$ from the MCMC samples for the four select schools. The red points are the percentage of item $i$ answered positively, $p_{\cdot i}$. The numbers reported on the top of each box plot indicates the posterior predictive p-values for the corresponding items.}
\end{figure}

As discussed in Section \ref{sec:other_topic}, we evaluate the goodness-of-fit of the proposed model by comparing the predicted $\hat{\bf p}_{\cdot i}$ from the posterior samples with ${\bf p}_{\cdot i}$ from the original data. Figure \ref{fig:gof} displays the boxplots of $\hat{\bf p}_{\cdot i} = (\hat{p}_{\cdot i}^{(1)}, \cdots, \hat{p}_{\cdot i}^{(m)})$ from the MCMC samples, where the red points indicate $p_{\cdot i}$ from the original data. The figure shows that the median of the posterior probability $\hat{p}_{\cdot i}$ is close to $p_{\cdot i}$ for all items in all four schools. The numbers reported on the top of each box plot indicate the posterior predictive p-values for the corresponding activities. As shown, the posterior predictive p-values range from 0.46 to 0.52 across all activities and schools. These results indicate that the proposed model fits reasonably well to the four school data under investigation. 

\section{Simulation Studies}

{
We conducted two simulation studies to evaluate the performance of the proposed approach in various data-generating conditions. In the first simulation study, we generate network data and item response data (item-level behavior data) using simple and bipartite stochastic blockmodels (SBMs) \citep{Holland1983, Decelle2011, Rohe2011, lei2015consistency, ZhZh16, abbe2017, LaRiSa17, Lee2019}, respectively. 
The goal is to explore how the overall social influence parameter ($\delta$) changes when the probabilities of connections within and between memberships vary, while the data are not directly generated from the proposed model. 
The second simulation study aims to assess the accuracy of the estimated parameters, where the data are generated from the proposed approach. In this section, we focus on discussing the first simulation study to reserve space. The details of the second simulation study are presented in Supplementary Material 4. 
}

\subsection{Study Design}\label{sec:setting}
{
We generated binary social network data of friendship ${\bf{Y}}_{n \times n}$ and binary item-level behavior data ${\bf X}_{n \times p}$ using the stochastic blockmodel. The stochastic blockmodel can effectively model various types of real-world networks and explore what happens in the network in practice \citep{Lu2019}. We considered four different data configurations, where the membership of respondents for network data and item response data is equal (in Scenarios 1.1 to 1.3), while the membership of respondents for two data is unequal (in Scenarios 2). 
}

{ 
In order to generate network data and item response data using simple and bipartite SBM, we need two connection probability matrices: $\mathbf{P}{\text{net}}$ for network data and $\mathbf{P}{\text{irt}}$ for item response data. For network data, $\mathbf{P}{\text{net}}$ includes both within- and between-group connection probabilities, where the within-group (or between-group) connection probability is the probability that nodes in the same (or different) group are connected. The connection probability matrix for item response data, $\mathbf{P}{\text{irt}}$, also includes within- and between-co-cluster connection probabilities, where the within-co-cluster (or between-co-cluster) connection probability is the probability of answering an item positively for a respondent when a respondent and an item are in the same (or different) co-cluster.
}

Our simulation design was motivated by the main empirical data under investigation. For example, the number of students was between 150 and 300 in most schools. {Therefore, we set the number of respondents to be in a similar range (between 150 and 300),  and the number of items to be in the range between 9 and 30 in the simulation. In addition, we observe that school networks showed two to four main clusters in empirical data analysis. Accordingly, we set two to four respondent clusters in the data simulation.
}

\paragraph*{Scenario 1.1} 
{
In Scenario 1.1, we consider three groups of items and respondents with an equal probability of the respondents and the items being assigned to each cluster.
}
In this basic scenario, the person clusters from the social network exactly match the person clusters from the item-response network. That is, people with only similar response behavior are assumed to be friends with each other in this scenario. 
{
Table \ref{tab:prob} shows $\mathbf{P}{\text{net}}$ and $\mathbf{P}{\text{irt}}$ for Scenario 1.1. The variables $a$ and $c$ were defined as the within-connection probabilities for network  and item response data, respectively, while $b$ and $d$ were defined as the between-connection probabilities for network  and item response data. We considered different values for these variables, ranging from 0.6 to 0.8 for $a$ and $c$ with an increment of 0.1, and from 0.1 to 0.4 for $b$ and $d$ with an increment of 0.1.
}

\begin{table}[htbp]
    \centering
    \begin{tabular}{cc}
    (a) Network Data & (b) Item response Data \\
    $\bf{P_{net}} = \begin{bmatrix}  a & b & b \\ b & a & b \\ b & b & a \end{bmatrix}$ &
    $\bf{P_{irt}} = \begin{bmatrix} c & d & d \\ d & c & d \\ d & d & c \end{bmatrix}$
    \end{tabular}
    \caption{\label{tab:prob}
    The connection probability matrices for generating network data and item response data, respectively.}
\end{table}

\paragraph*{Scenario 1.2} 
In Scenario 1.2, we consider four-person clusters for the social network space. For the item-respondent network space, we consider four-person clusters and three-item clusters.
{
$\mathbf{P}{\text{net}}$ and $\mathbf{P}{\text{irt}}$ for Scenario 1.2 are shown in Table \ref{tab:prob12}. This scenario is similar to Scenario 1.1, except for the addition of a fourth respondent group. The values for $a$, $b$, $c$, and $d$ are the same as in Scenario 1.1. The probability of answering all items positively for the respondents in the fourth-person group is denoted as $e$. $e$ is set from 0 to 0.1 with an increment of 0.05.
}
\begin{table}[htbp]
    \centering
    {
    \begin{tabular}{cc}
    (a) Network Data & (b) Item response Data \\
    $\bf{P_{net}} = \begin{bmatrix}  a & b & b & b \\ b & a & b & b \\ b & b & a & b \\ b & b & b & a \end{bmatrix}$ &
    $\bf{P_{irt}} = \begin{bmatrix} c & d & d  \\ d & c & d  \\ d & d & c \\ e & e & e  \end{bmatrix}$
    \end{tabular}
    \caption{
    \label{tab:prob12}
    The connection probability matrices for generating network data and item response data, respectively, in Scenario 1.2.
    }
    }
\end{table}

\paragraph*{Scenario 1.3} 

In Scenario 1.3, we consider three-person clusters for the social network space as in Scenario 1.1. For the item-respondent network space, we consider three-person clusters and four-item clusters. 
{
$\mathbf{P}{\text{net}}$ and $\mathbf{P}{\text{irt}}$ for Scenario 1.3 are shown in Table \ref{tab:prob13}. This scenario is similar to Scenario 1.1, except for the addition of a fourth item group. The values for $a$, $b$, $c$, and $d$ are the same as in Scenario 1.1. The probability of answering the fourth cluster item positively for all the respondents is denoted as $e$. $e$ is set from 0 to 0.1 with an increment of 0.05.
}
\begin{table}[h]
    \centering
    {
    \begin{tabular}{cc}
    (a) Network Data & (b) Item response Data \\
    $\bf{P_{net}} = \begin{bmatrix}  a & b & b \\ b & a & b \\ b & b & a \end{bmatrix}$ &
    $\bf{P_{irt}} = \begin{bmatrix} c & d & d & e \\ d & c & d & e \\ d & d & c & e \end{bmatrix}$
    \end{tabular}
    \caption{\label{tab:prob13}
    The connection probability matrices for generating network data and item response data, respectively, in Scenario 1.3.}
    }
\end{table}

\paragraph*{Scenario 2} 

{
Unlike Scenarios 1.1 to 1.3, we assume that the cluster membership of respondents in social network mismatches the cluster membership of respondents in the item-respondent network data.} Specifically, for the social network, we consider two-person clusters, while three-person clusters and three-item clusters in the item-respondent network. We generate the item response data as in Scenario 1.1. For the social network data, we randomly generated the number of respondents between 75 and 150 per cluster. 
{
The respondents included in the third cluster of the item response network randomly change their membership to the first or second cluster. $\mathbf{P}{\text{net}}$ and $\mathbf{P}{\text{irt}}$ for Scenario 2 are shown in Table \ref{tab:prob2}. Scenario 2 is similar to Scenario 1.1 except for the number of person memberships. $a$, $b$, $c$, and $d$ are set the same as in Scenario 1.1.
}
\begin{table}[h]
    \centering
    {
    \begin{tabular}{cc}
    (a) Network Data & (b) Item response Data \\
    $\bf{P_{net}} = \begin{bmatrix}  a & b \\ b & a  \end{bmatrix}$ &
    $\bf{P_{irt}} = \begin{bmatrix} c & d & d \\ d & c & d \\ d & d & c \end{bmatrix}$
    \end{tabular}
    \caption{\label{tab:prob2}
    The connection probability matrices for generating network data and item response data, respectively, in Scenario 2.}
    }
\end{table}

\paragraph*{Summary}
In Scenarios 1.1 to 1.3, the {the memberships of respondents for network data and for item response data are matched.} Scenario 1.1 is the simplest setting where the number of person groups is the same as the number of items. This is no longer true in Scenarios 1.2 and 1.3. In Scenario 2, {the memberships of respondents for network data and for item response data are not matched.} Note that there is no change in the item-item dependence structures in Scenario 2. 

\subsection{Estimation}

For each scenario, 200 datasets were generated. For each simulated dataset, we applied the proposed approach to estimate the social influence parameter with the MCMC algorithm described in Section \ref{sec:estimation}. We used 30,000 iterations, with the first 5,000 iterations being discarded as a burn-in process. From the remaining 25,000 iterations, 5,000 samples were collected in a time space of 5 iterations. We adjusted the jumping rules for the proposal distributions to achieve ideal acceptance rates (20\% to 40\%), identical to those in real data analysis. Details of the jumping rules are given in Supplementary Material 1.

\subsection{Result}

{
For comparison with existing models, we focused on evaluating the overall social influence parameter ($\delta$) together with the probability of the social influence parameter ($\omega$). This is because the existing models do not estimate differential social influence, as in our model, while all models in the comparison include overall social influence parameters. As in the empirical study, we considered the network autocorrelation model and the linear-in-mean model for comparison. }
{
The SAOM was not considered here as the SAOM requires network and item response measures at both time points, while in the current simulation, we generated one set of network measures (from time $t$) and one set of item-level behavioral measures (from time $t+1$). 
}

{
Table \ref{tab:five_summary1} summarizes the estimates of $\delta$ and $\omega$ in each scenario under the rows labeled ``Original''. When the respondents' memberships of the social network exactly match their clusters from the item response (Scenarios 1.1 to 1.3), all $\delta$ are significant, and all $\omega$ are greater than 0.5, indicating the presence of overall social influence effects, as expected by the simulation design. 
The estimated social influences appear higher in Scenario 1.2 compared to Scenario 1.1. This makes sense given that Scenario 1.2 has an additional respondent group that influences the behavioral items. 
}
{
In Scenario 2, where the respondents' clusters of the social network and the item response are intentionally mismatched, $\delta$ is slightly lower than in Scenarios 1.1 to 1.3, and $\omega$ is lower than 0.5 in some simulated data sets. 
}

{
To assess the performance of the proposed model in the absence of social influences, we randomly generated item response data by setting $c$ = 0.5, $d$ = 0.5, and $e$ = 0.5.  This setting is labeled ``Random IR'' in Table \ref{tab:five_summary1}. As expected, both $\delta$ and $\omega$ in ``Random IR'' are close to zero, implying no social influence of a network on behavioral responses in all scenarios.}

{
In summary, the results above provide evidence that the proposed approach can capture social influence's absence as well as presence with sufficient precision, regardless of the match or mismatch between the two network configurations. Tables 1-8 in Supplementary Material 3 give details of the $\delta$ and $\omega$ estimates. 
}

\begin{table}[htbp]
\centering
{
\footnotesize{
\begin{tabular}{c|rrrrrr|rrrrrr} 
\hline
& \multicolumn{6}{c|}{$\delta$}& \multicolumn{6}{c}{$\omega$}  \\  
 & Min & 25\% & Median & Mean & 75\% & Max & Min & 25\% & Median & Mean & 75\% & Max\\ \hline
Scenario 1.1& \multicolumn{6}{|c|}{} & \multicolumn{6}{c}{}\\
Original & 0.325 & 0.696 & 0.916 & 1.023 & 1.313 & 1.785 & 0.710 & 0.962 & 0.985 & 0.973 & 0.995 & 0.998 \\
Random IR & 0.042 & 0.049 & 0.052 & 0.051 & 0.054 & 0.055 & 0.020 & 0.032 & 0.037 & 0.036 & 0.041 & 0.043 \\ \hline
Scenario 1.2 & \multicolumn{6}{|c|}{} & \multicolumn{6}{c}{}\\
Original & 0.435 & 0.976 & 1.192 & 1.231 & 1.459 & 2.137 & 0.804 & 0.986 & 0.993 & 0.986 & 0.996 & 0.999 \\ 
Random IR & 0.039 & 0.045 & 0.047 & 0.047 & 0.049 & 0.050 & 0.016 & 0.024 & 0.028 & 0.027 & 0.030 & 0.034 \\  \hline
Scenario 1.3 & \multicolumn{6}{|c|}{} & \multicolumn{6}{c}{}\\
Original & 0.316 & 0.681 & 0.927 & 1.007 & 1.291 & 1.798 & 0.697 & 0.962 & 0.985 & 0.972 & 0.995 & 0.998 \\
Random IR & 0.040 & 0.046 & 0.048 & 0.047 & 0.049 & 0.052 & 0.017 & 0.026 & 0.029 & 0.028 & 0.031 & 0.037 \\ \hline
Scenario 2 & \multicolumn{6}{|c|}{} & \multicolumn{6}{c}{}\\
Original & 0.174 & 0.428 & 0.572 & 0.596 & 0.763 & 1.014 & 0.345 & 0.859 & 0.939 & 0.887 & 0.974 & 0.989 \\
Random IR &  0.044 & 0.047 & 0.049 & 0.049 & 0.050 & 0.052 & 0.022 & 0.027 & 0.031 & 0.030 & 0.033 & 0.036 \\
\hline
\end{tabular}
}
}
{
\caption{
\label{tab:five_summary1}
The mean and five-number summaries of the posterior means for $\delta$ and $\omega$ in Scenarios 1.1 to 1.3, and 2 across ``Original'' and ``Random IR'' conditions. 
}
}
\end{table}

\begin{landscape}
\begin{figure}[p]
    \centering
    \footnotesize{
    \begin{tabular}{cc|ccc|ccc|ccc}
    \multicolumn{2}{c|}{} & \multicolumn{3}{c|}{Proposed Model} & \multicolumn{3}{c|}{Linear-in-means Model} & \multicolumn{3}{c}{Network Autocorrelation Model} \\ 
    \multicolumn{2}{c|}{} & (a) Weak & (b) Moderate & (c) Strong & (a) Weak & (b) Moderate & (c) Strong  & (a) Weak &  (b) Moderate & (c) Strong  \\ \hline
    \multirow{8}{*}{Network}
    & Scenario 1.1 &   
    \includegraphics[width=0.1\textwidth, valign = m]{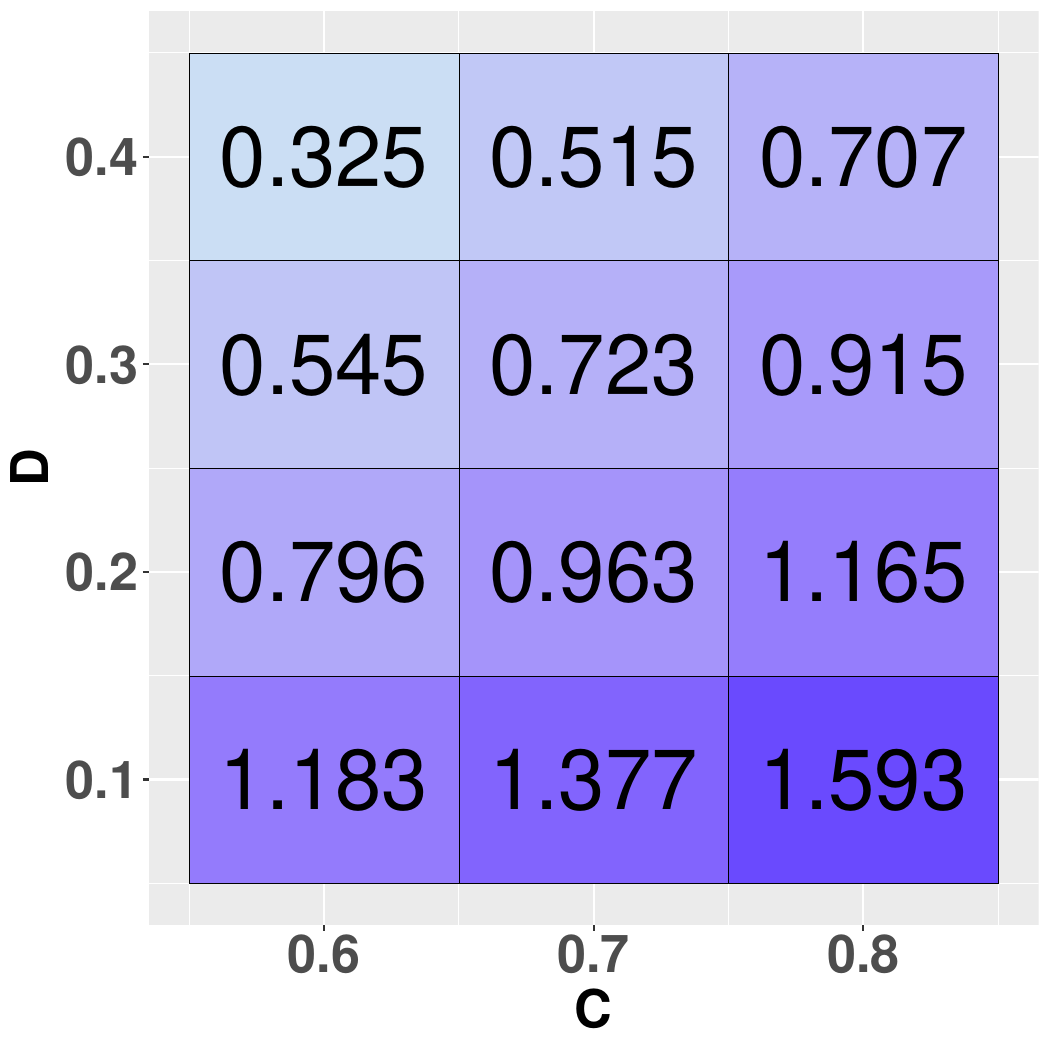} &
    \includegraphics[width=0.1\textwidth, valign = m]{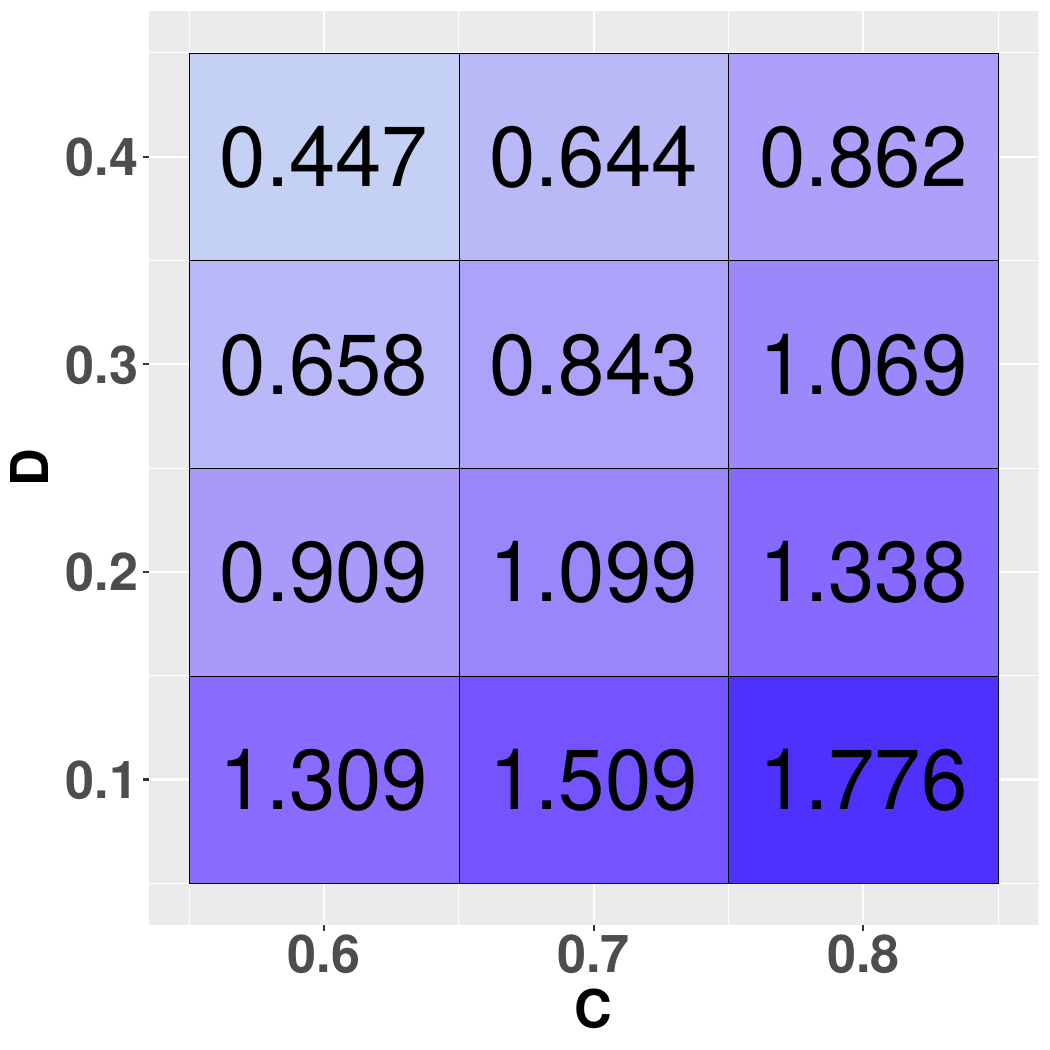} &
    \includegraphics[width=0.1\textwidth, valign = m]{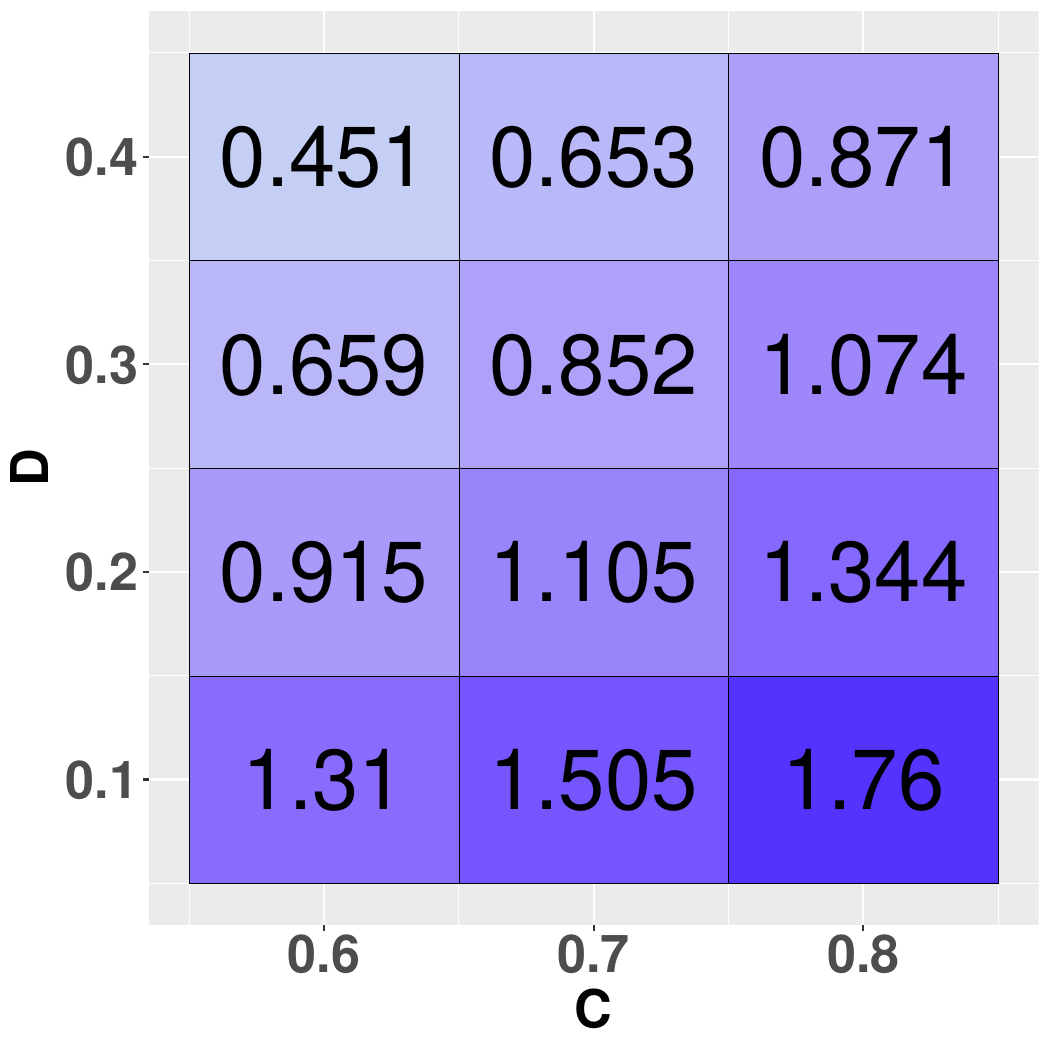} &
    \includegraphics[width=0.1\textwidth, valign = m]{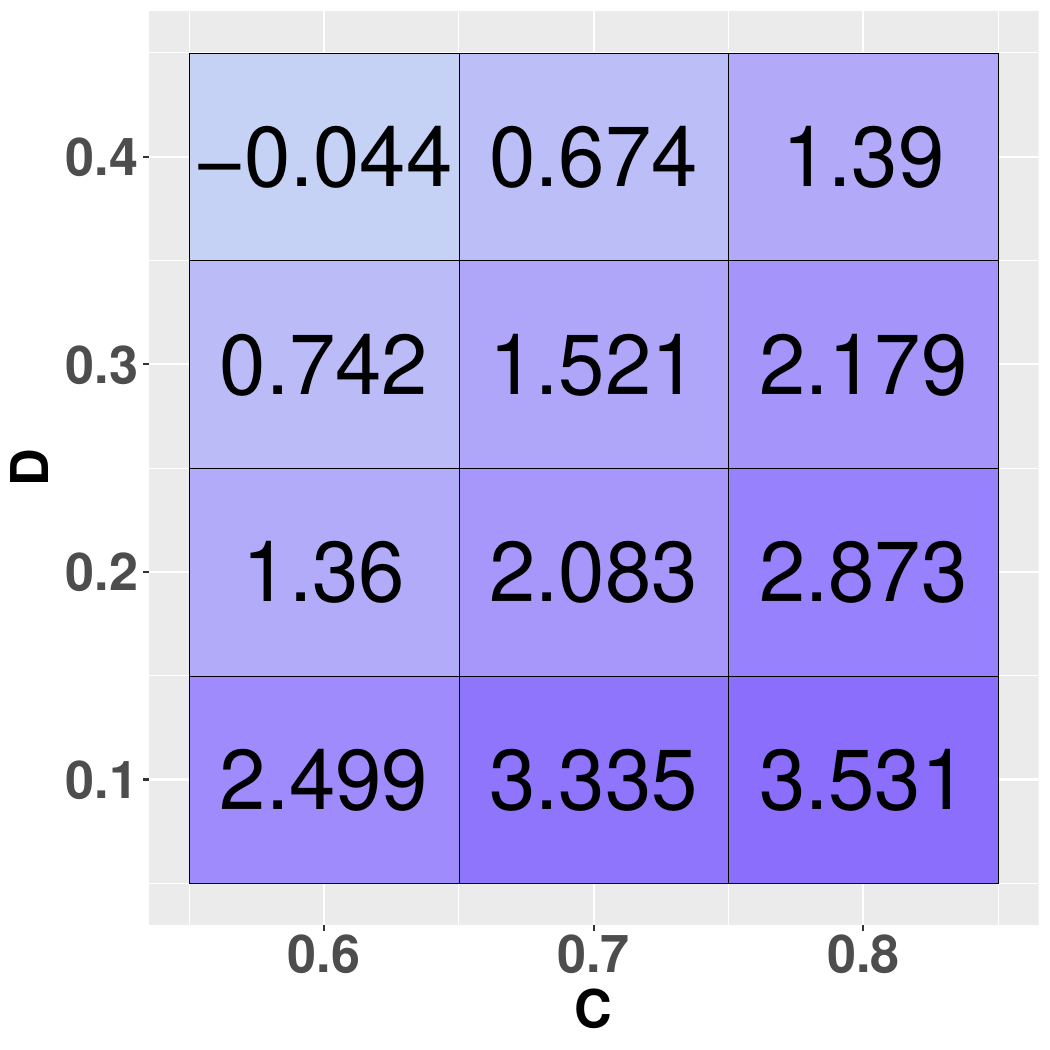} &
    \includegraphics[width=0.1\textwidth, valign = m]{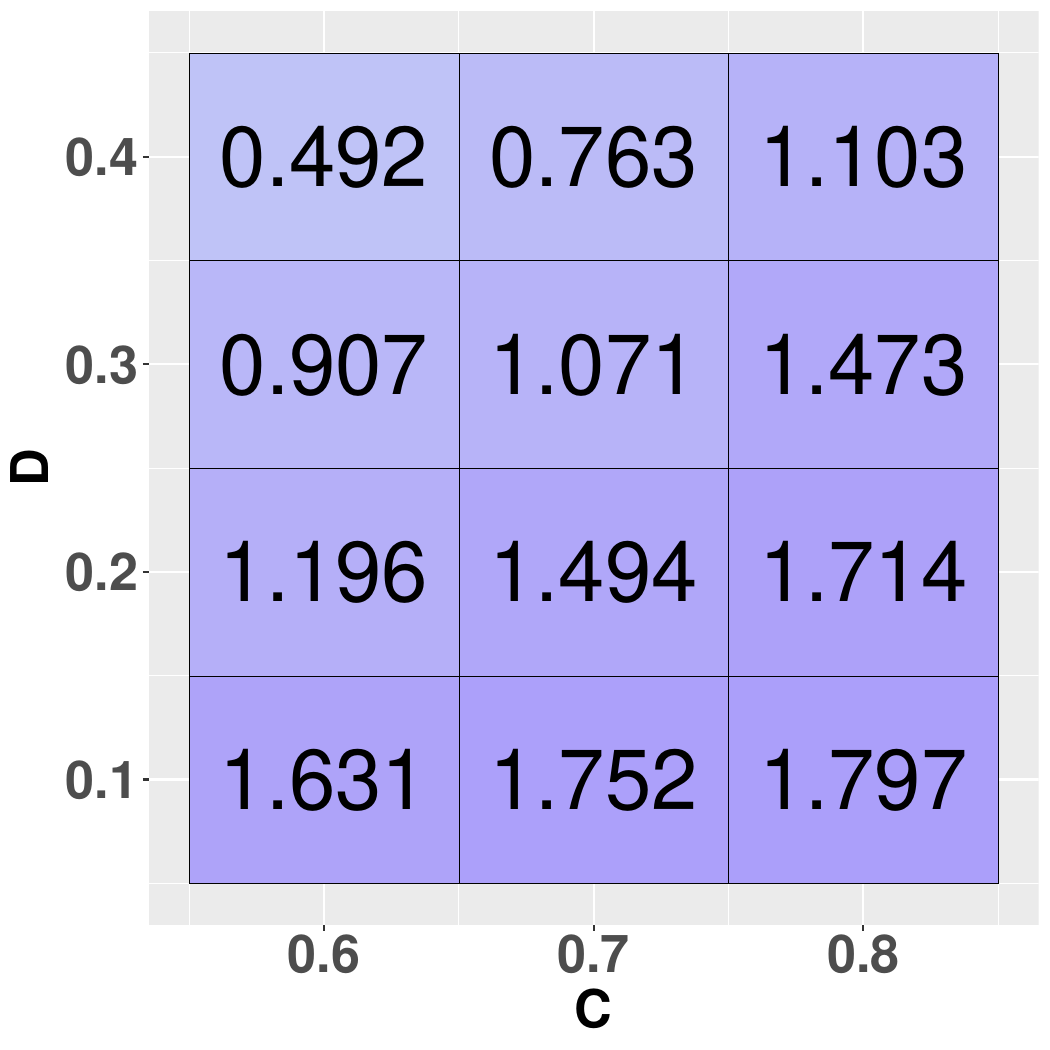} &
    \includegraphics[width=0.1\textwidth, valign = m]{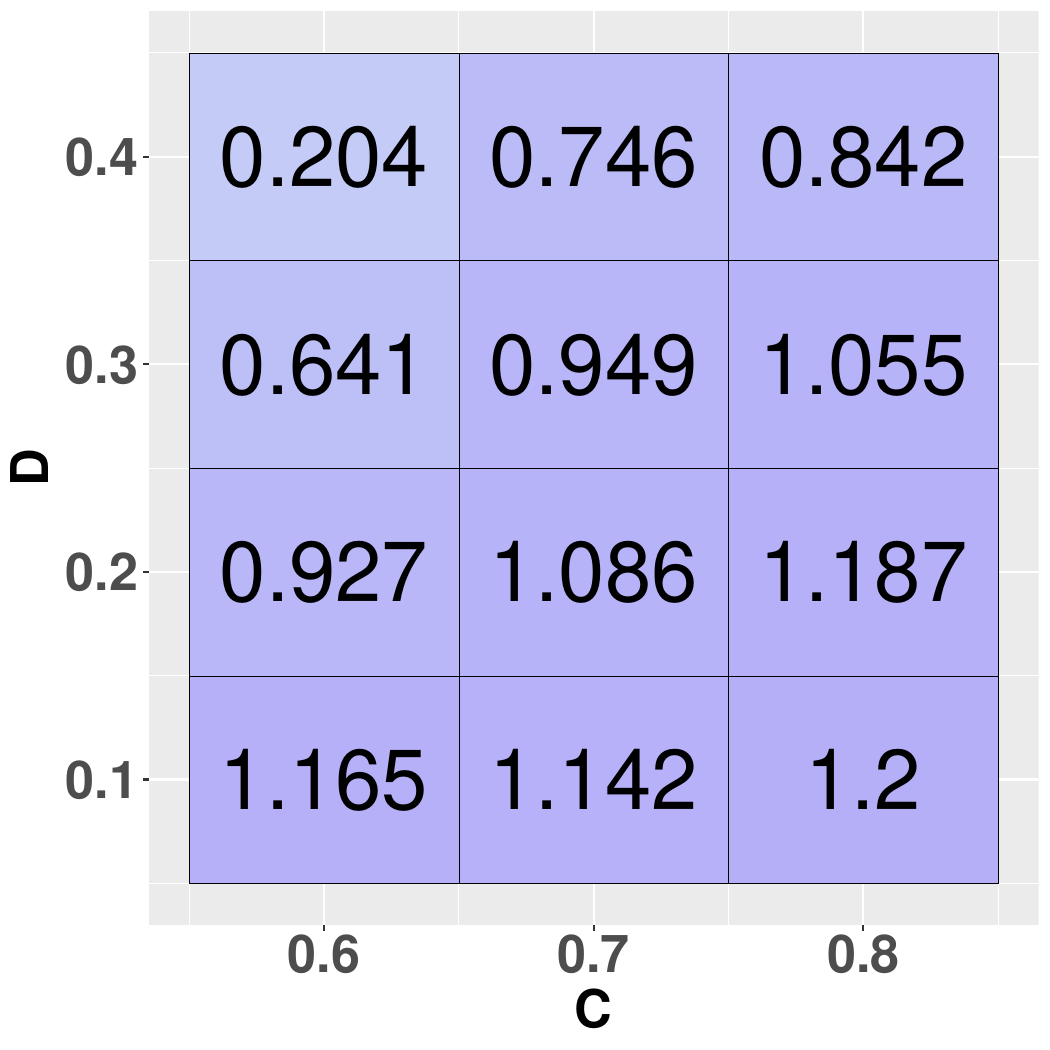} &
    \includegraphics[width=0.1\textwidth, valign = m]{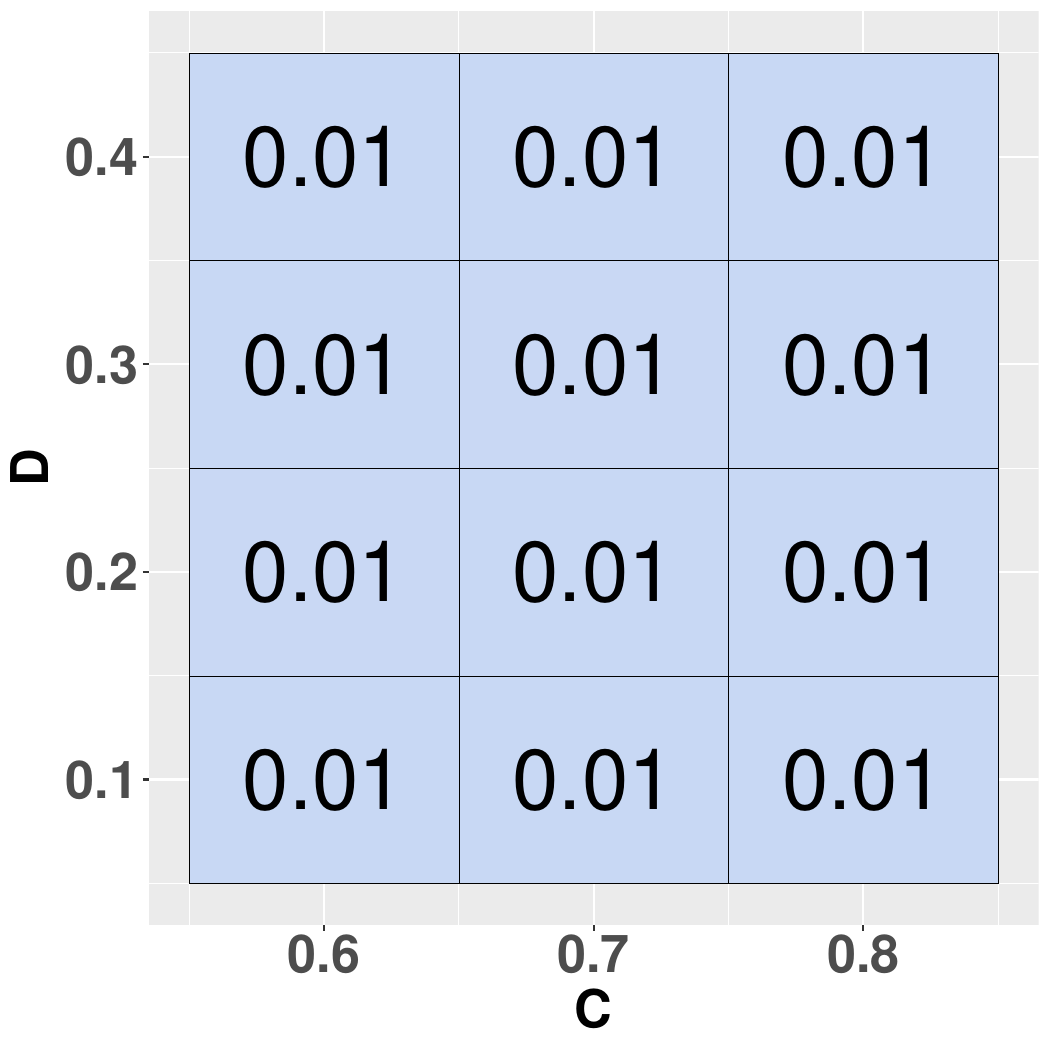} &
    \includegraphics[width=0.1\textwidth, valign = m]{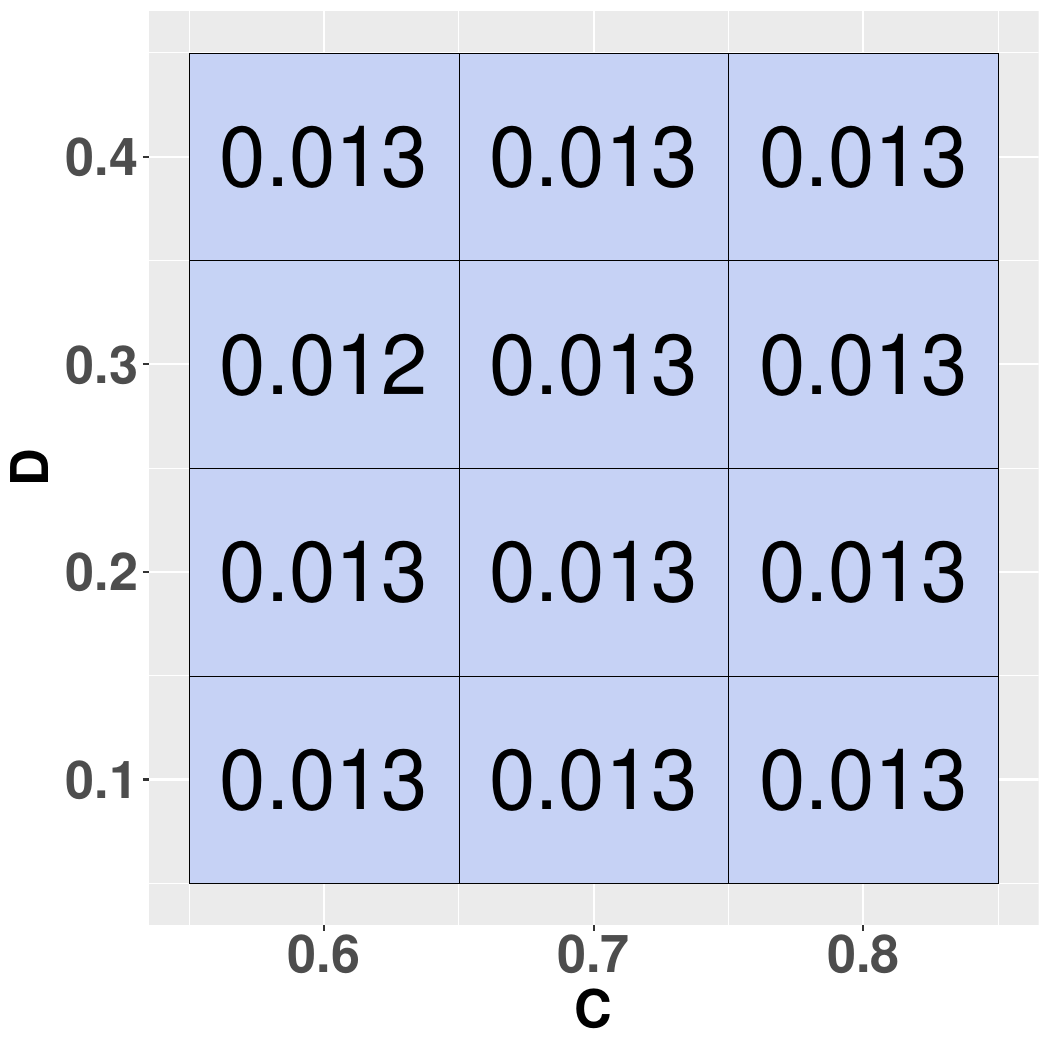} &
    \includegraphics[width=0.1\textwidth, valign = m]{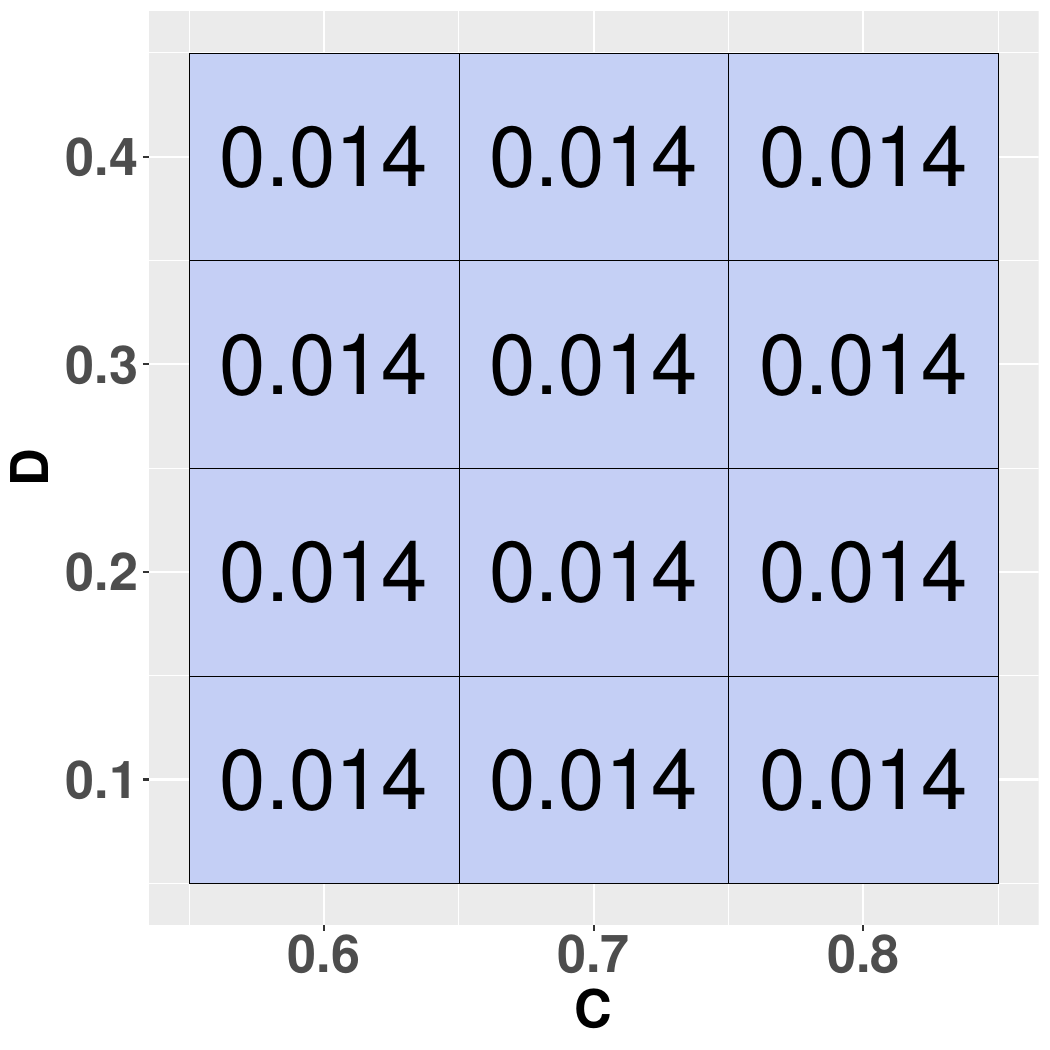} \\ 
    & Scenario 1.2 &  
    \includegraphics[width=0.1\textwidth, valign = m]{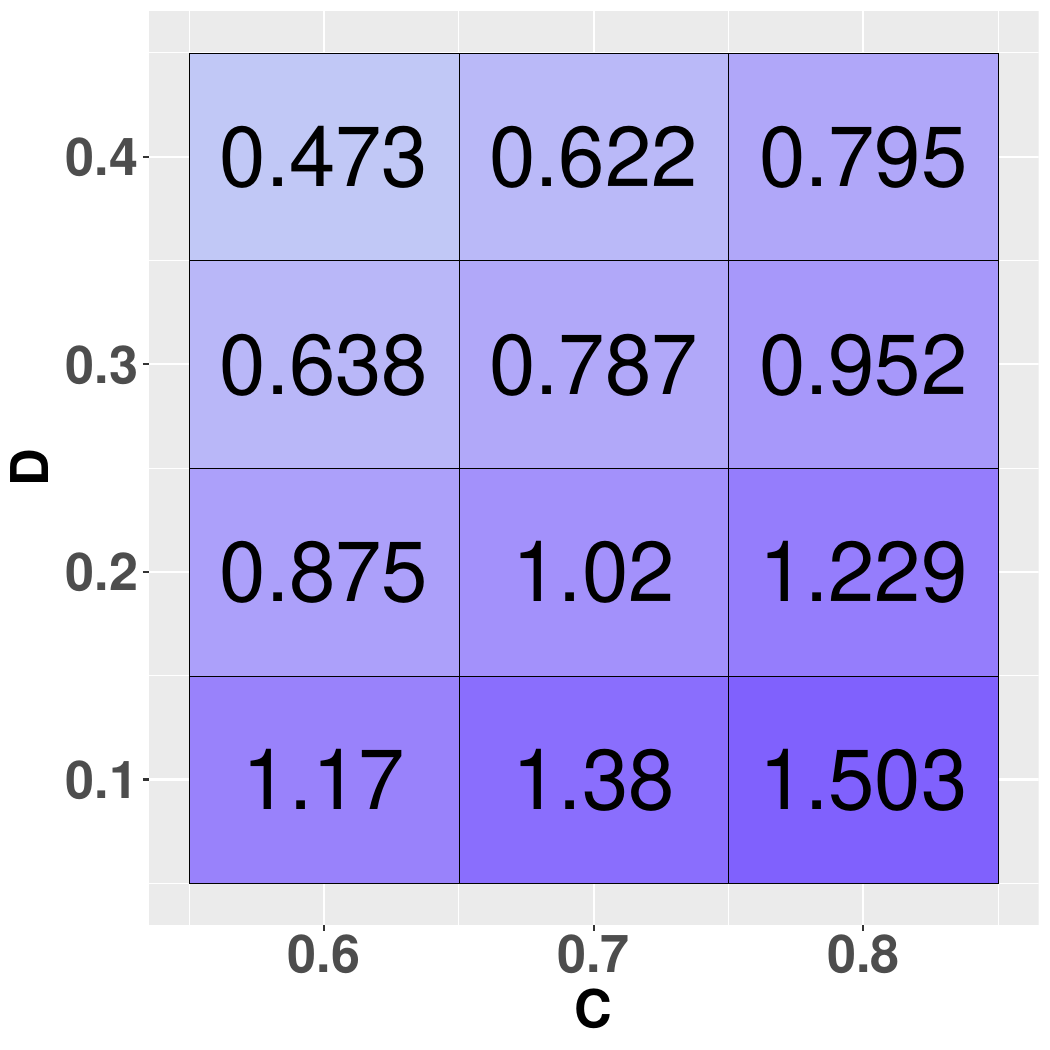} &
    \includegraphics[width=0.1\textwidth, valign = m]{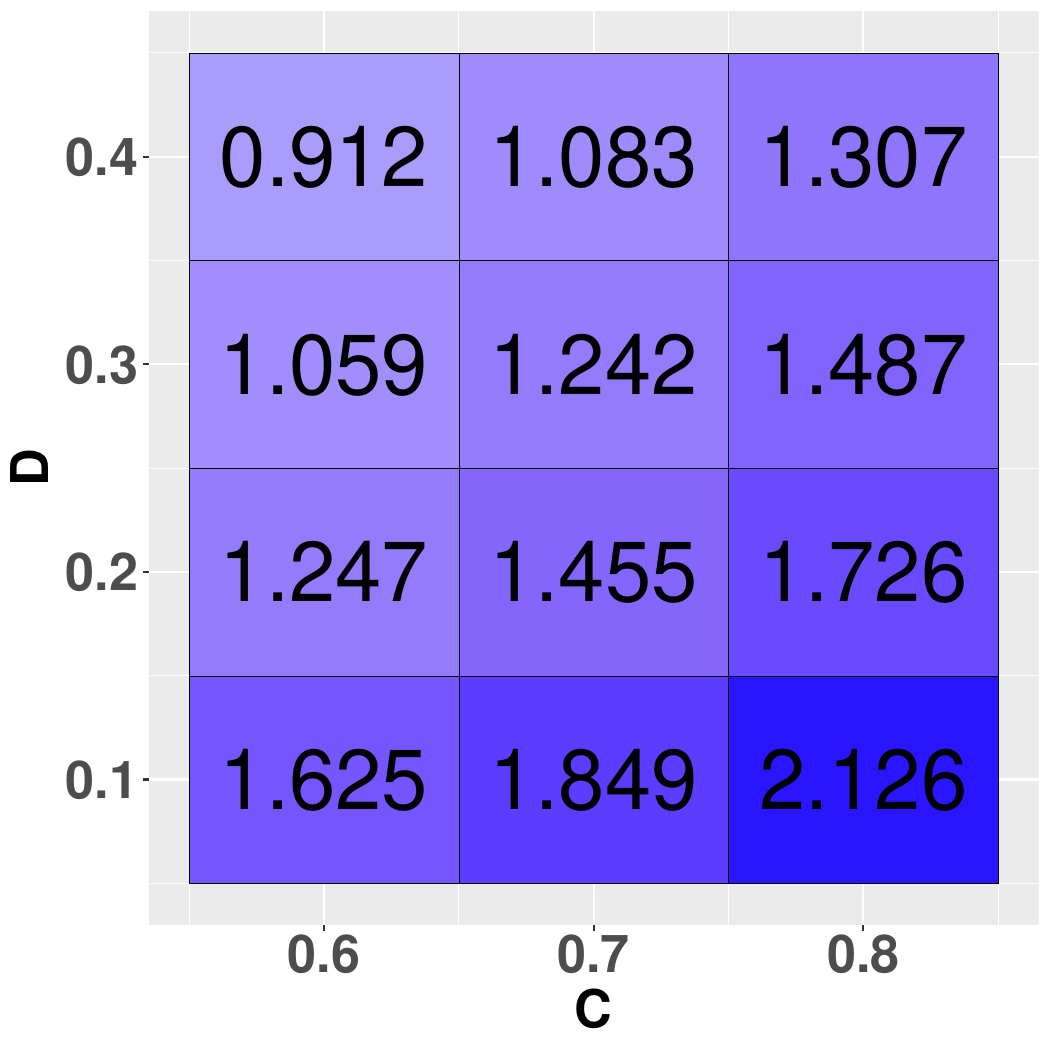} &
    \includegraphics[width=0.1\textwidth, valign = m]{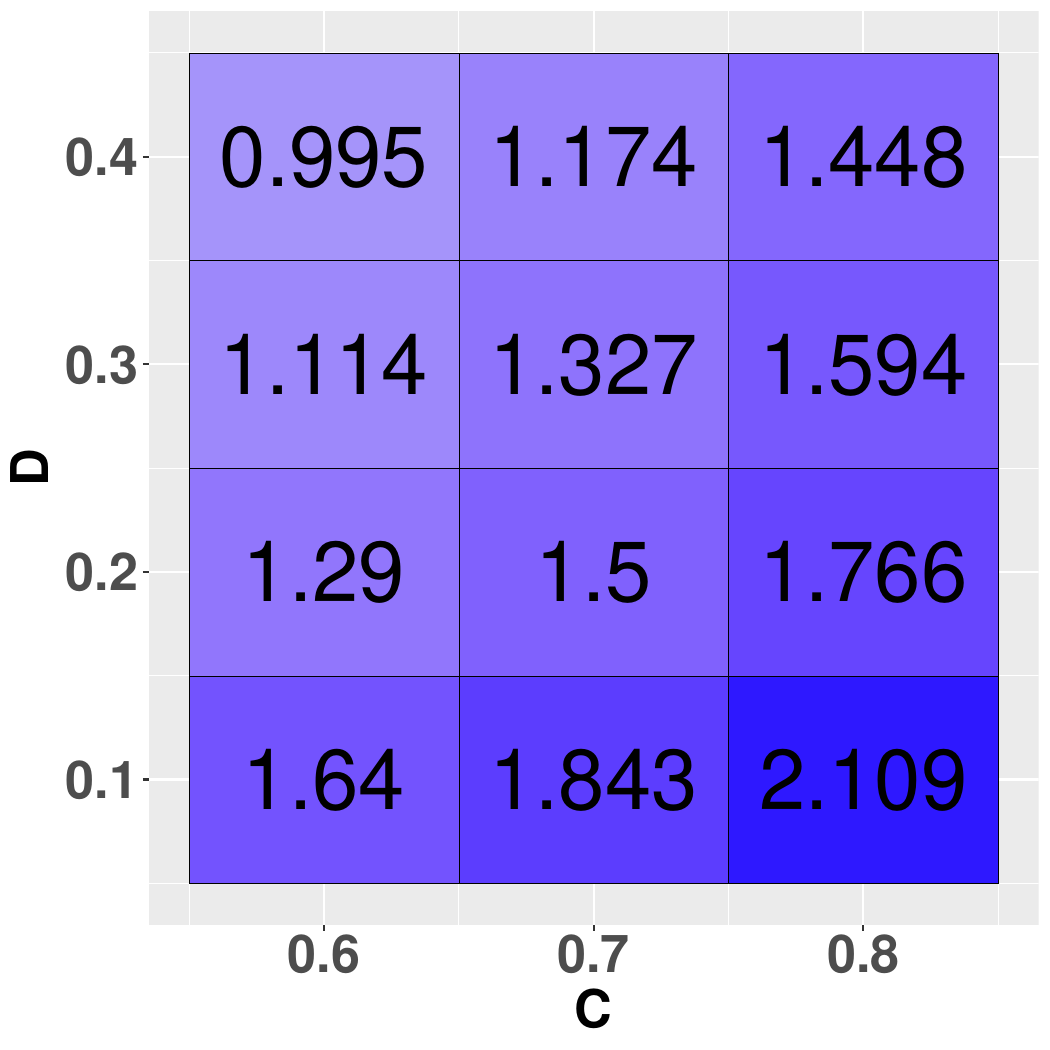} &
    \includegraphics[width=0.1\textwidth, valign = m]{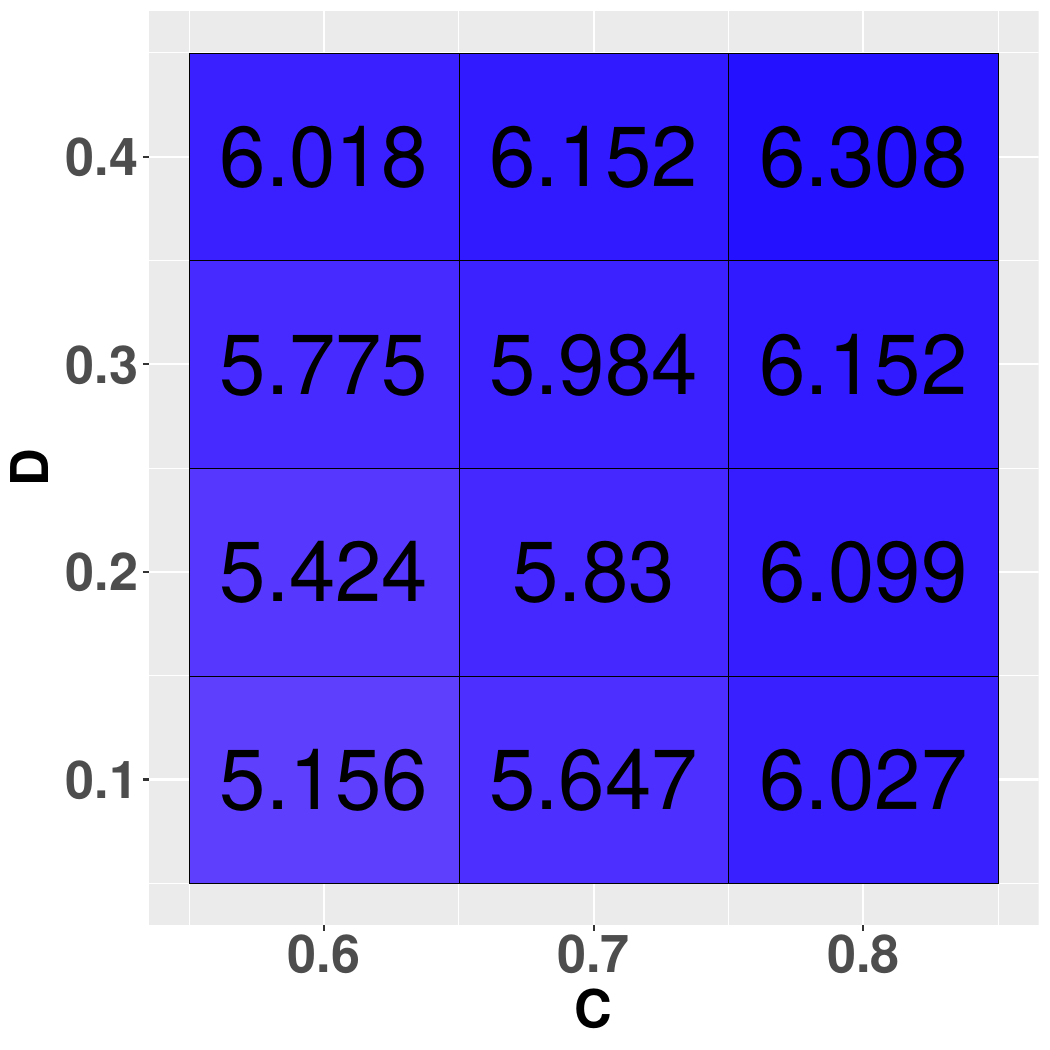} &
    \includegraphics[width=0.1\textwidth, valign = m]{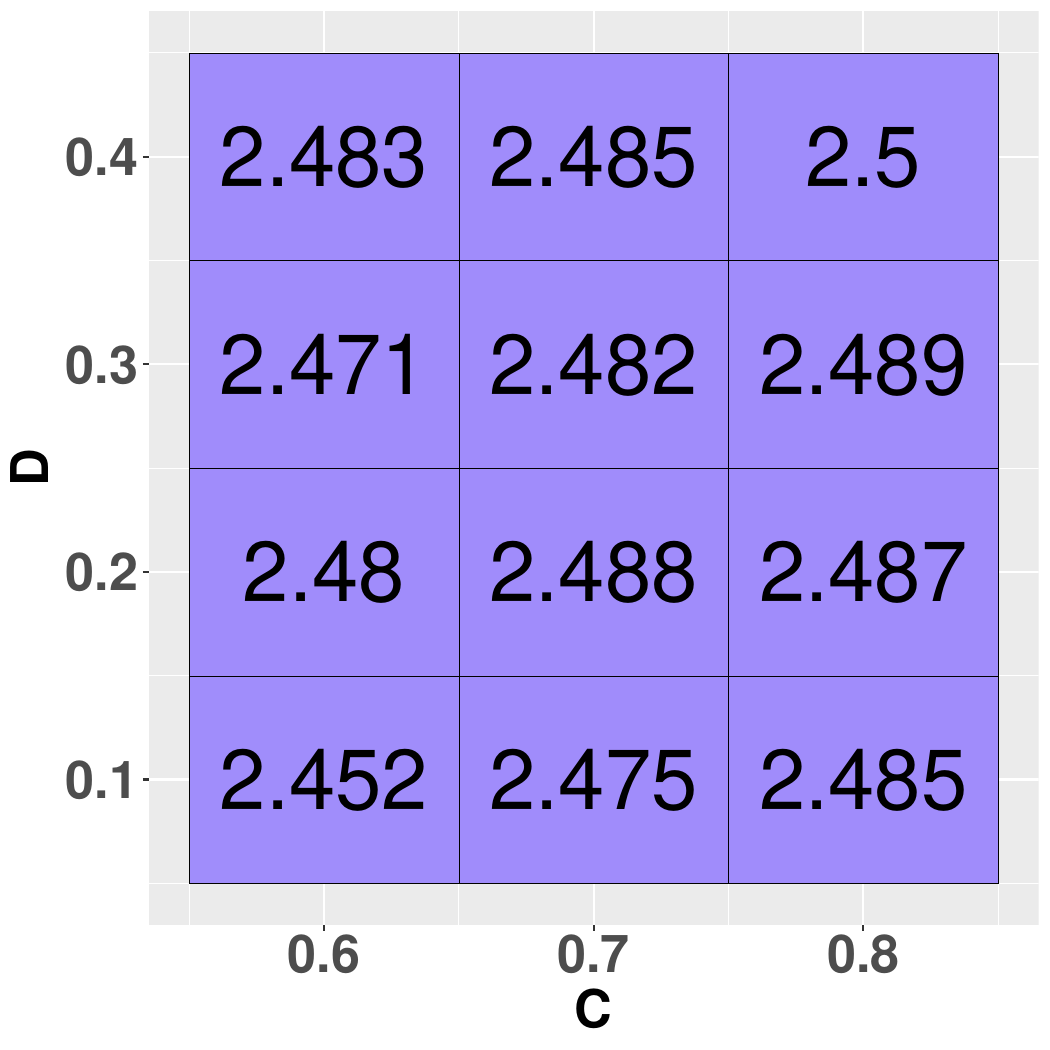} &
    \includegraphics[width=0.1\textwidth, valign = m]{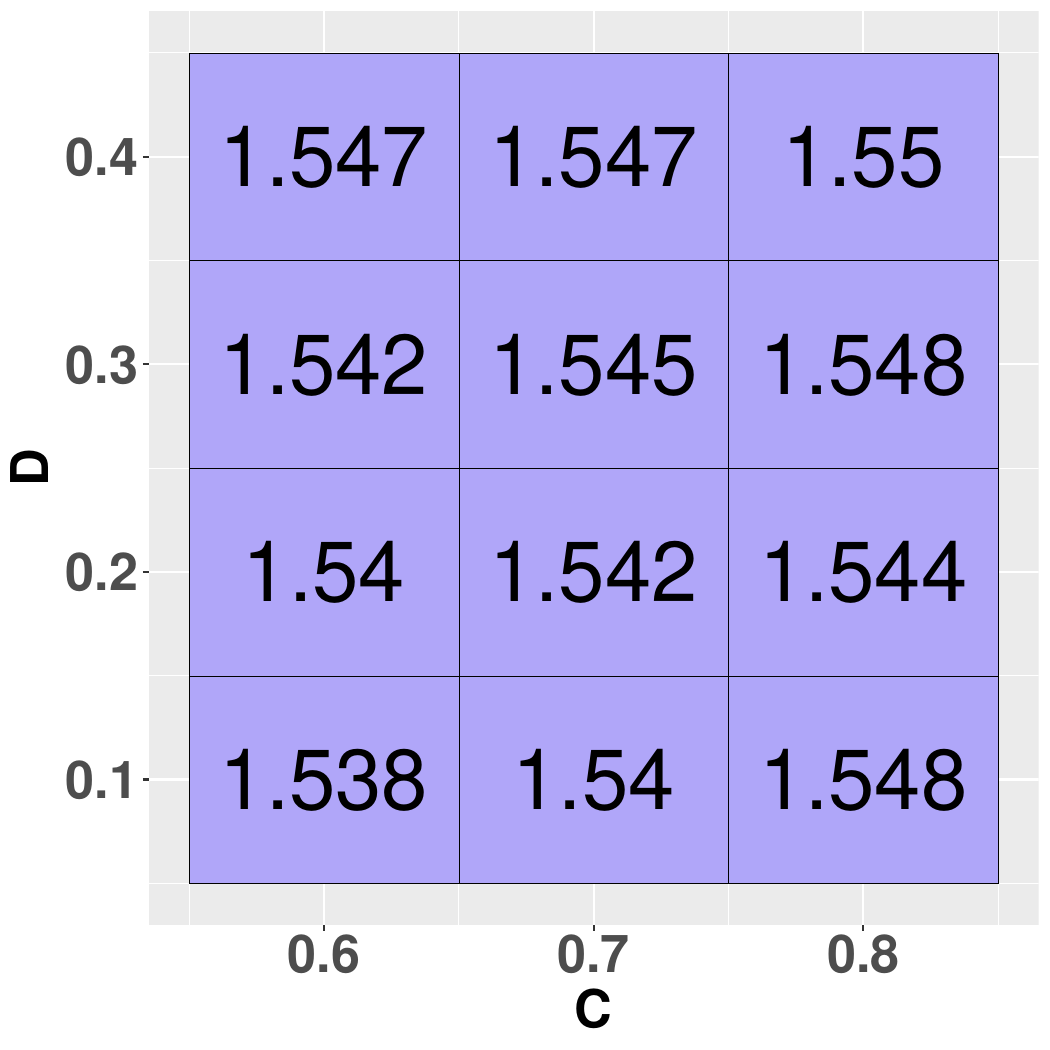} &
    \includegraphics[width=0.1\textwidth, valign = m]{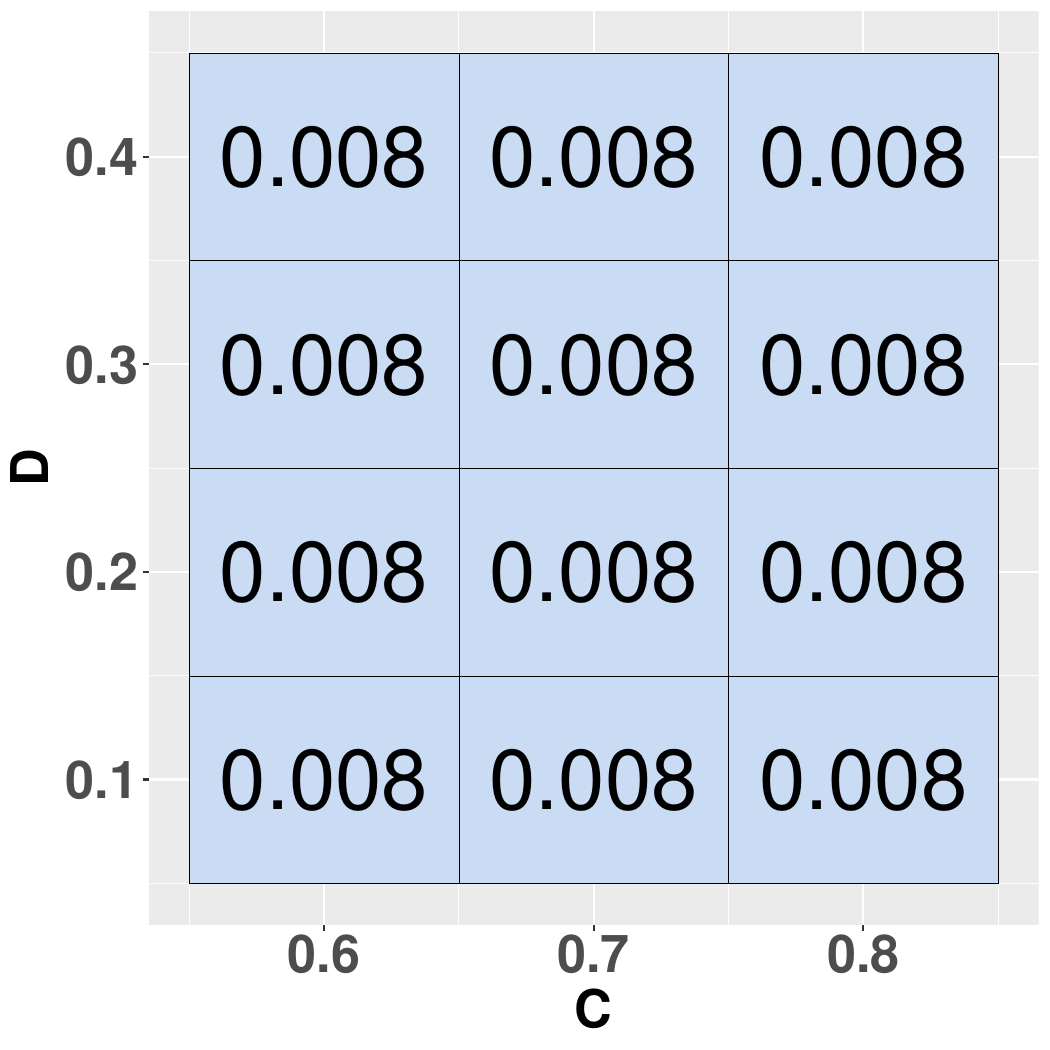} &
    \includegraphics[width=0.1\textwidth, valign = m]{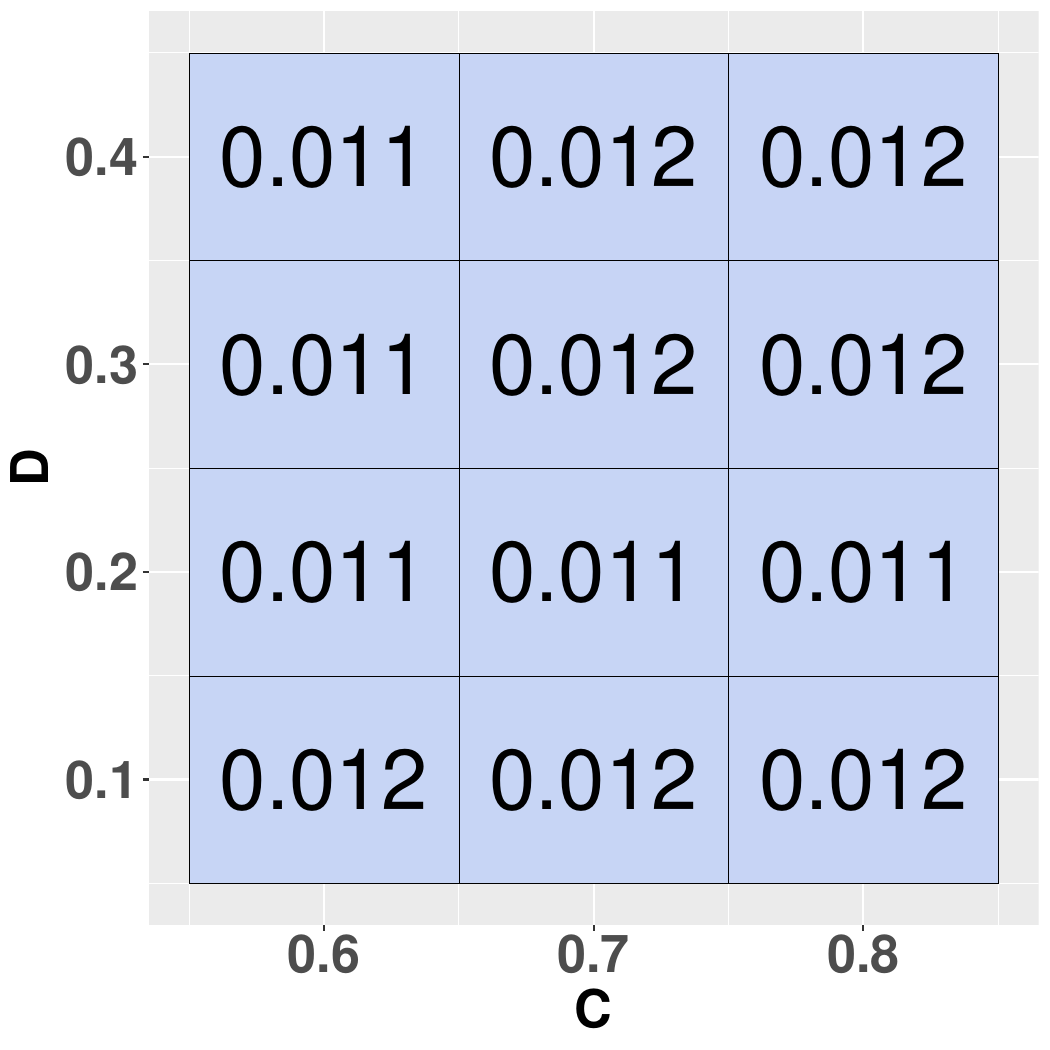} &
    \includegraphics[width=0.1\textwidth, valign = m]{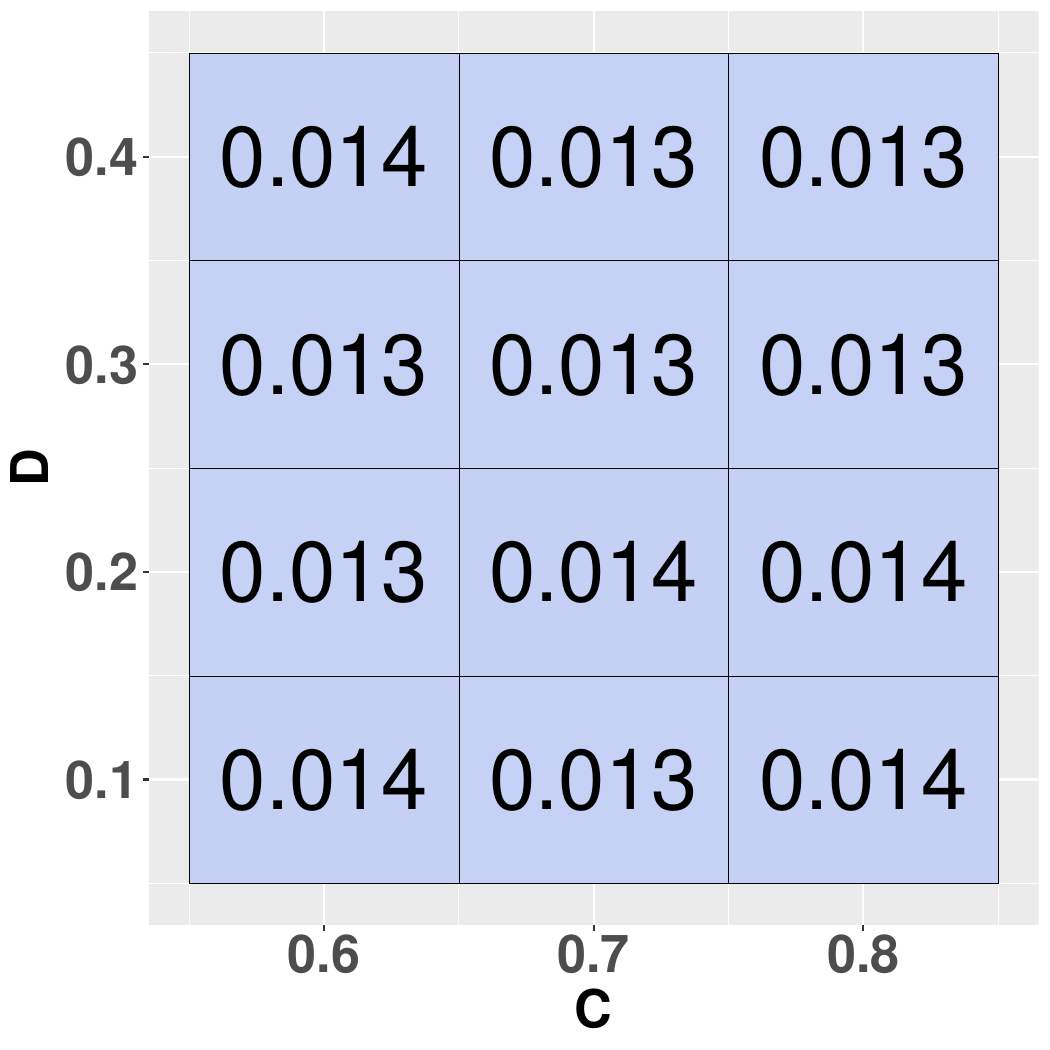} \\ 
    & Scenario 1.3 &  
    \includegraphics[width=0.1\textwidth, valign = m]{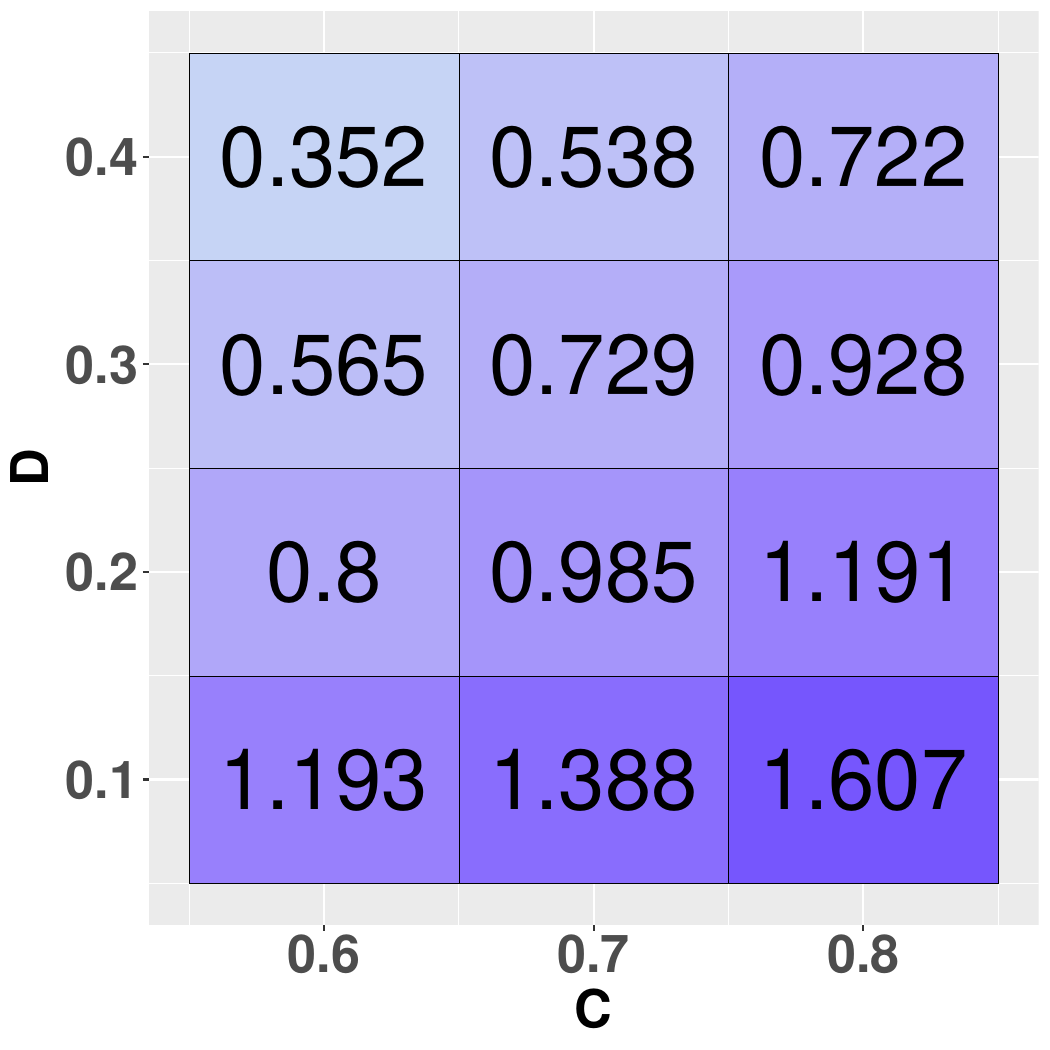} &
    \includegraphics[width=0.1\textwidth, valign = m]{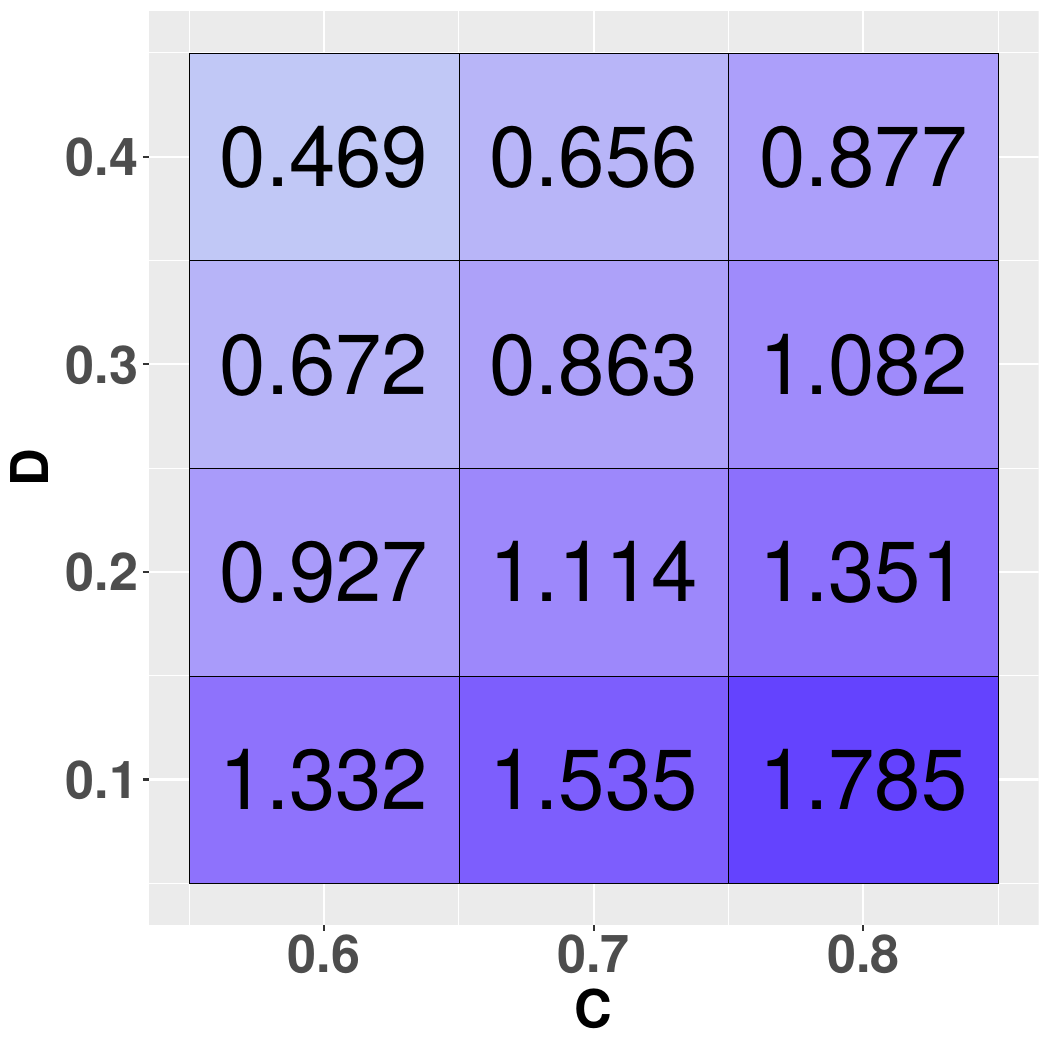} &
    \includegraphics[width=0.1\textwidth, valign = m]{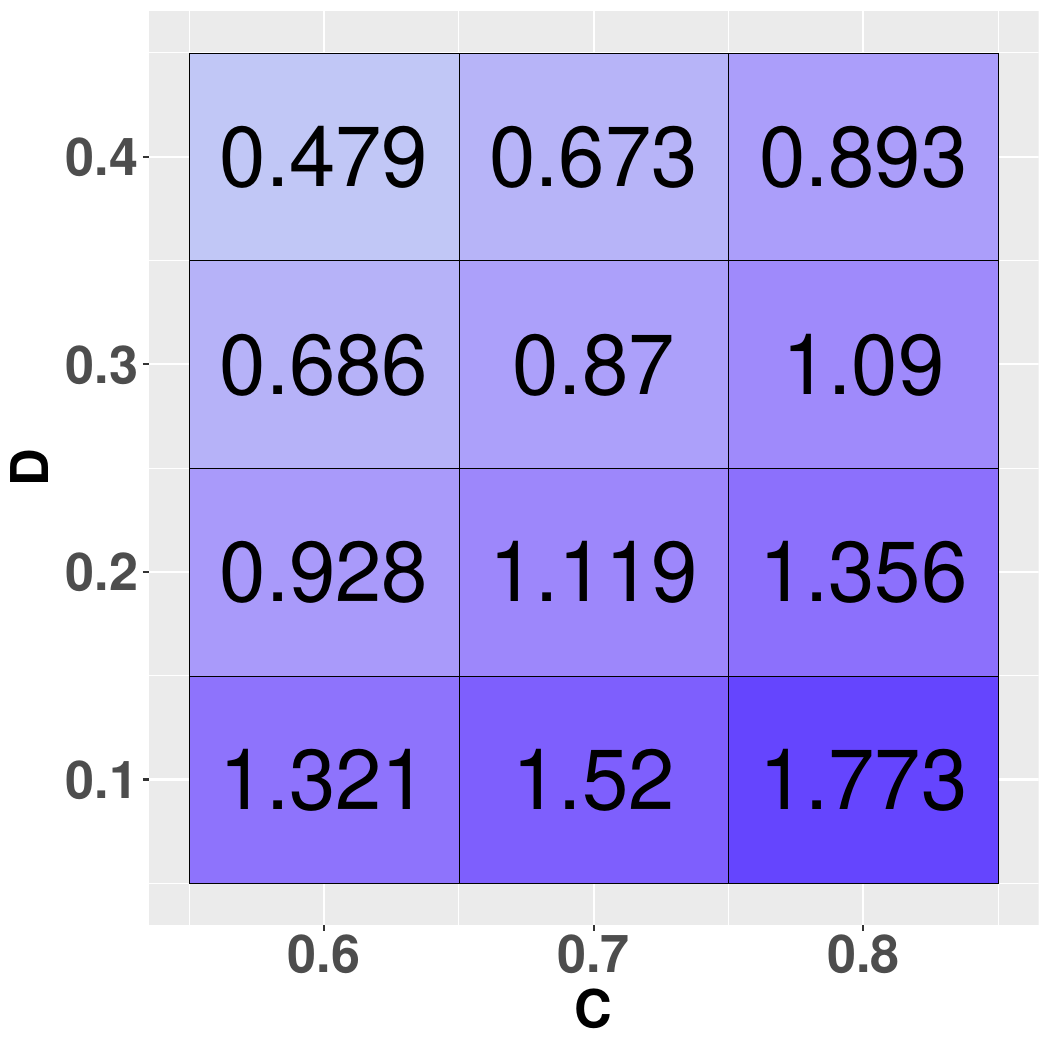} &
    \includegraphics[width=0.1\textwidth, valign = m]{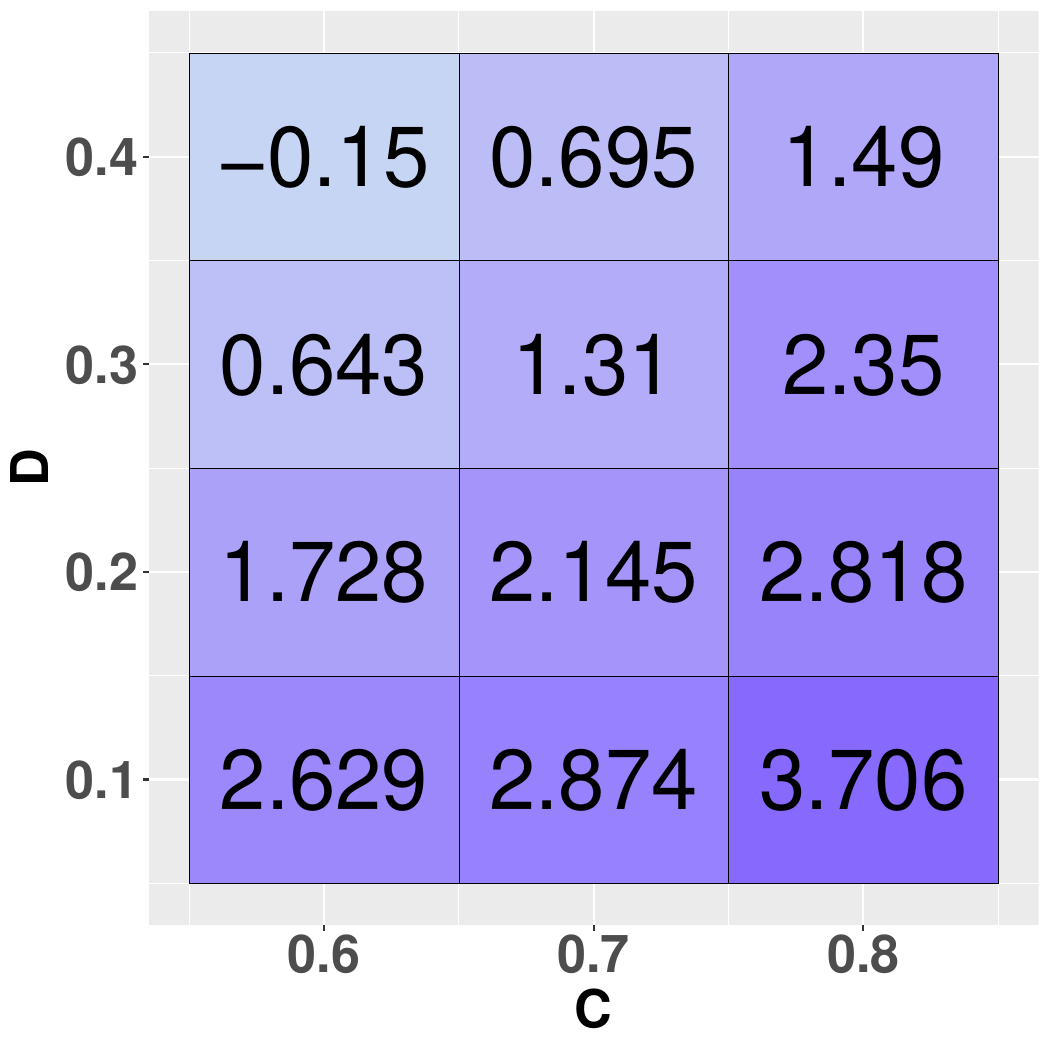} &
    \includegraphics[width=0.1\textwidth, valign = m]{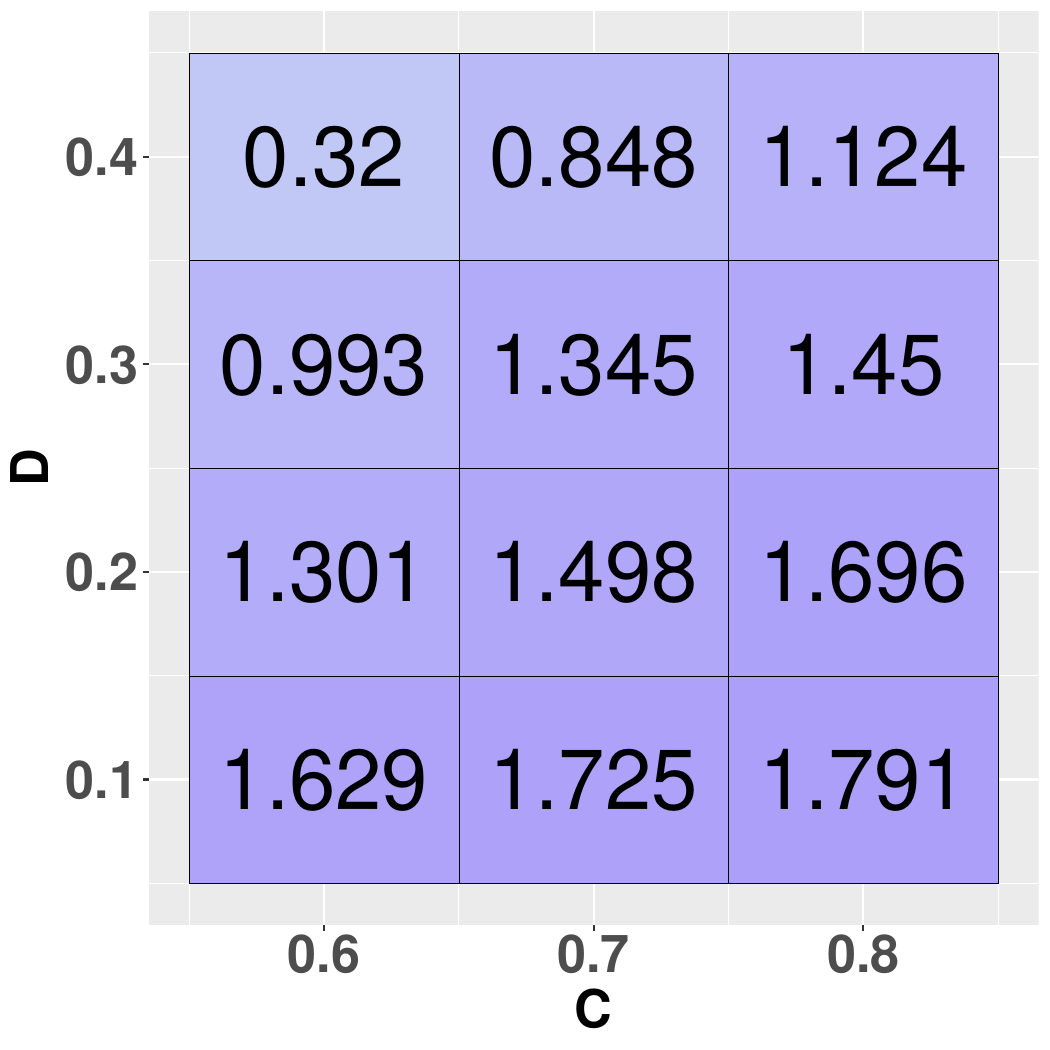} &
    \includegraphics[width=0.1\textwidth, valign = m]{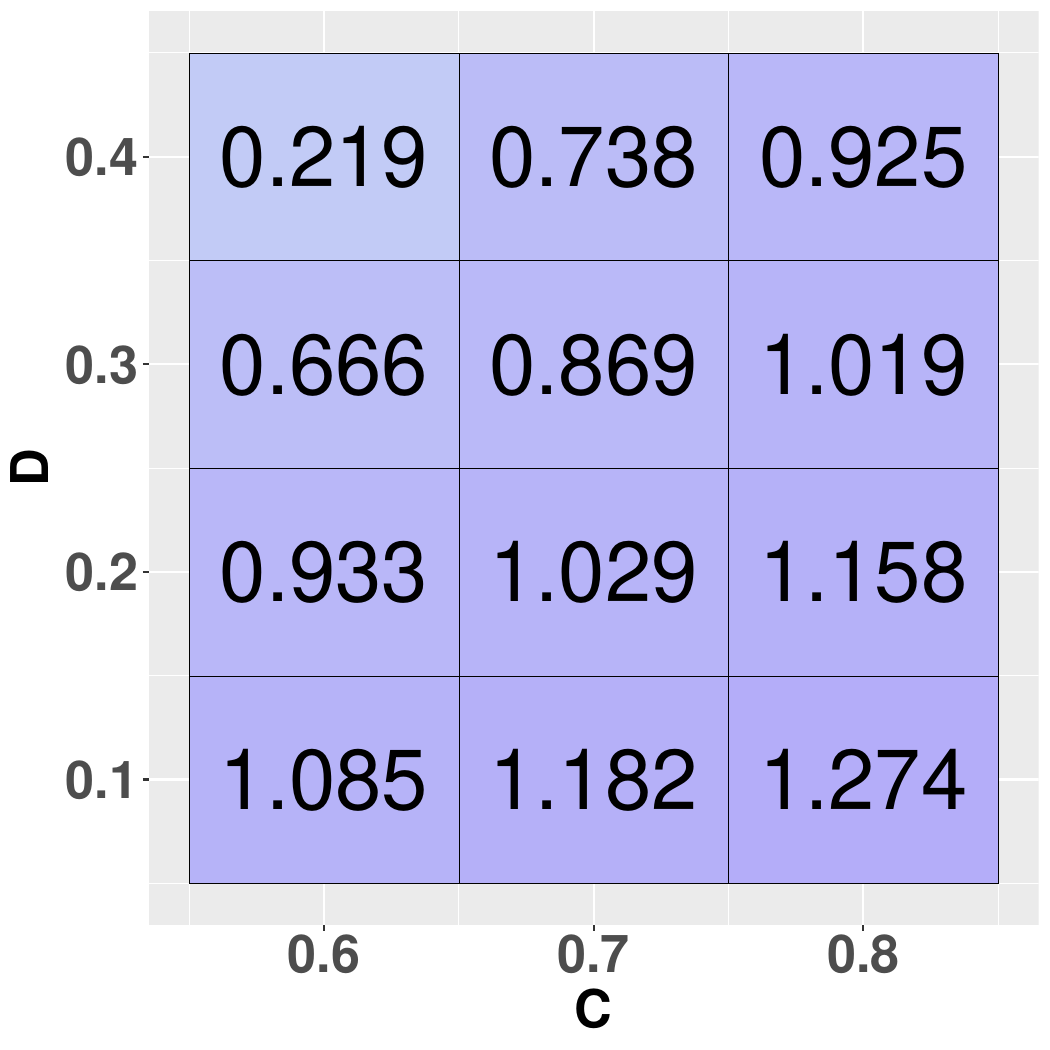} &
    \includegraphics[width=0.1\textwidth, valign = m]{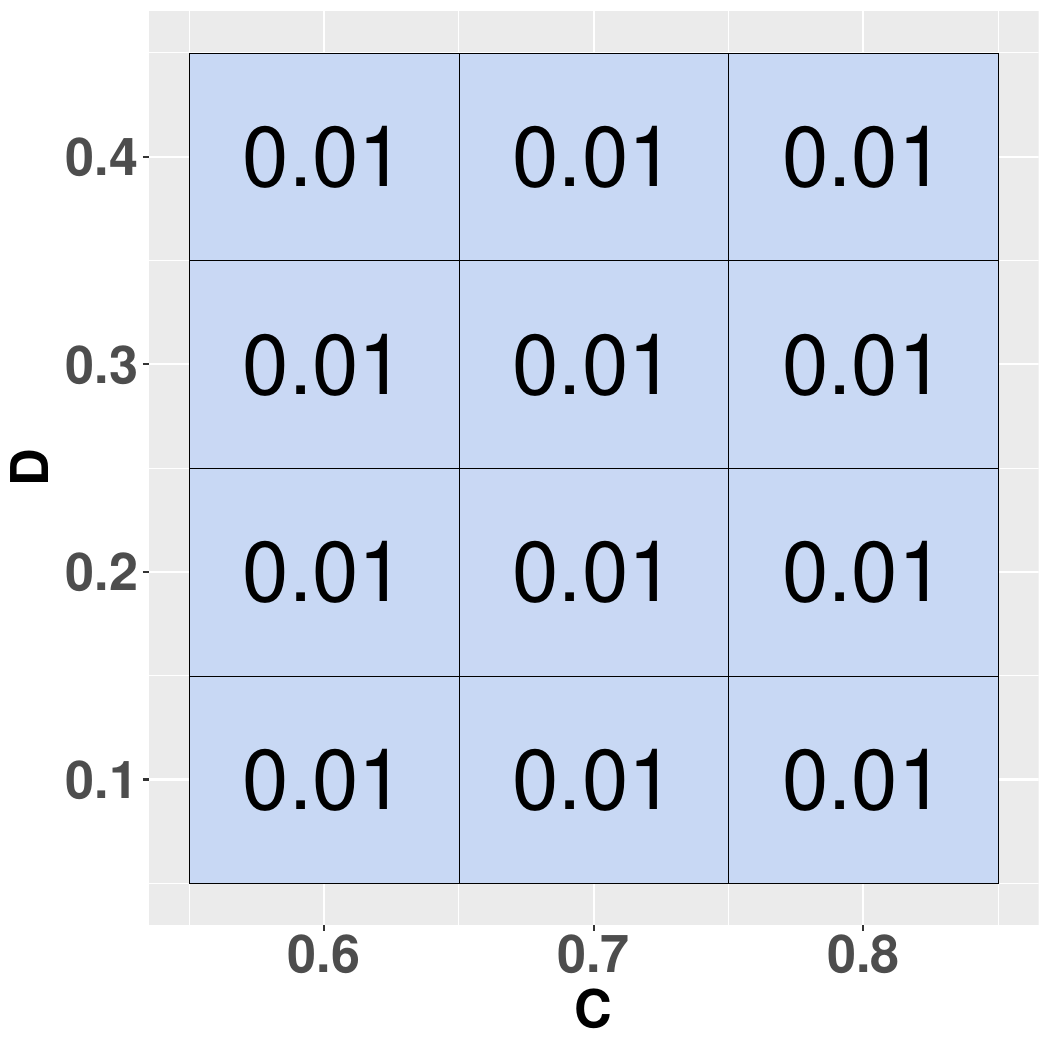} &
    \includegraphics[width=0.1\textwidth, valign = m]{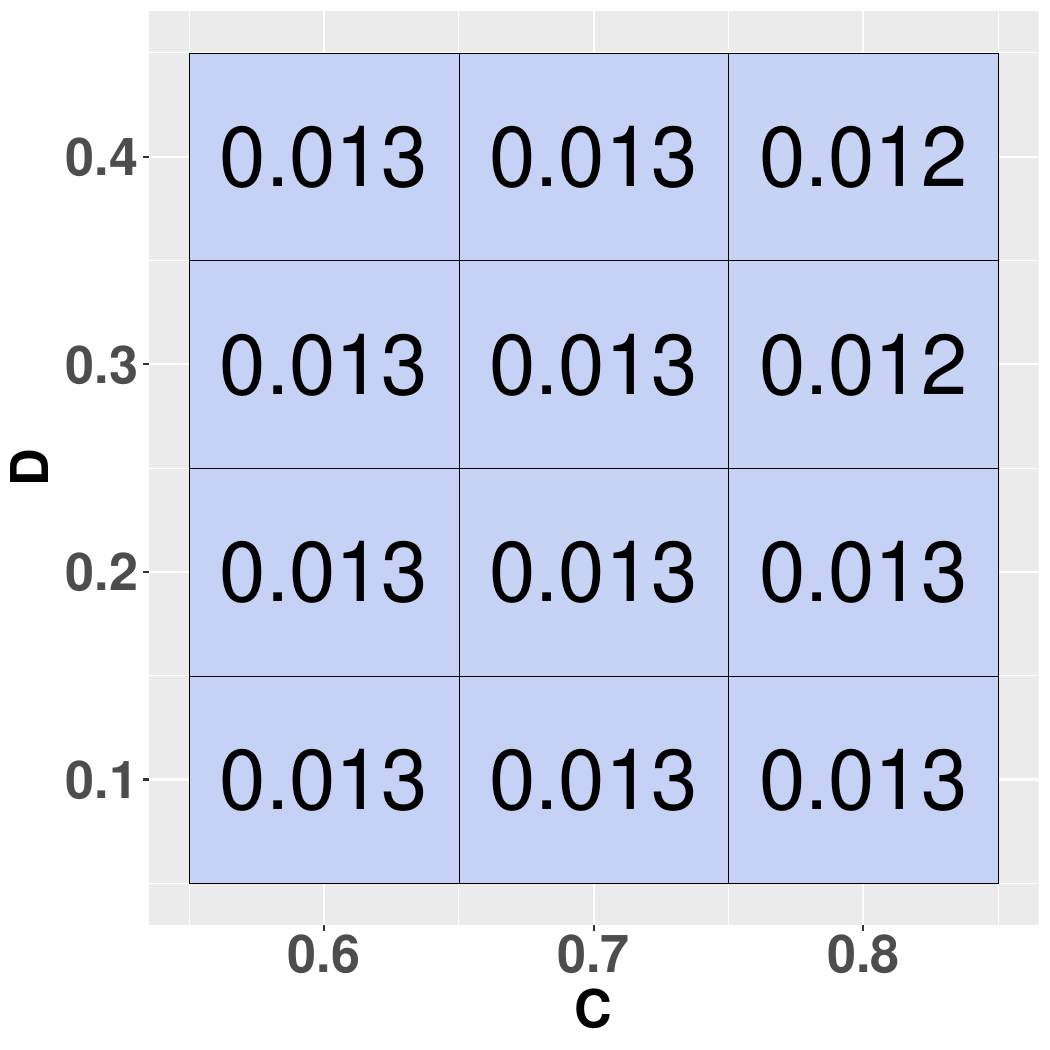} &
    \includegraphics[width=0.1\textwidth, valign = m]{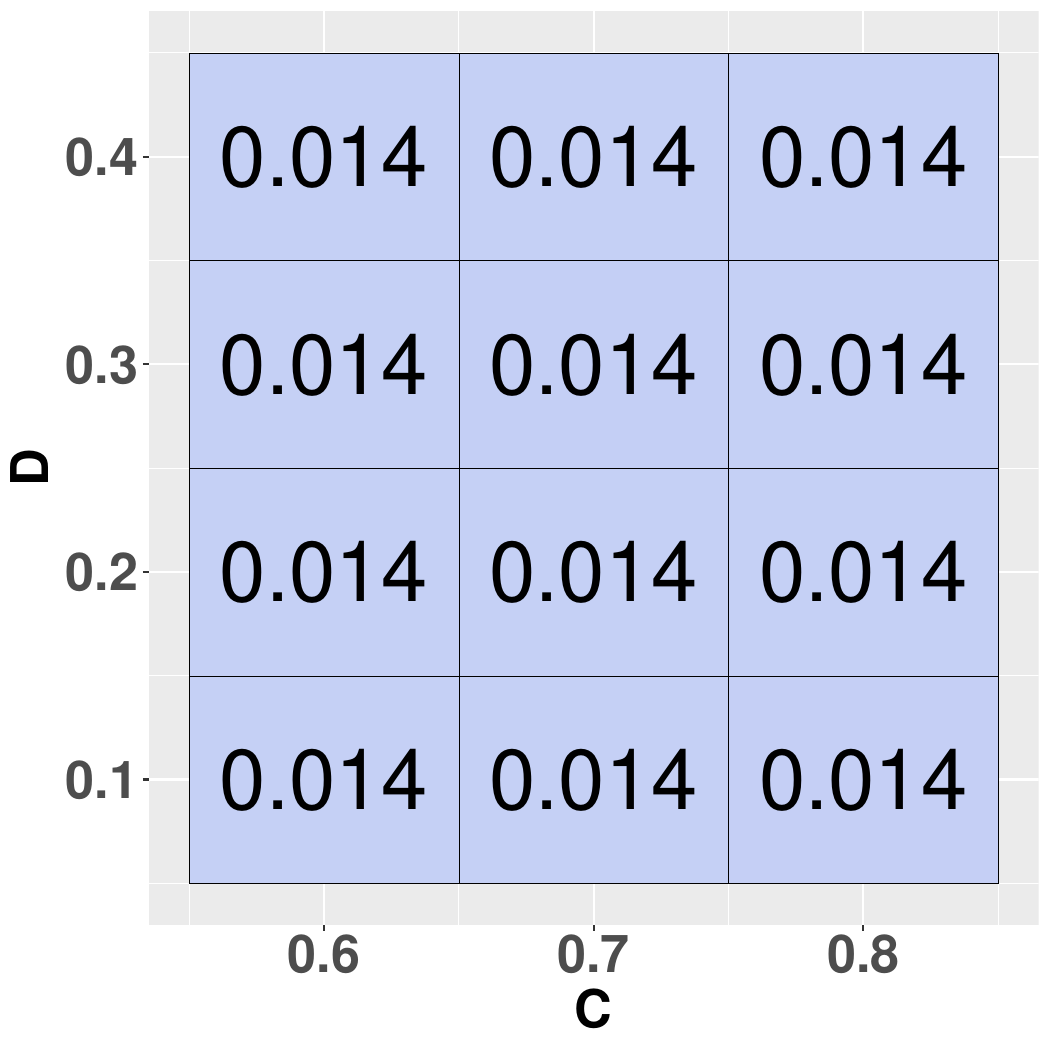} \\ 
    & Scenario 2 &  
    \includegraphics[width=0.1\textwidth, valign = m]{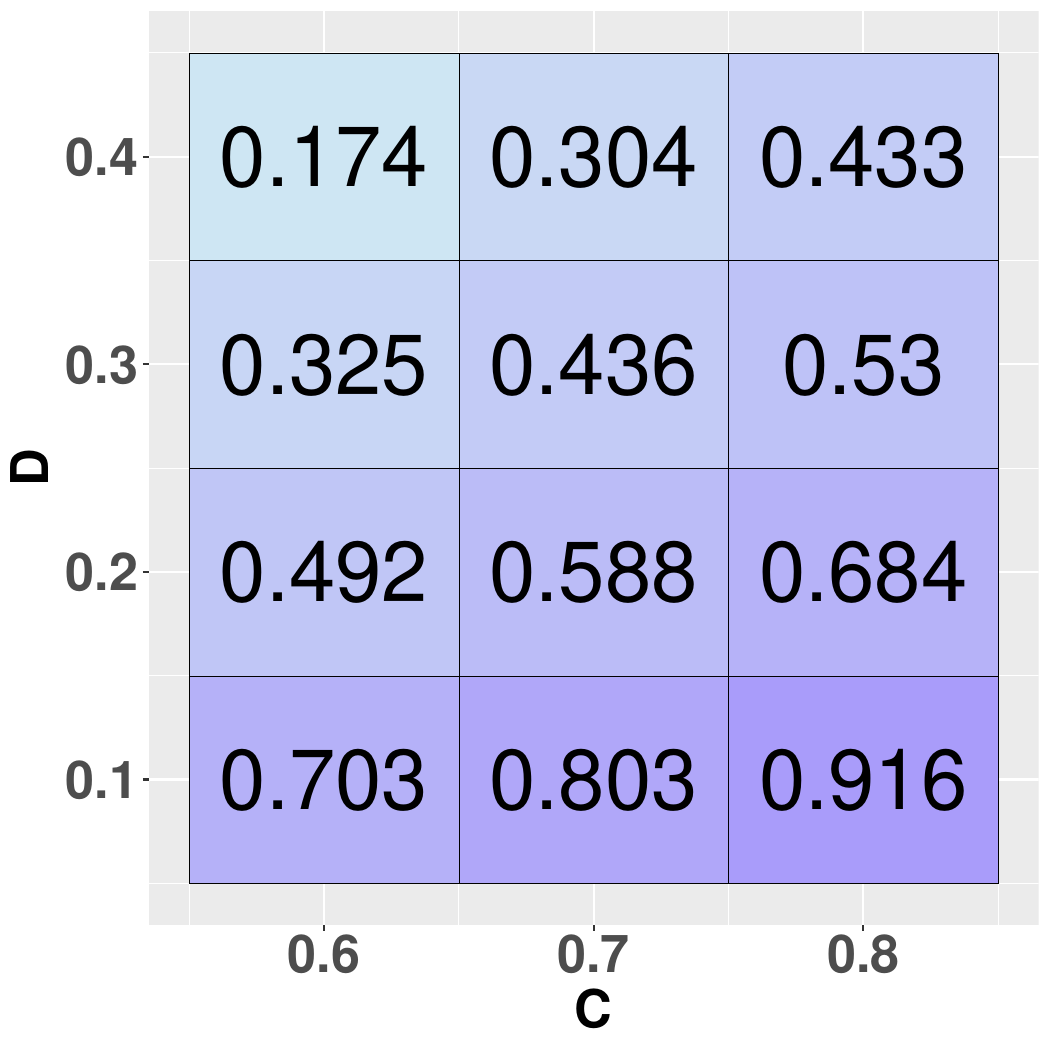} &
    \includegraphics[width=0.1\textwidth, valign = m]{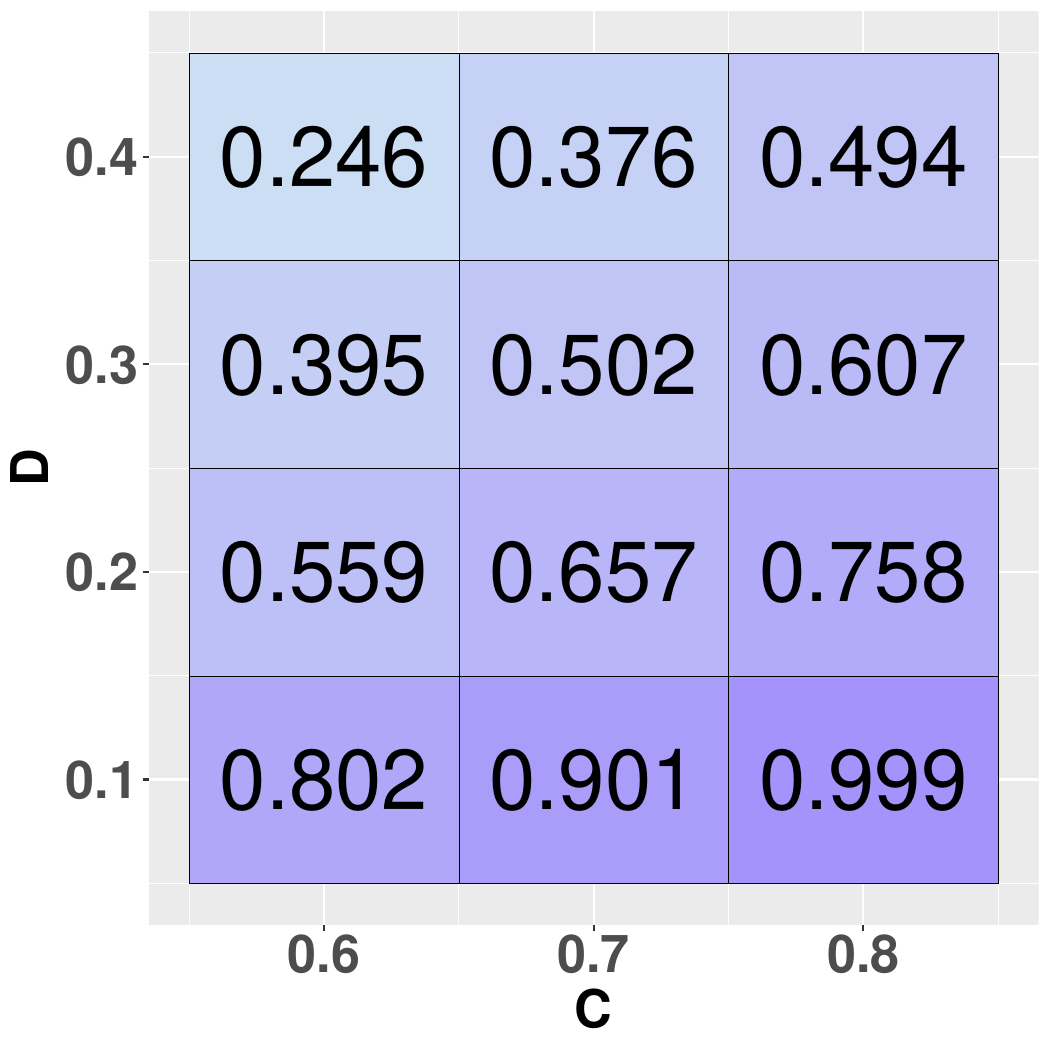} &
    \includegraphics[width=0.1\textwidth, valign = m]{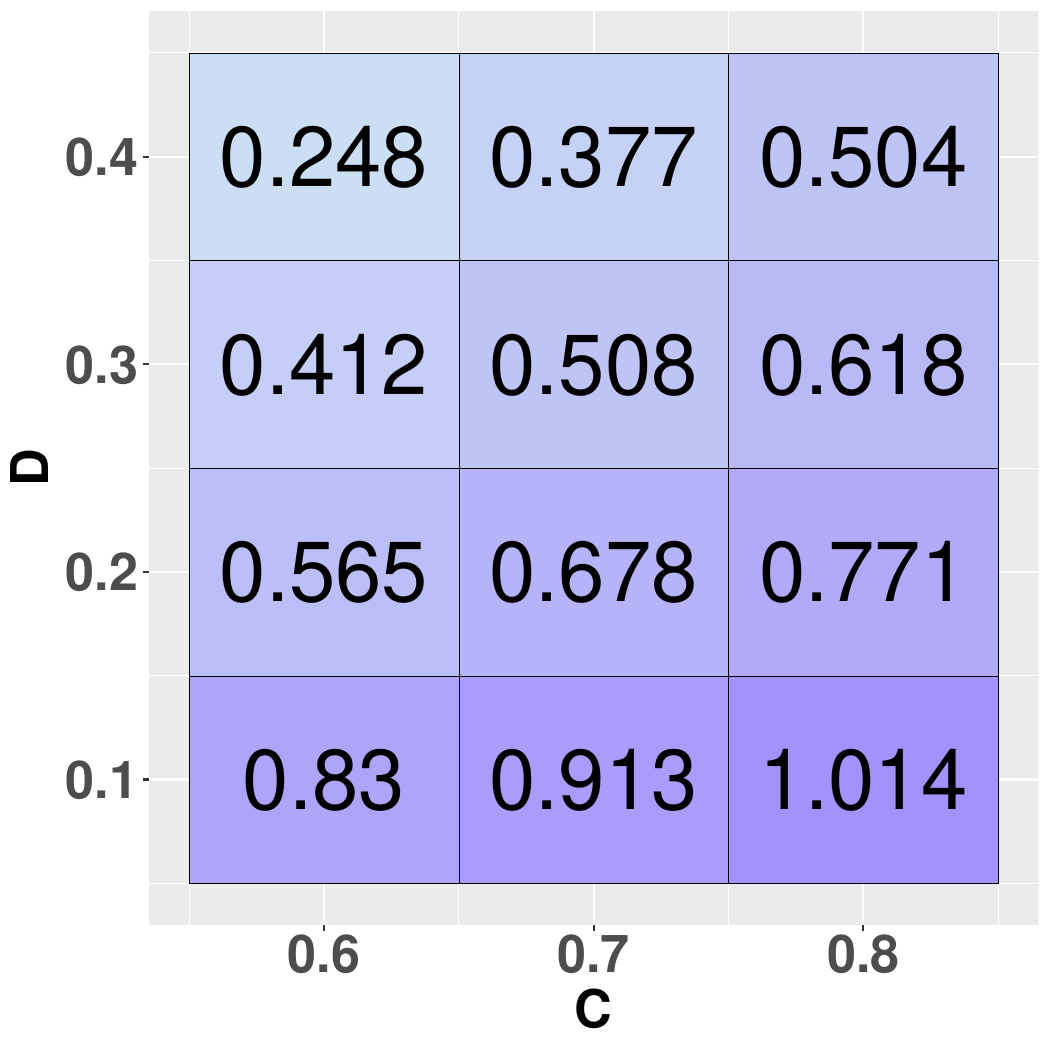} &
    \includegraphics[width=0.1\textwidth, valign = m]{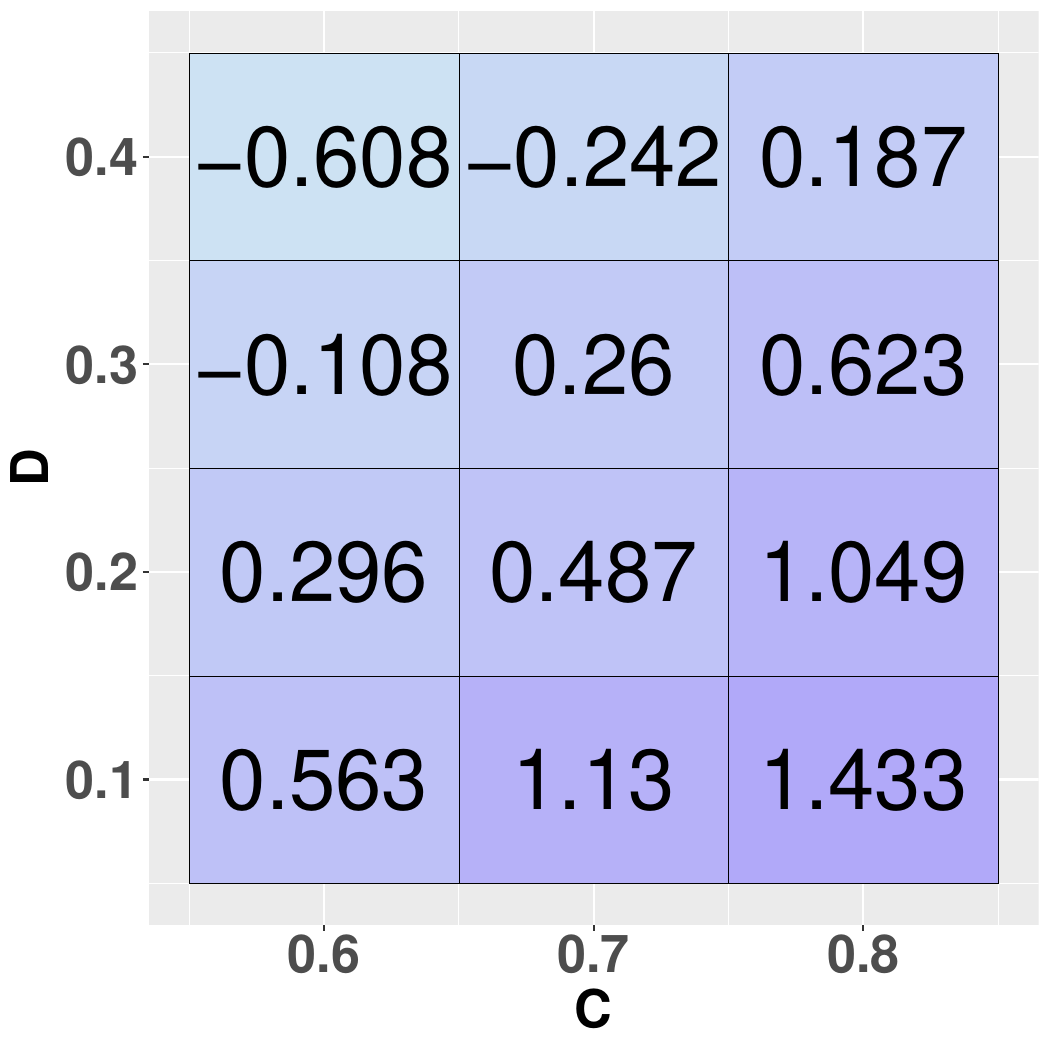} &
    \includegraphics[width=0.1\textwidth, valign = m]{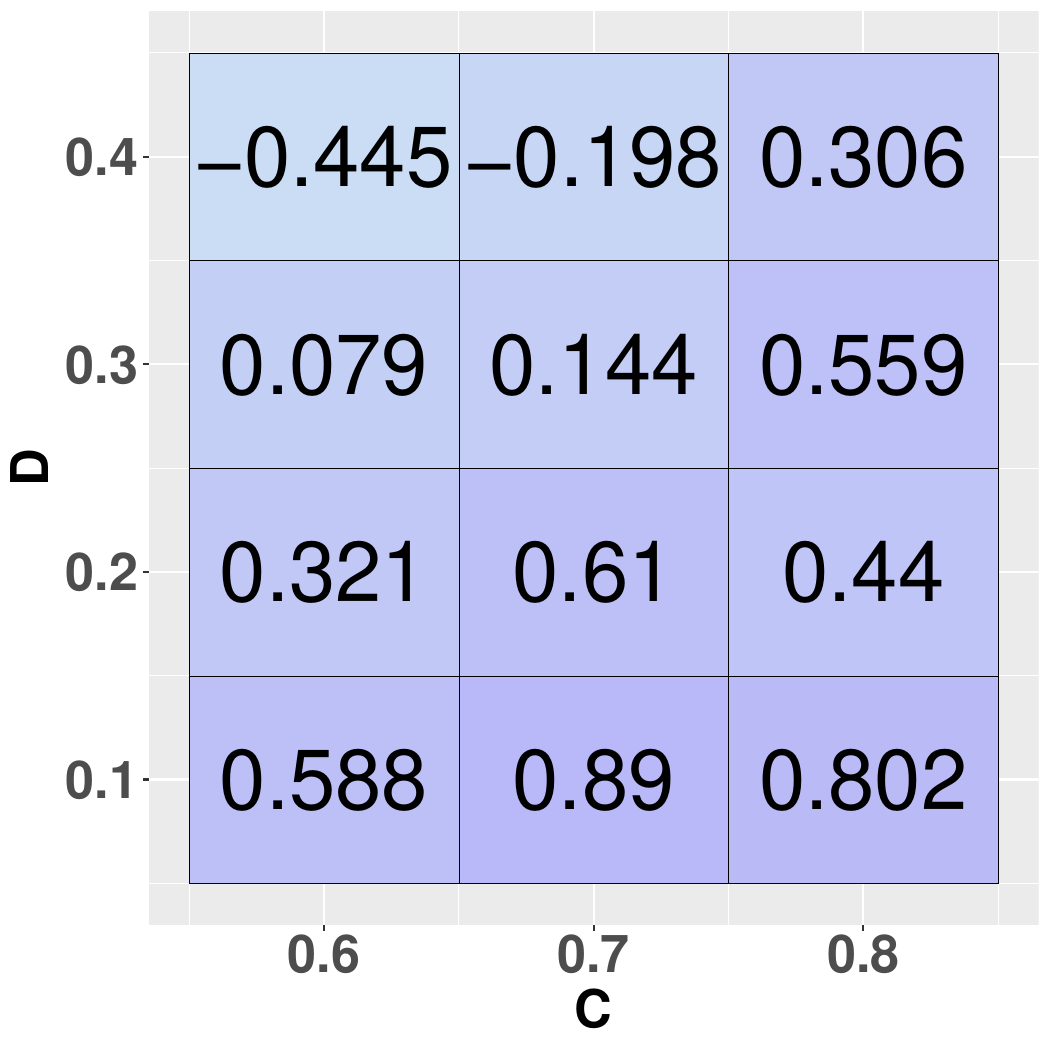} &
    \includegraphics[width=0.1\textwidth, valign = m]{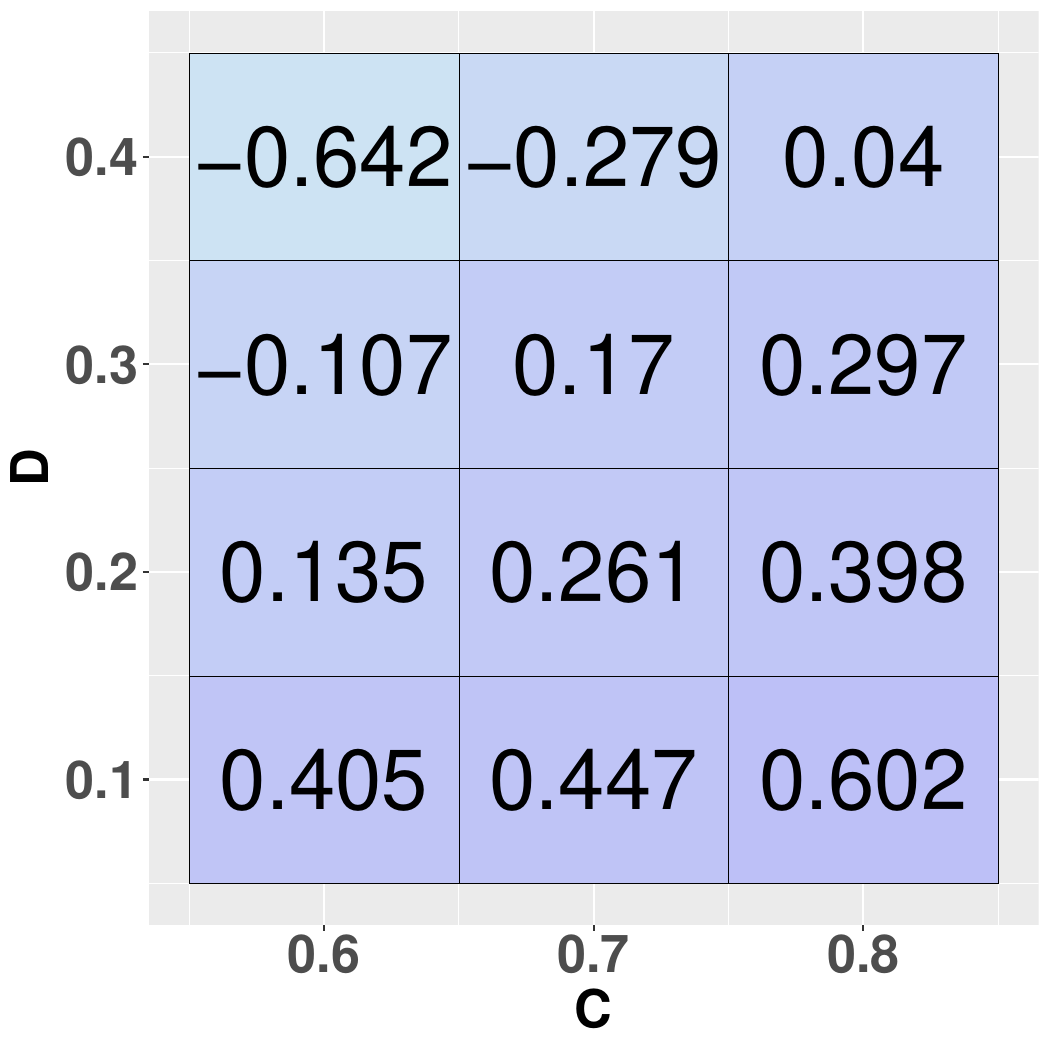} &
    \includegraphics[width=0.1\textwidth, valign = m]{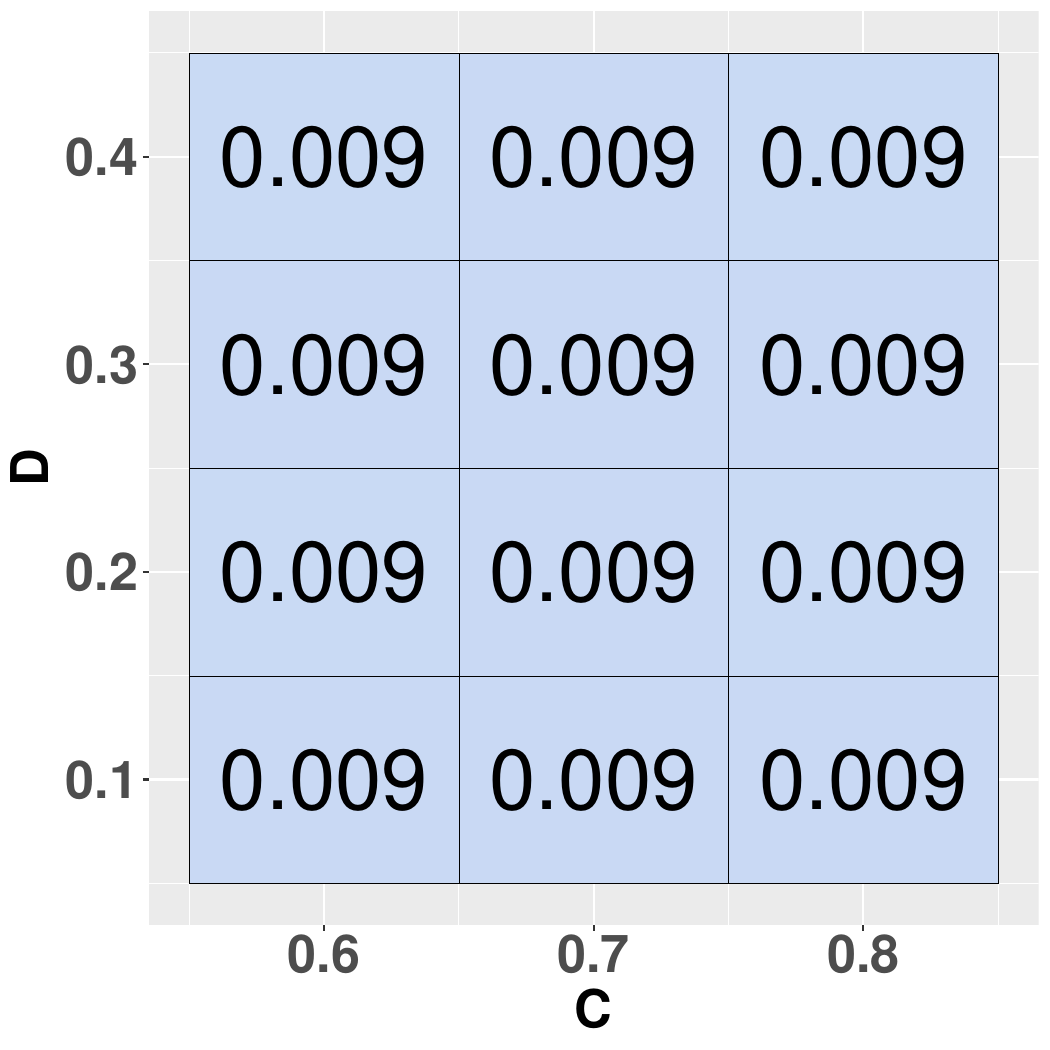} &
    \includegraphics[width=0.1\textwidth, valign = m]{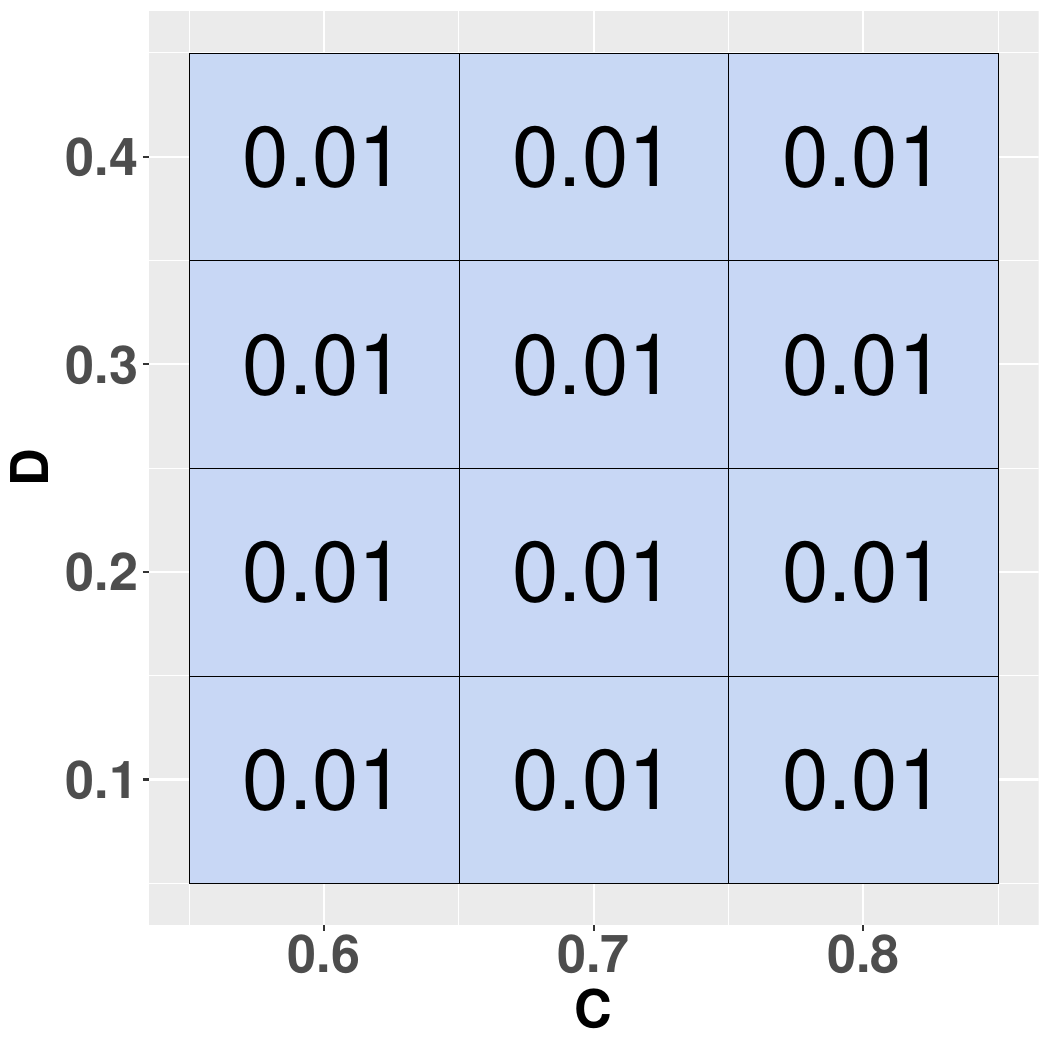} &
    \includegraphics[width=0.1\textwidth, valign = m]{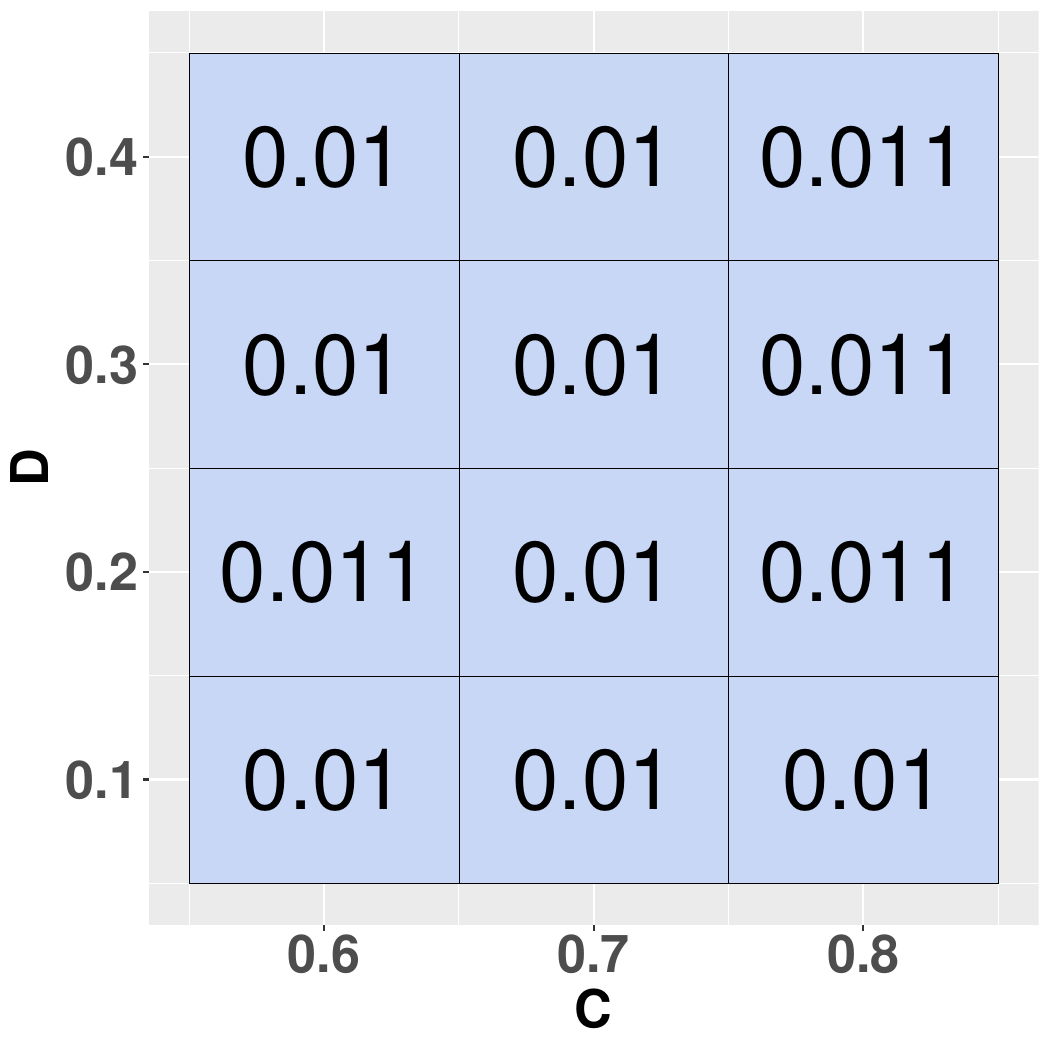} \\ \hline
    \multirow{8}{*}{Item Response} 
    & Scenario 1.1 &   
    \includegraphics[width=0.1\textwidth, valign = m]{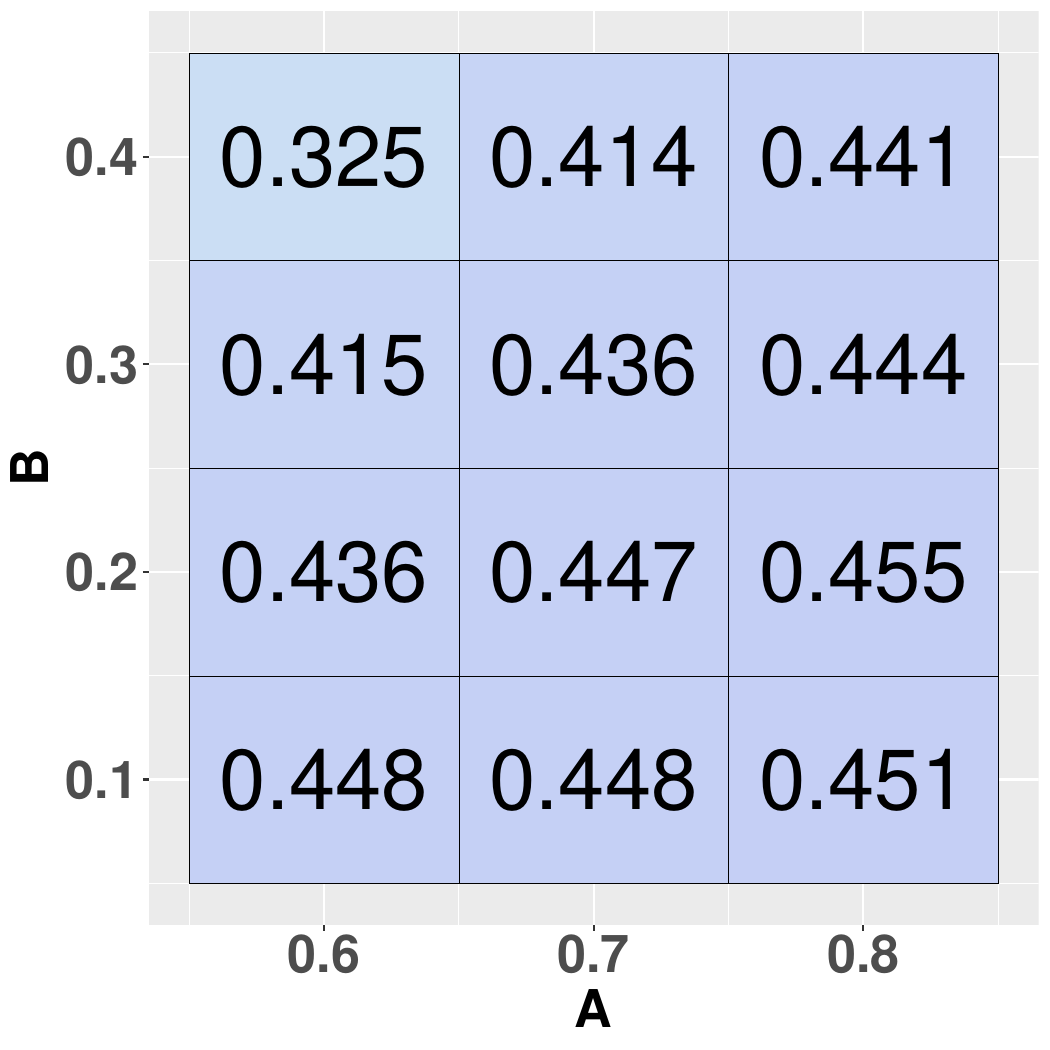} &
    \includegraphics[width=0.1\textwidth, valign = m]{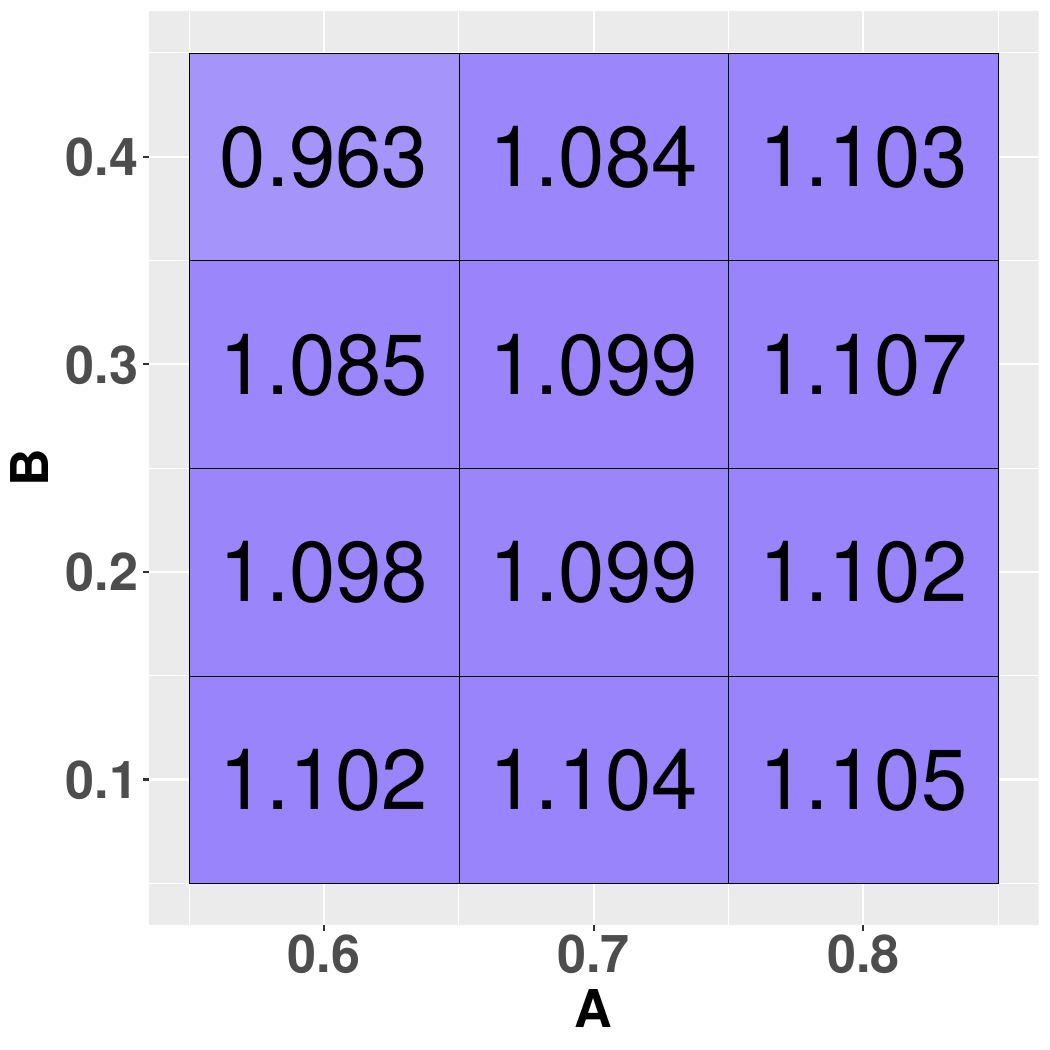} &
    \includegraphics[width=0.1\textwidth, valign = m]{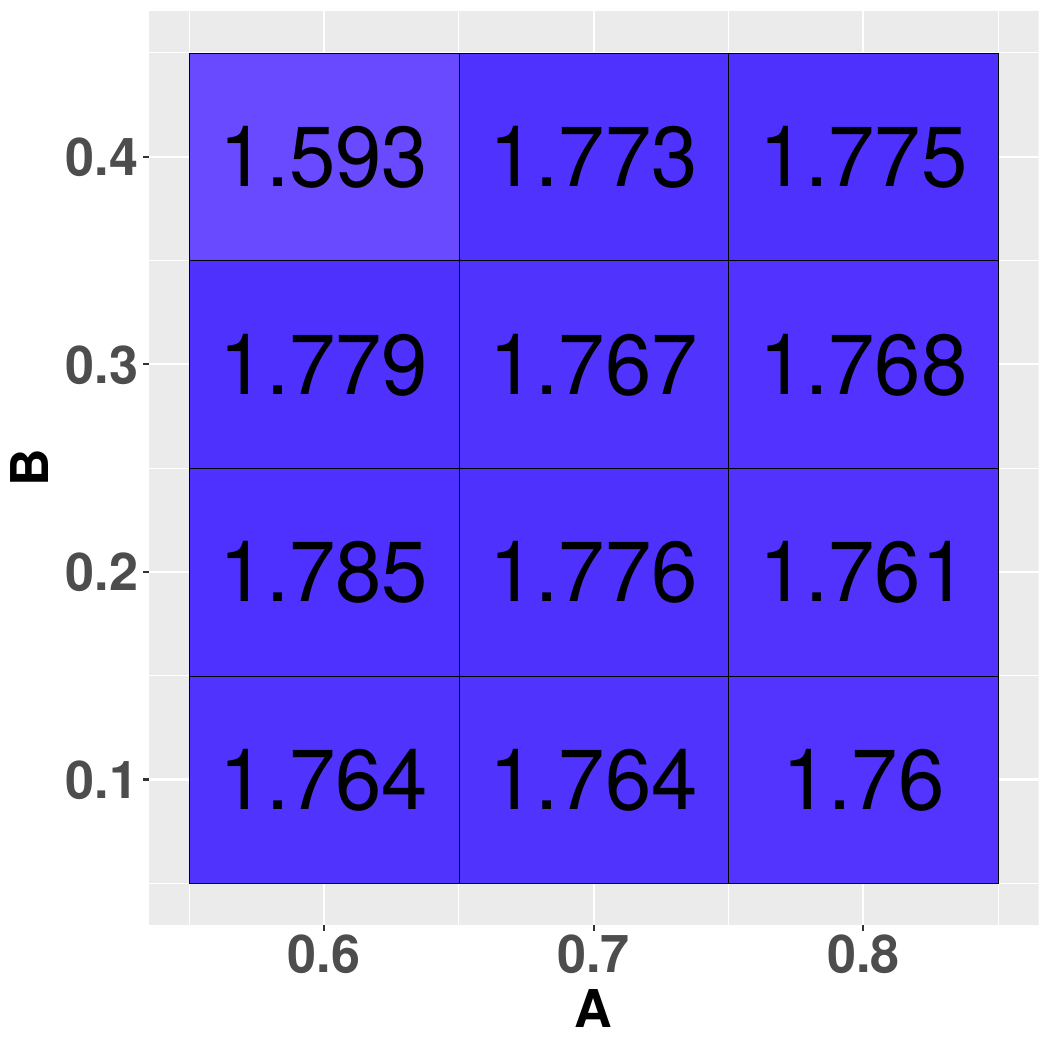} &
    \includegraphics[width=0.1\textwidth, valign = m]{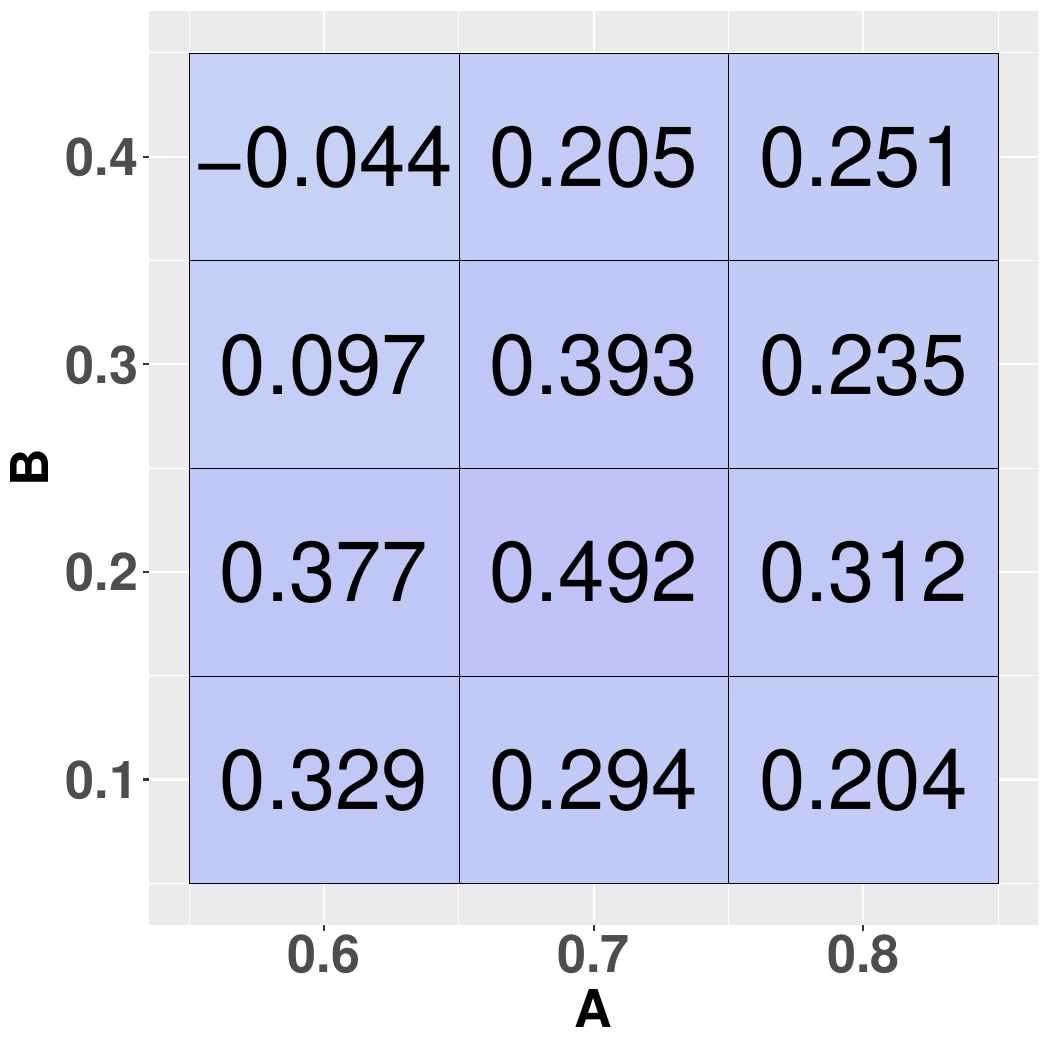} &
    \includegraphics[width=0.1\textwidth, valign = m]{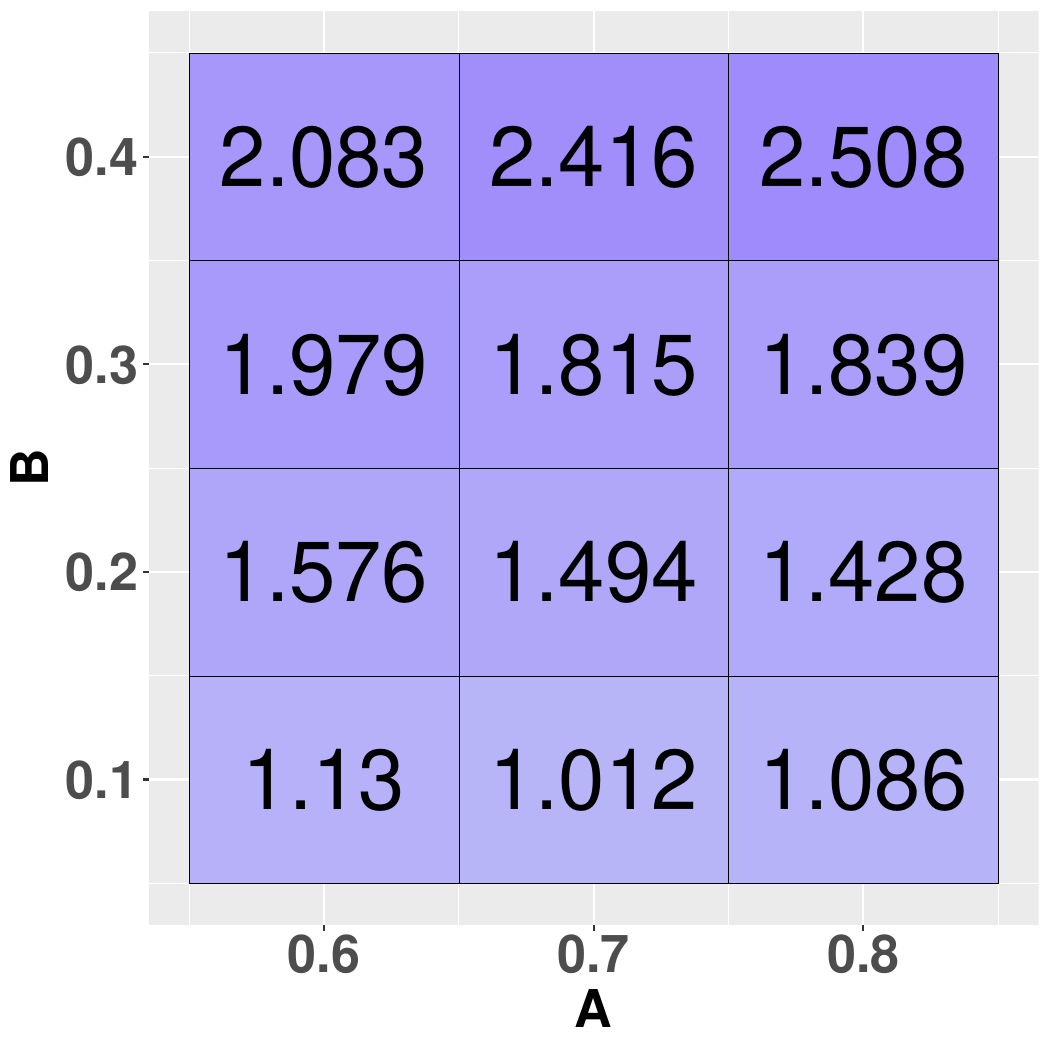} &
    \includegraphics[width=0.1\textwidth, valign = m]{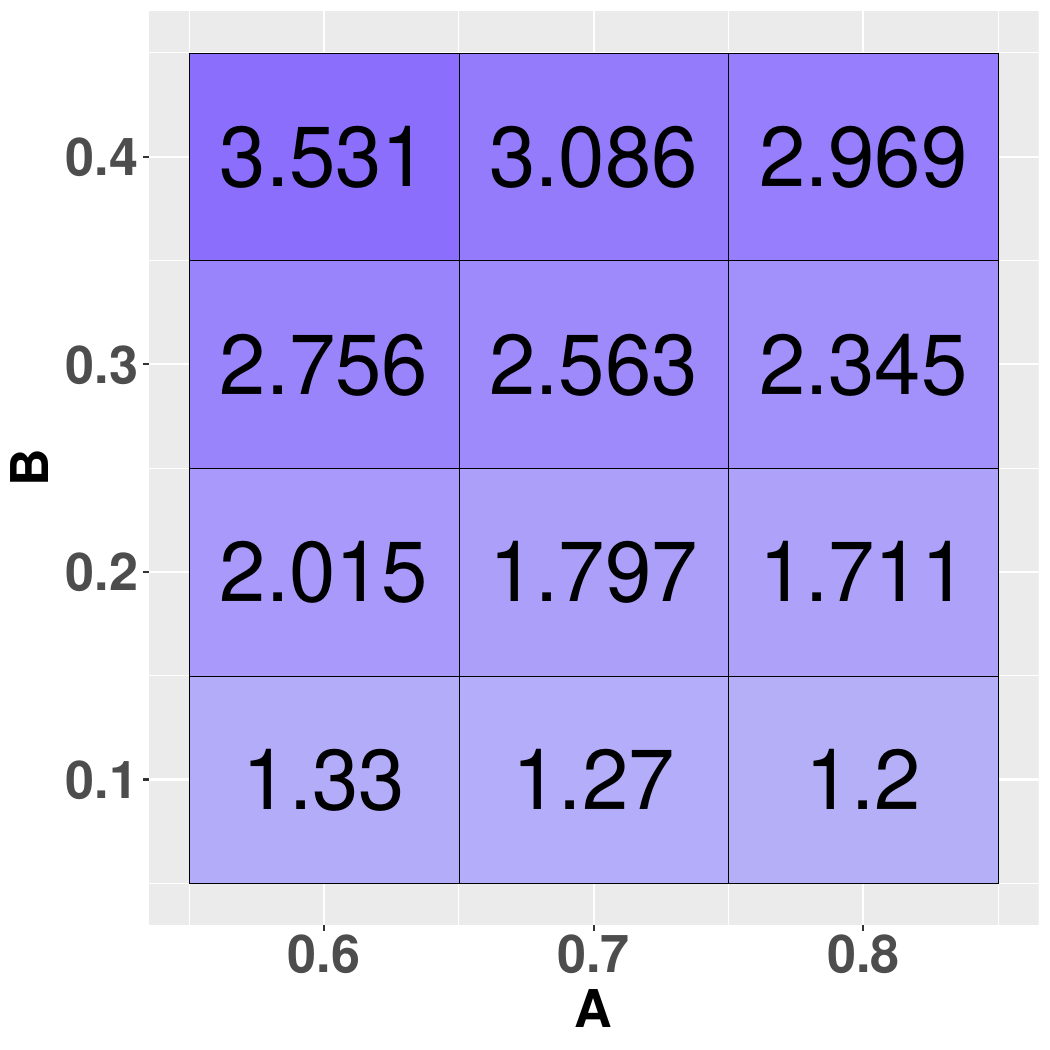} &
    \includegraphics[width=0.1\textwidth, valign = m]{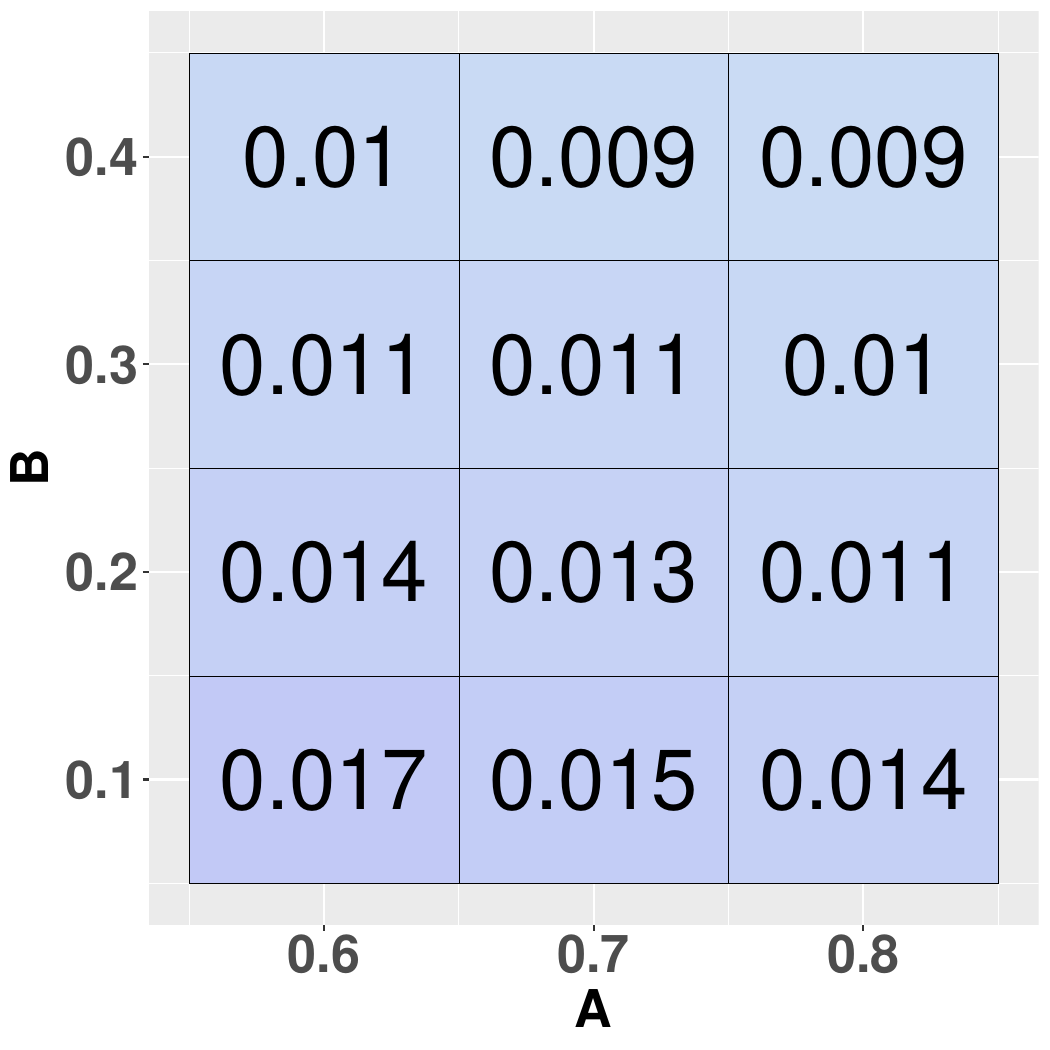} &
    \includegraphics[width=0.1\textwidth, valign = m]{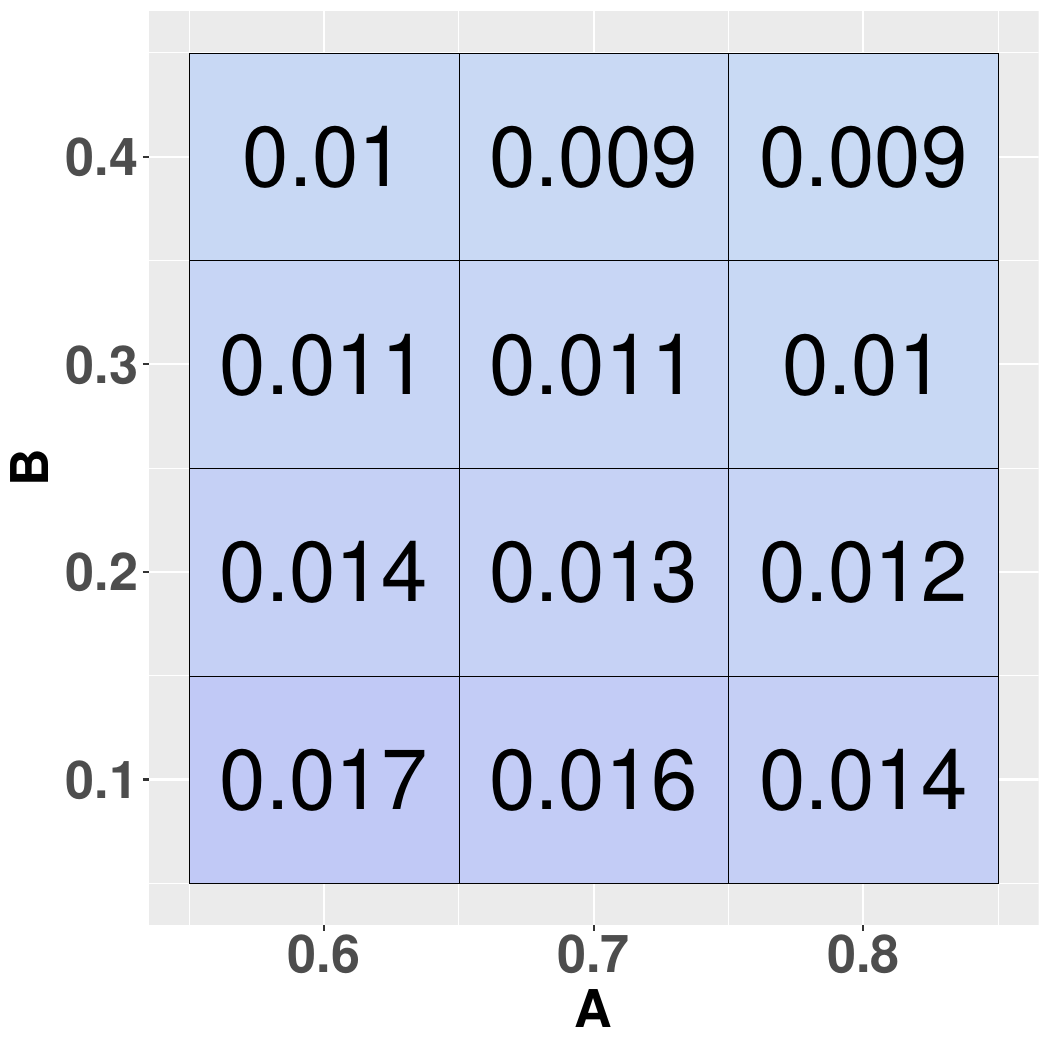} &
    \includegraphics[width=0.1\textwidth, valign = m]{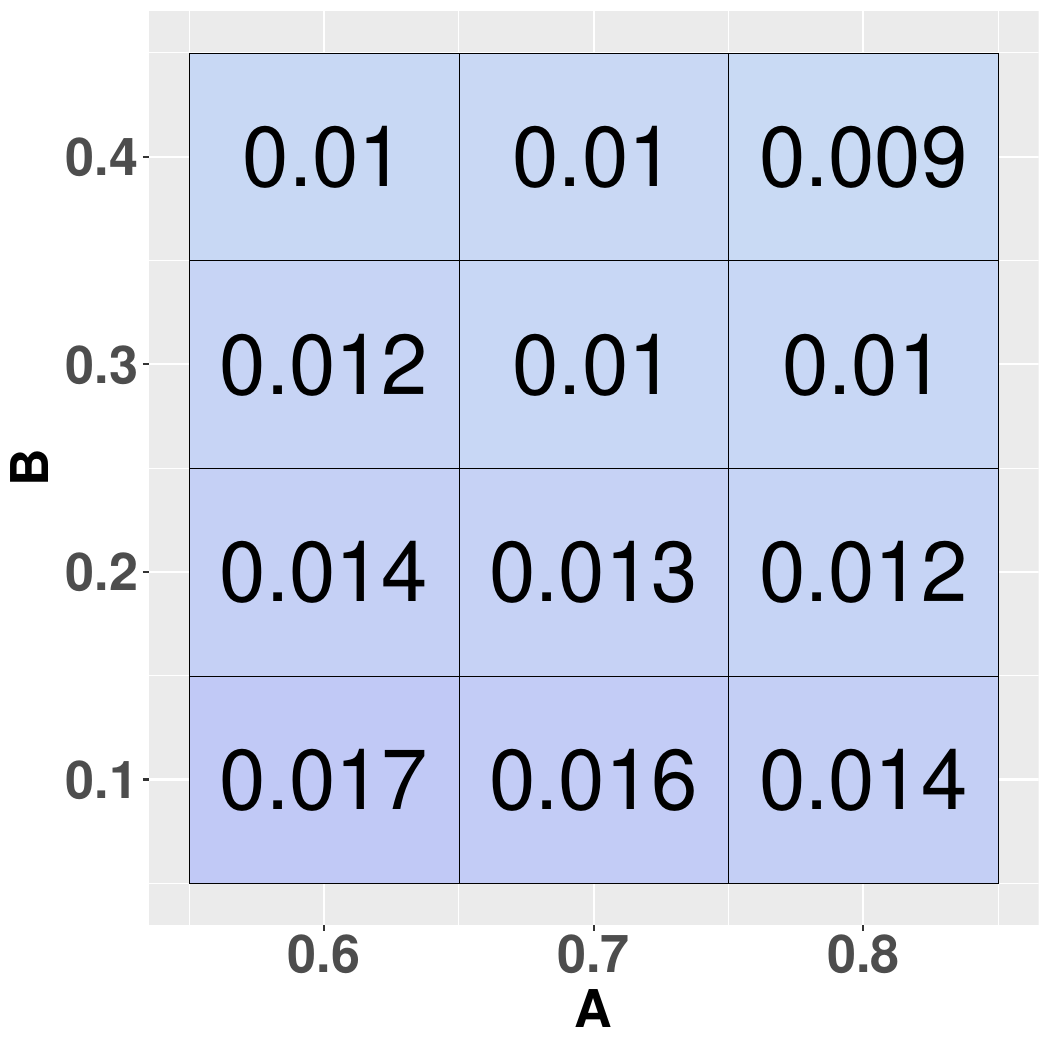} \\ 
    & Scenario 1.2 &  
    \includegraphics[width=0.1\textwidth, valign = m]{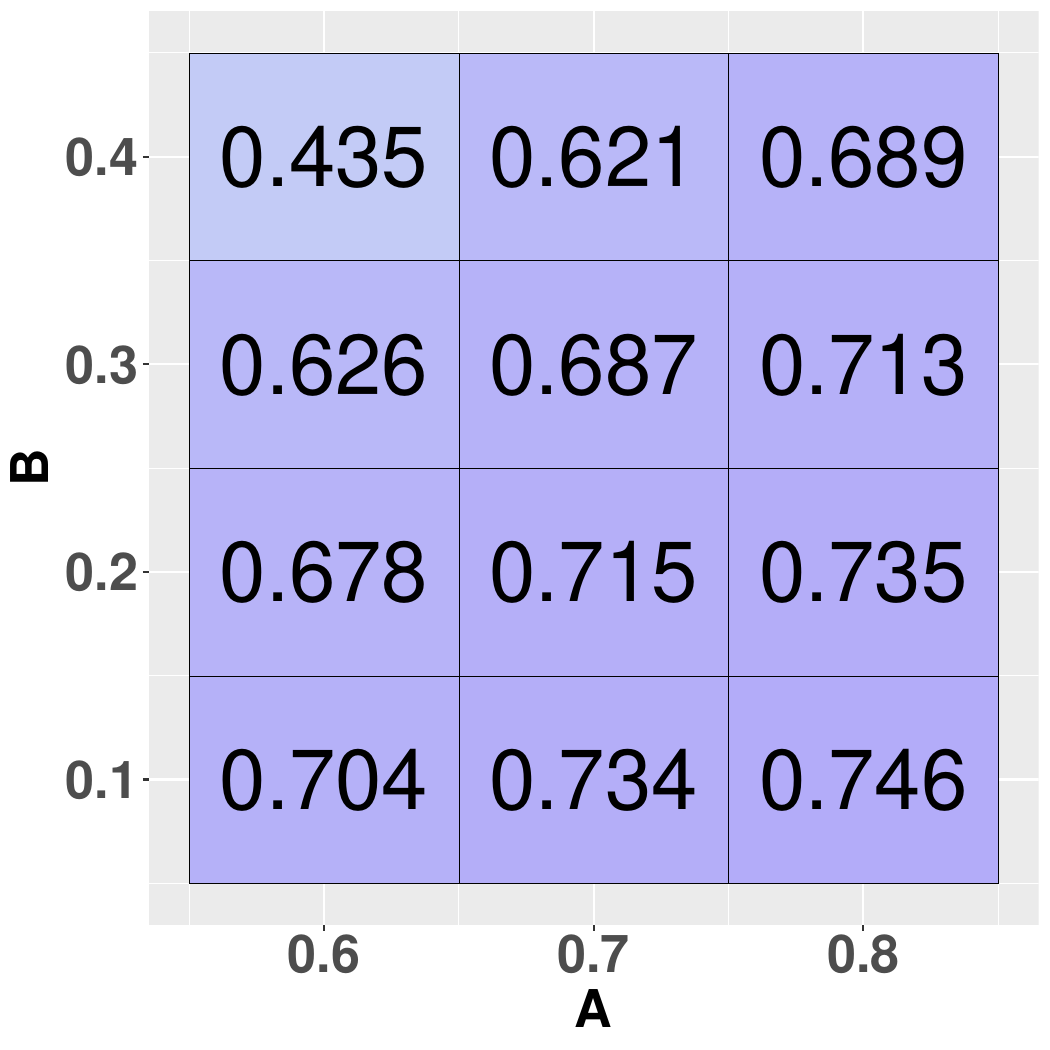} &
    \includegraphics[width=0.1\textwidth, valign = m]{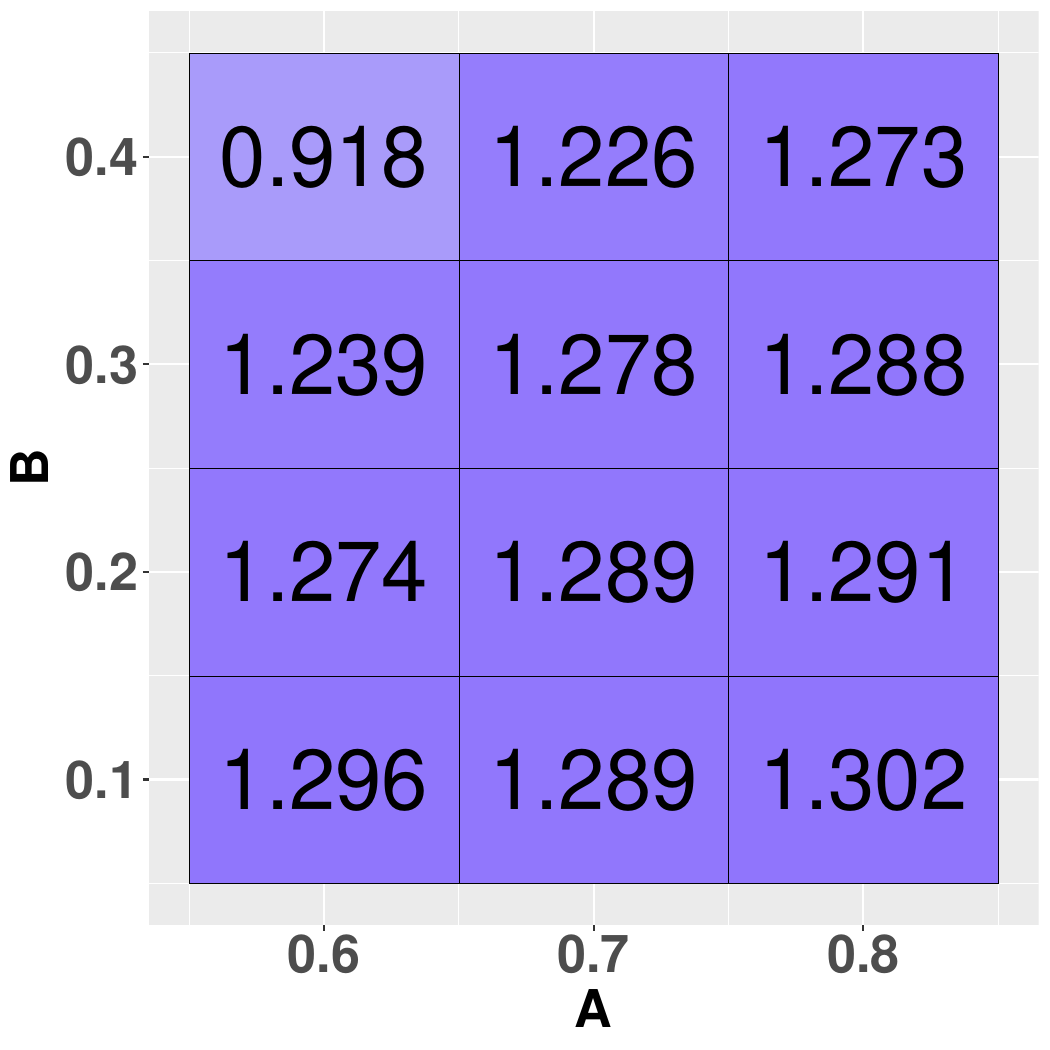} &
    \includegraphics[width=0.1\textwidth, valign = m]{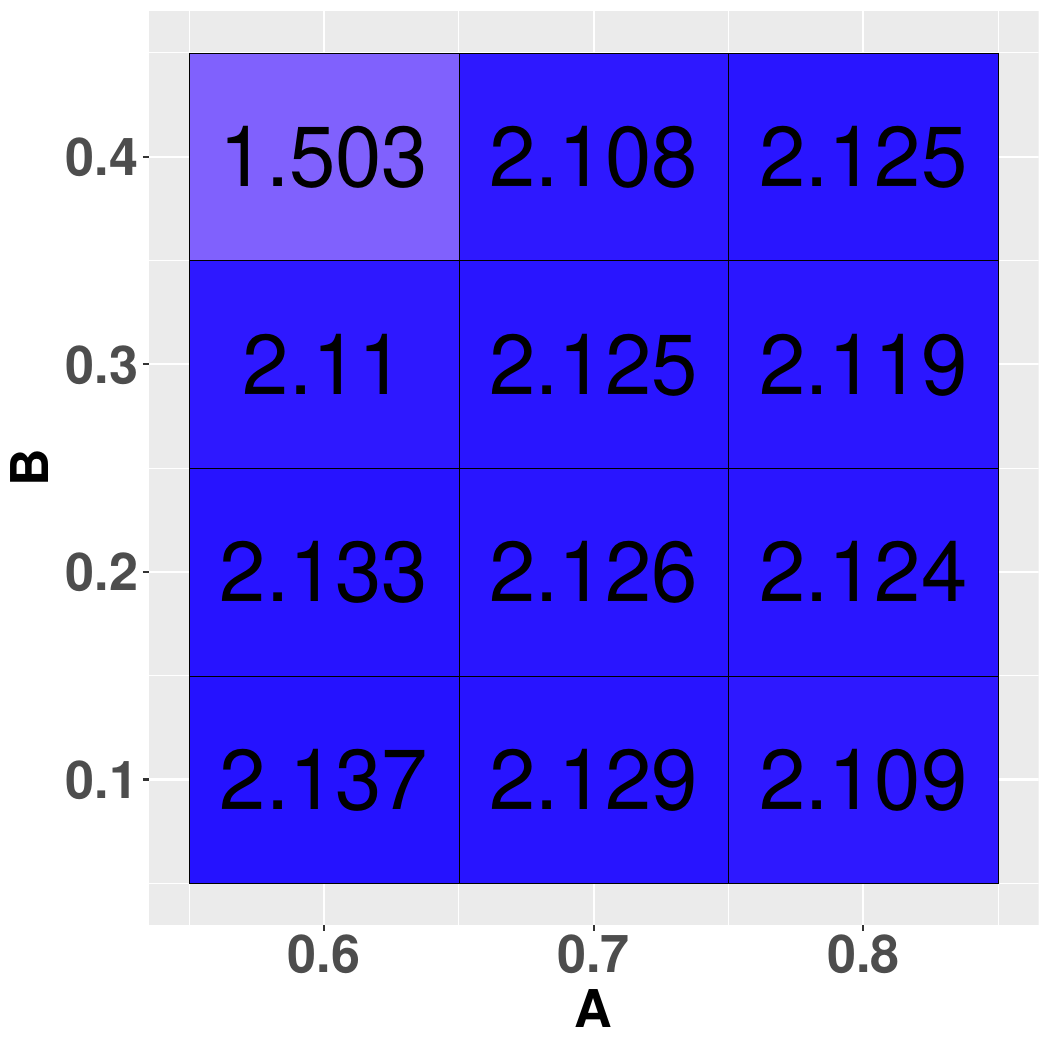} &
    \includegraphics[width=0.1\textwidth, valign = m]{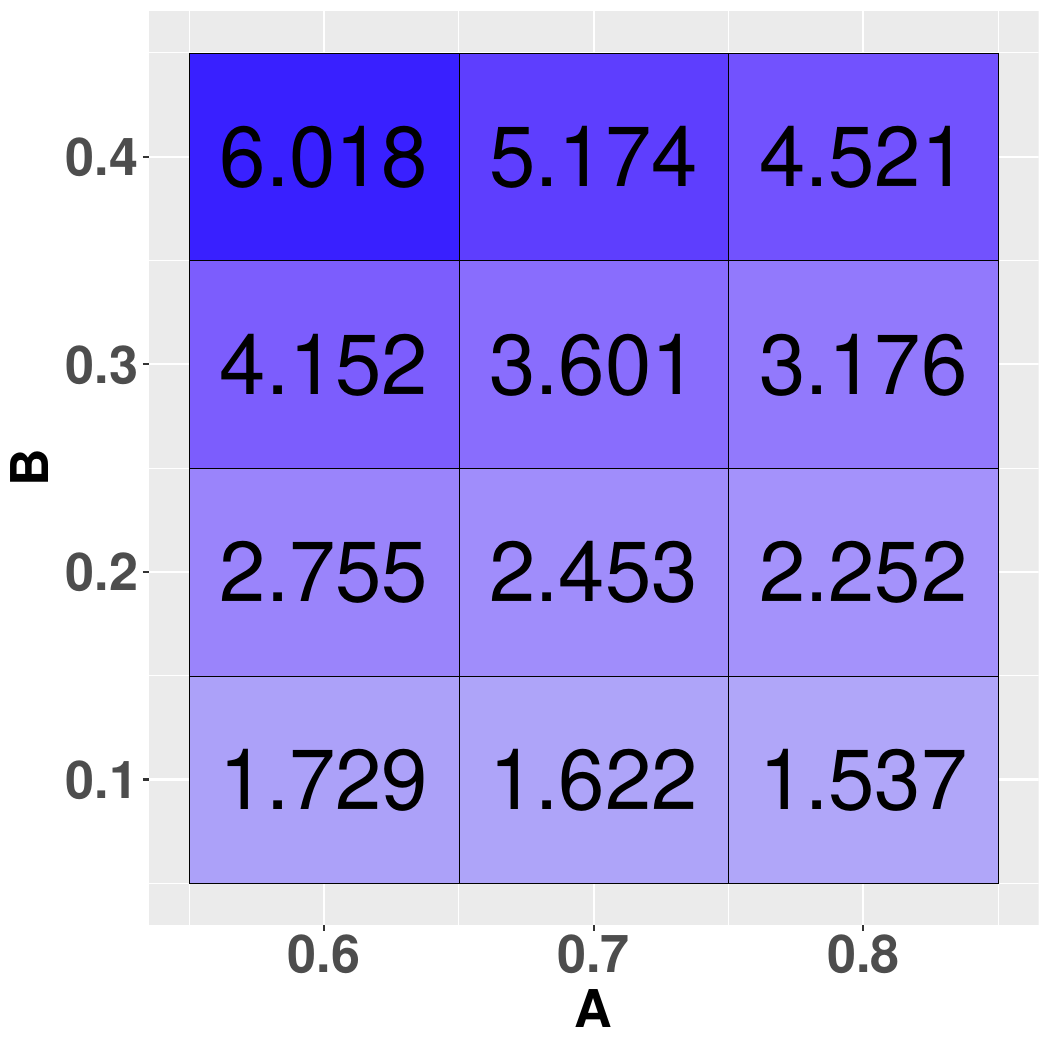} &
    \includegraphics[width=0.1\textwidth, valign = m]{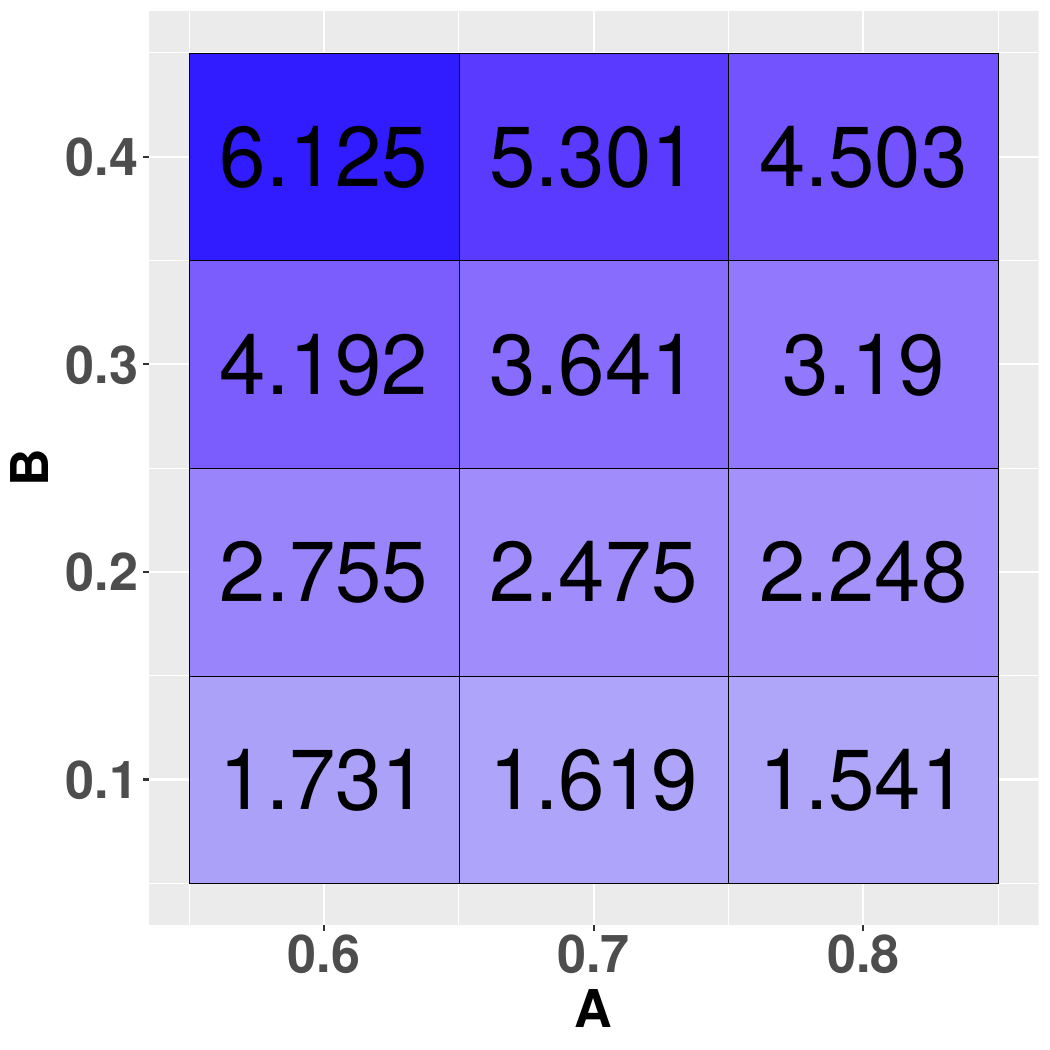} &
    \includegraphics[width=0.1\textwidth, valign = m]{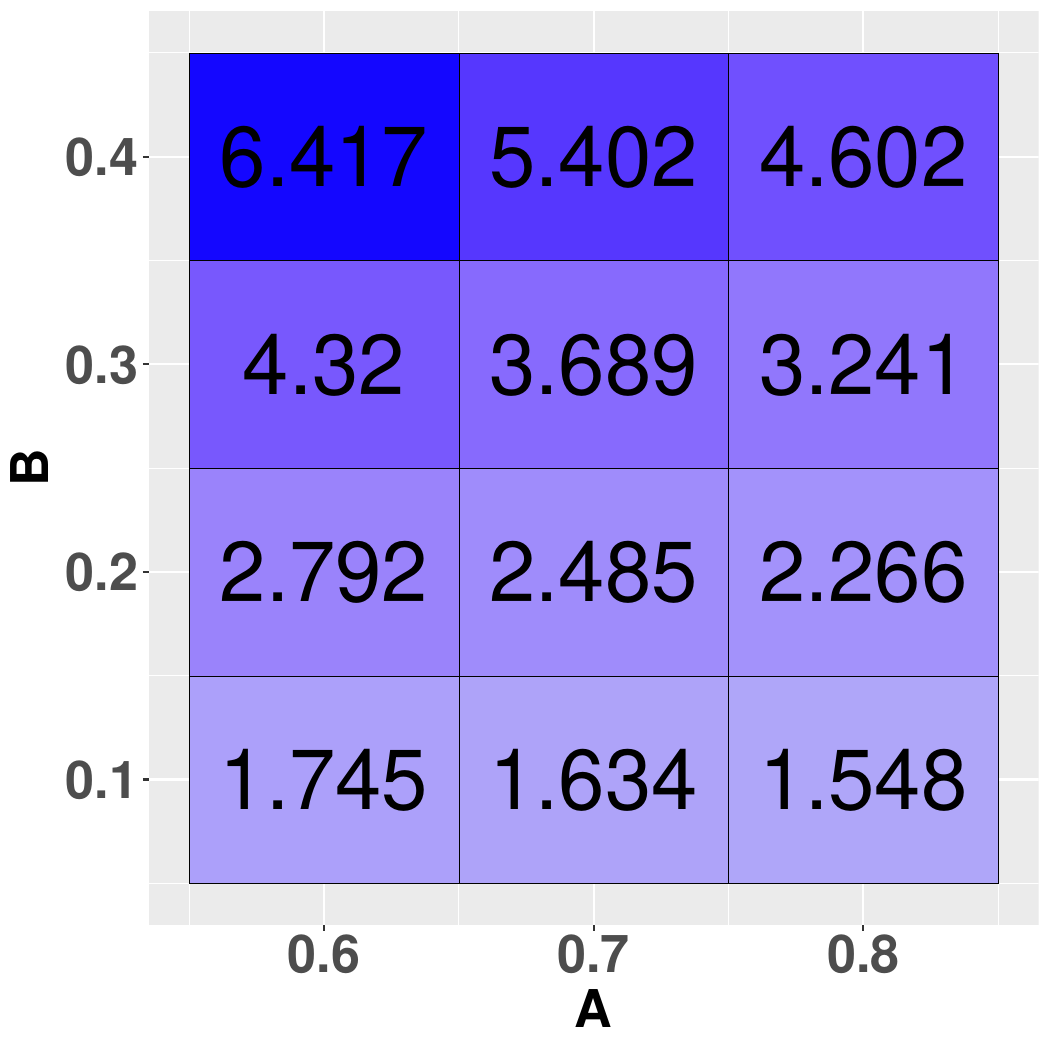} & 
    \includegraphics[width=0.1\textwidth, valign = m]{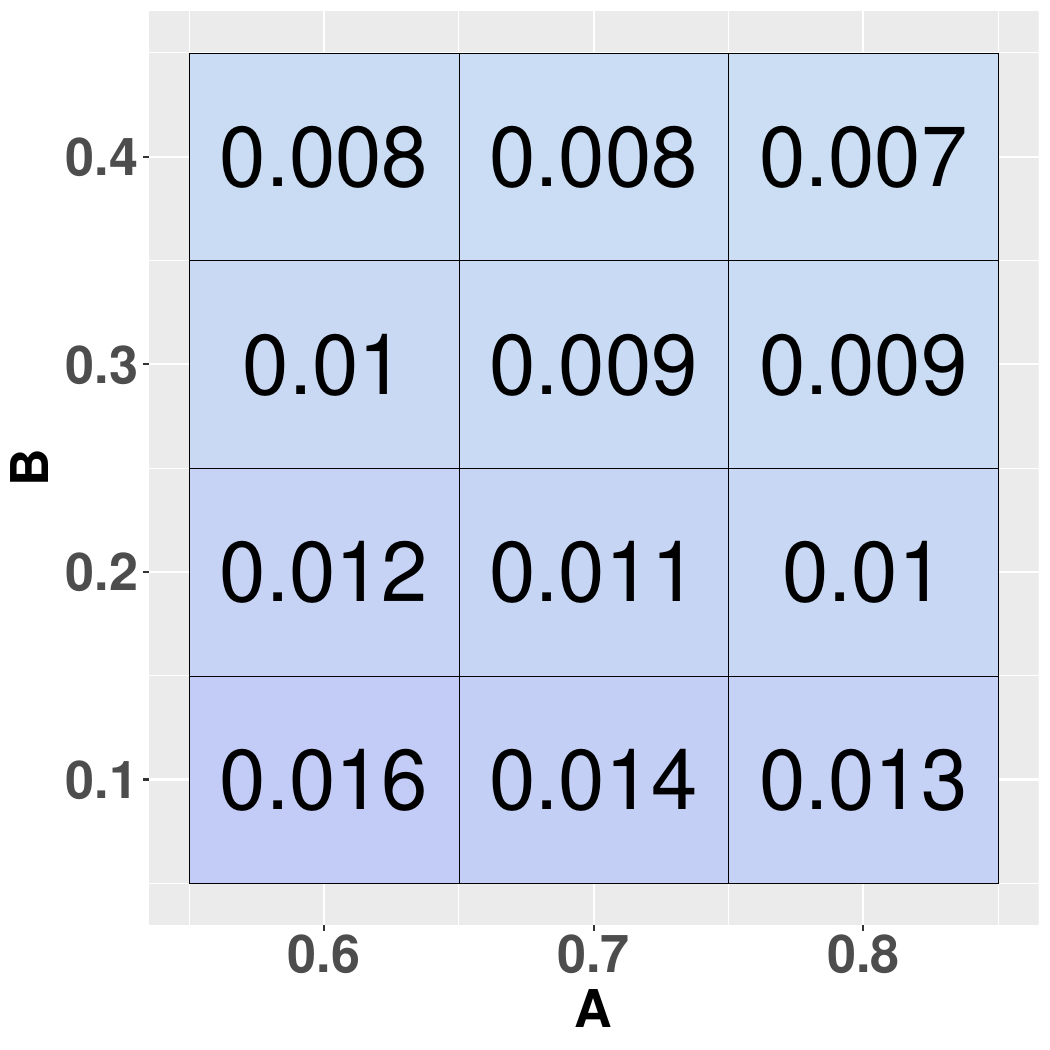} &
    \includegraphics[width=0.1\textwidth, valign = m]{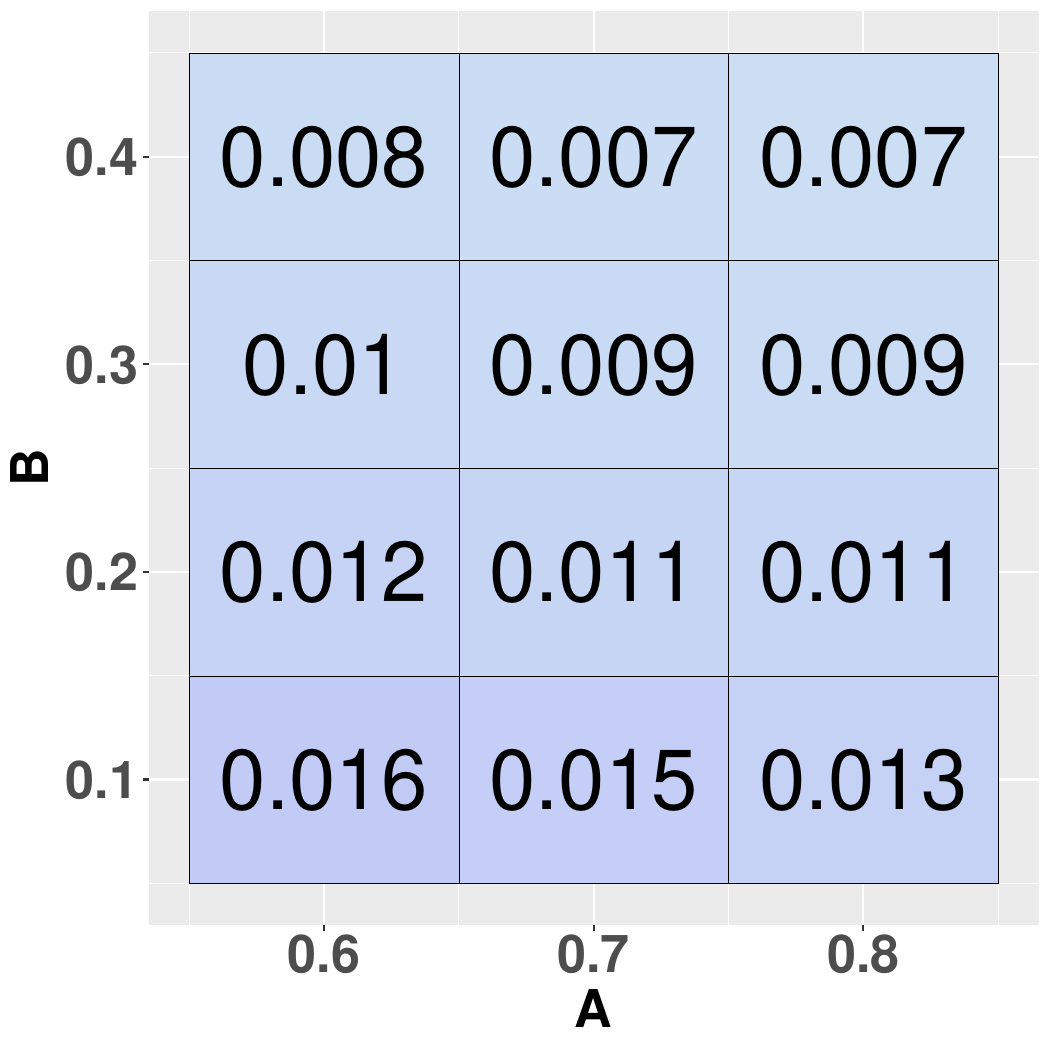} &
    \includegraphics[width=0.1\textwidth, valign = m]{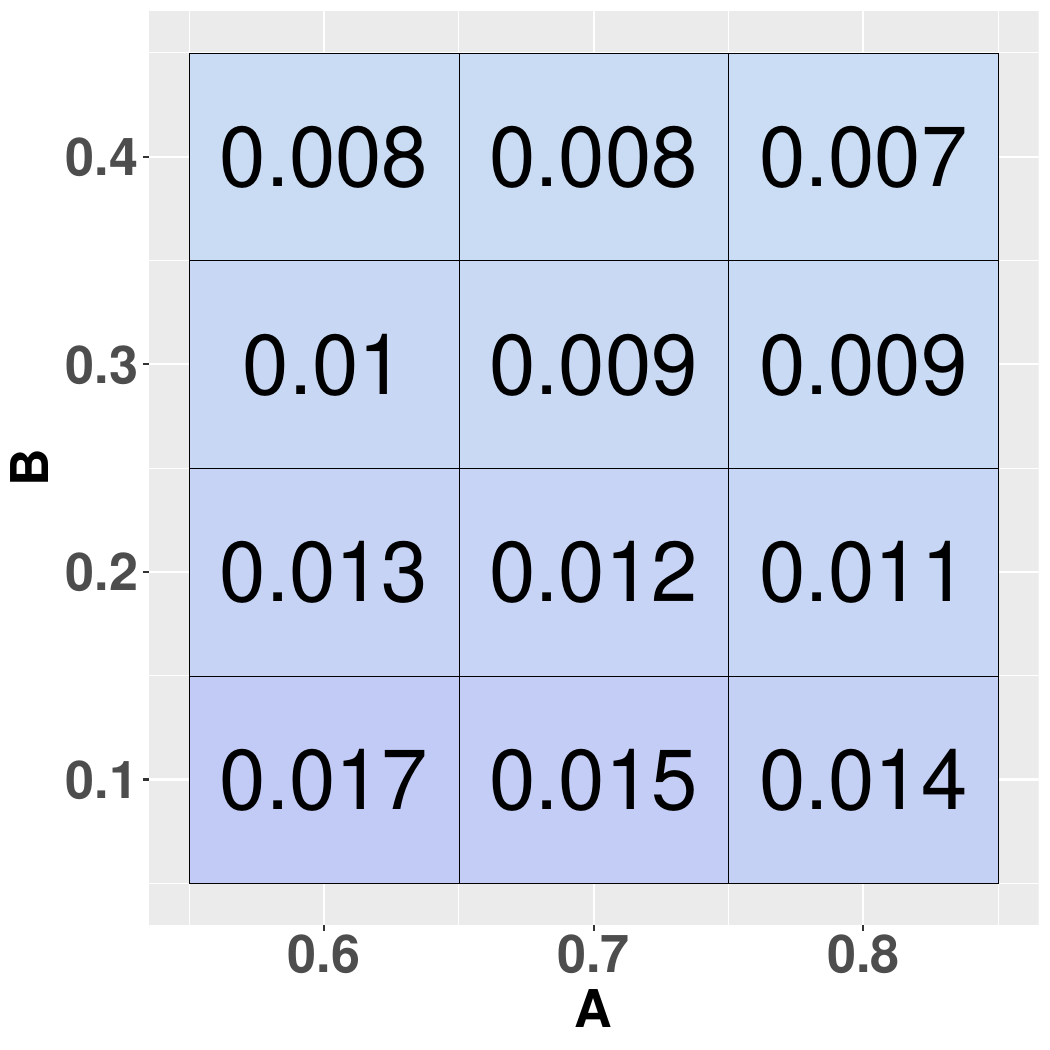} \\ 
    & Scenario 1.3 &  
    \includegraphics[width=0.1\textwidth, valign = m]{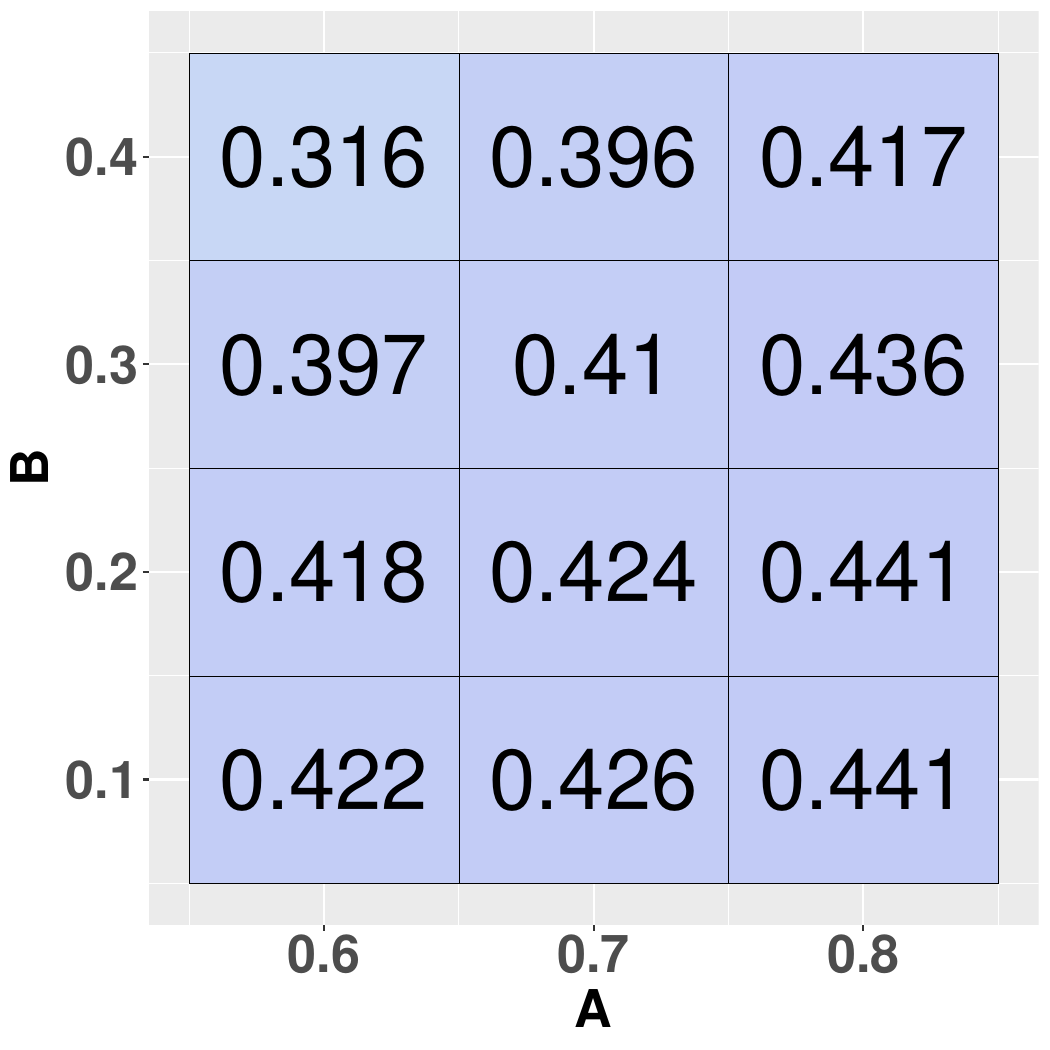} &
    \includegraphics[width=0.1\textwidth, valign = m]{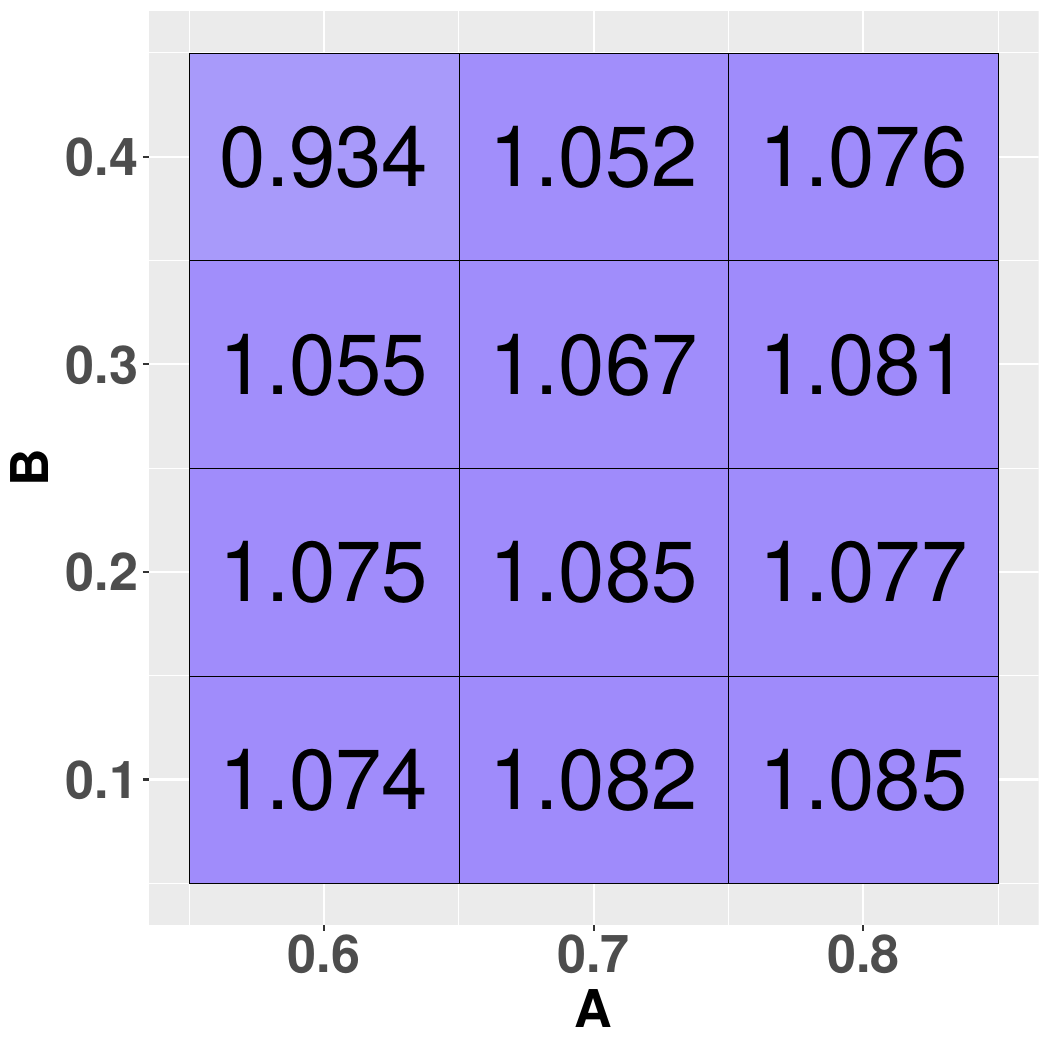} &
    \includegraphics[width=0.1\textwidth, valign = m]{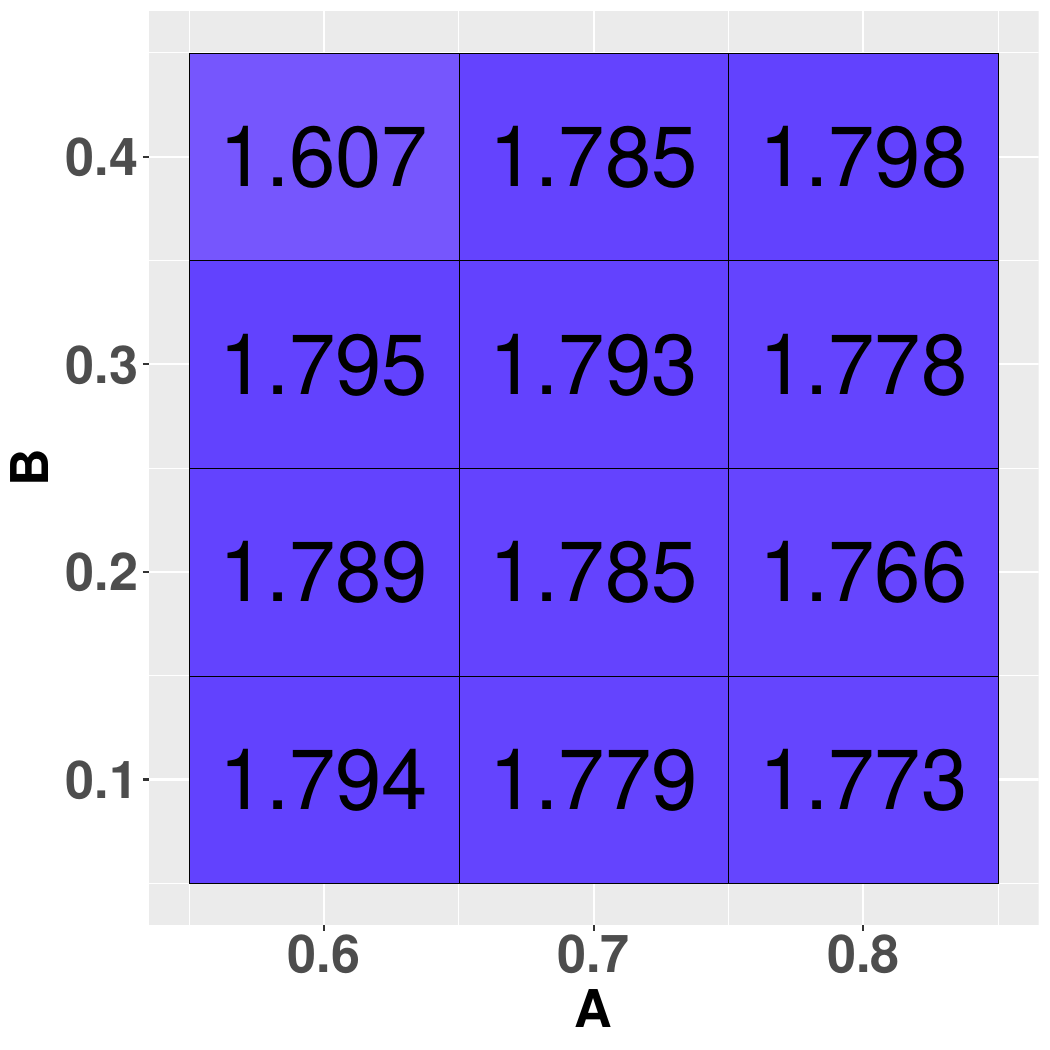} &
    \includegraphics[width=0.1\textwidth, valign = m]{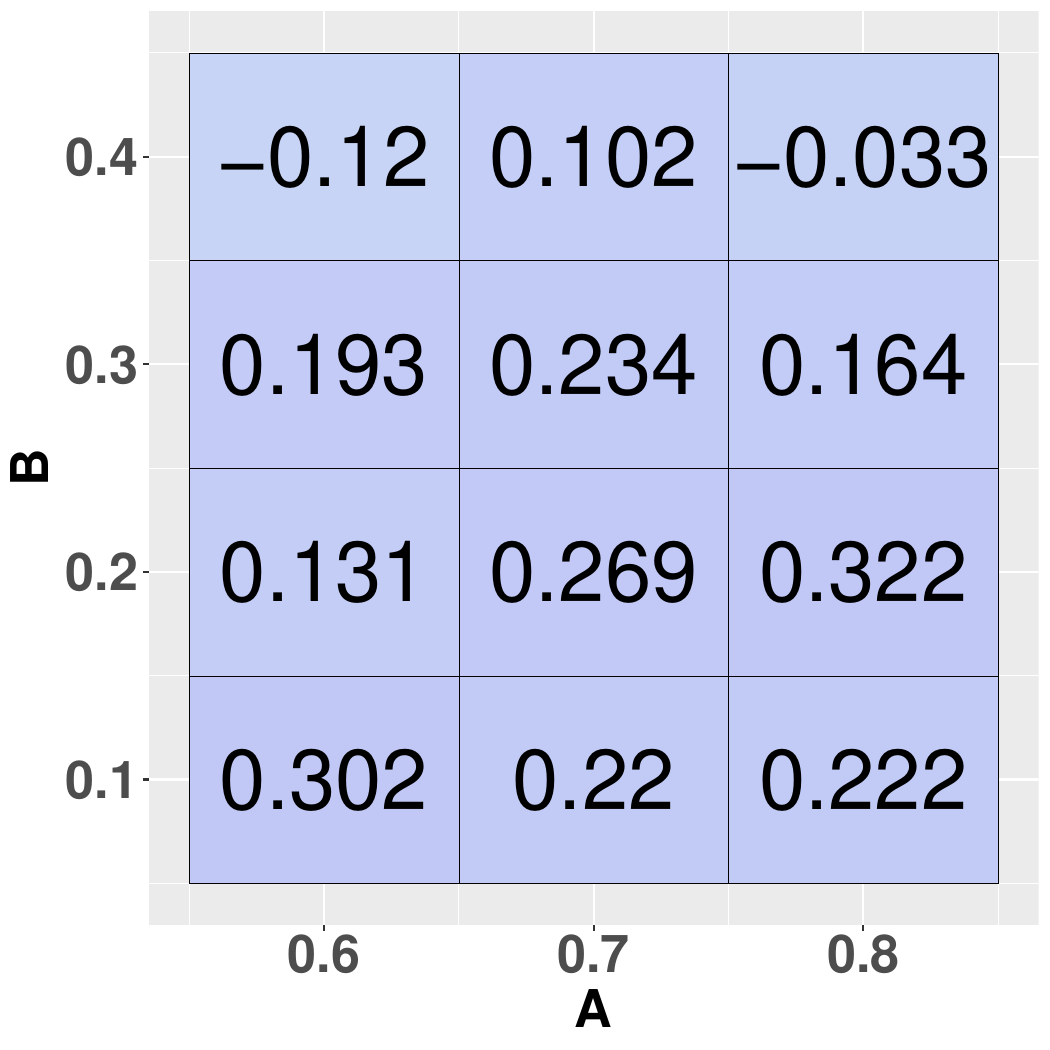} &
    \includegraphics[width=0.1\textwidth, valign = m]{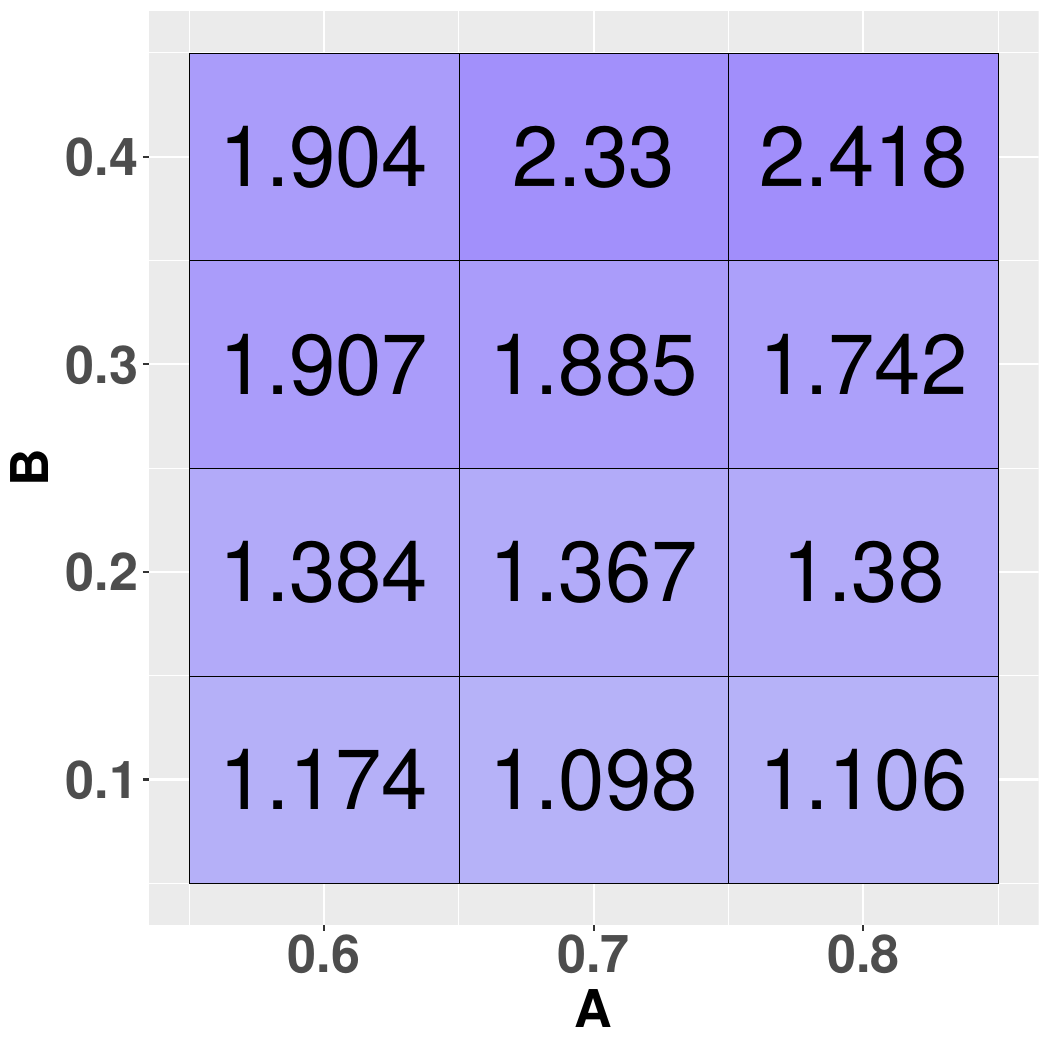} &
    \includegraphics[width=0.1\textwidth, valign = m]{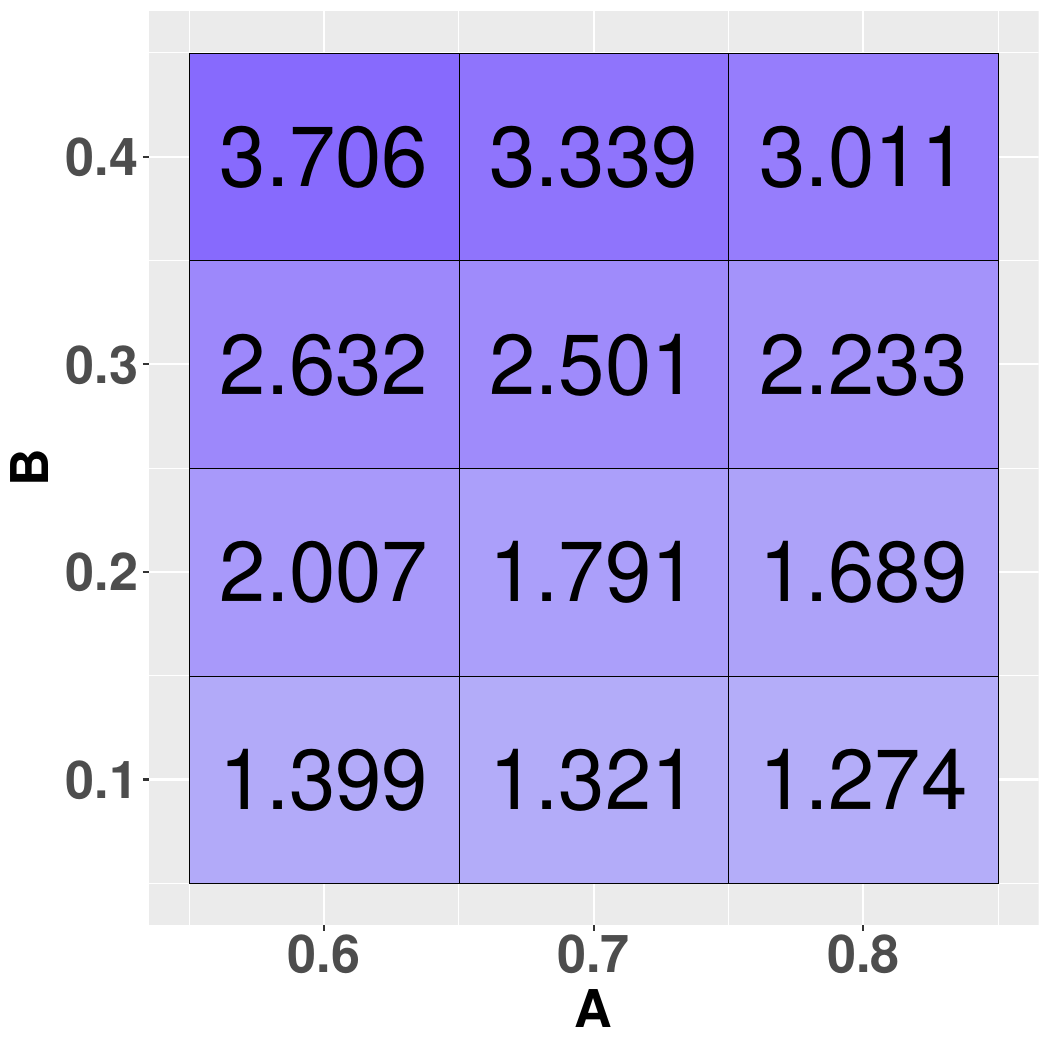} &
    \includegraphics[width=0.1\textwidth, valign = m]{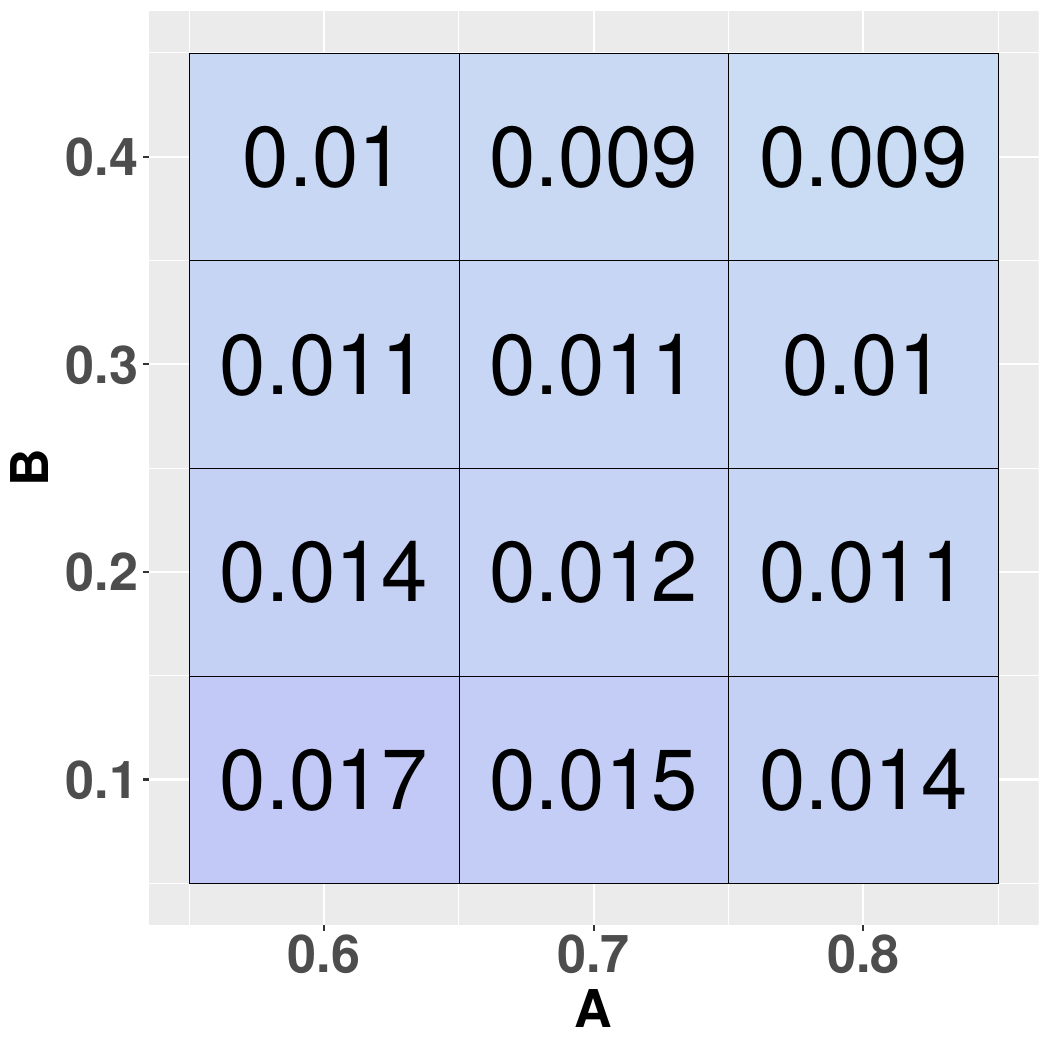} &
    \includegraphics[width=0.1\textwidth, valign = m]{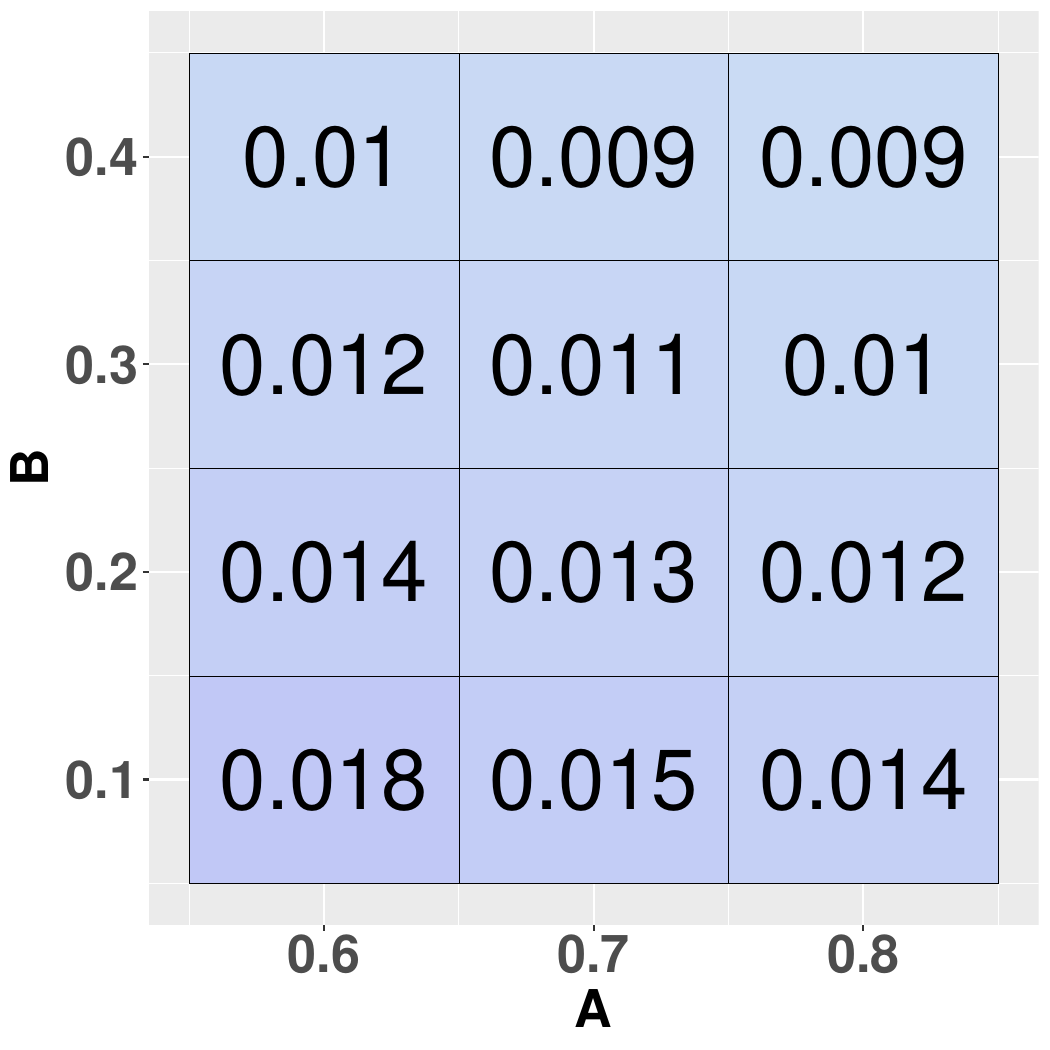} &
    \includegraphics[width=0.1\textwidth, valign = m]{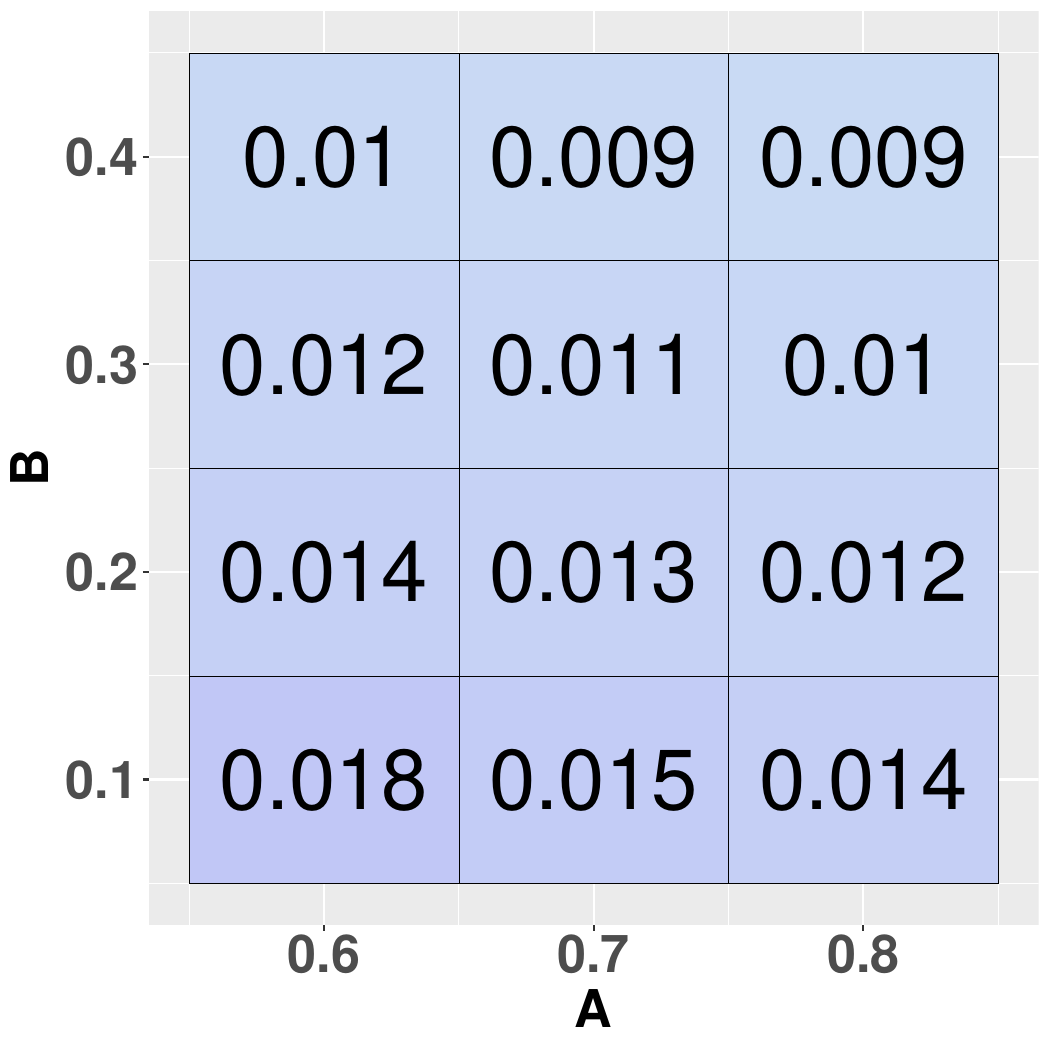} \\ 
    & Scenario 2 &  
    \includegraphics[width=0.1\textwidth, valign = m]{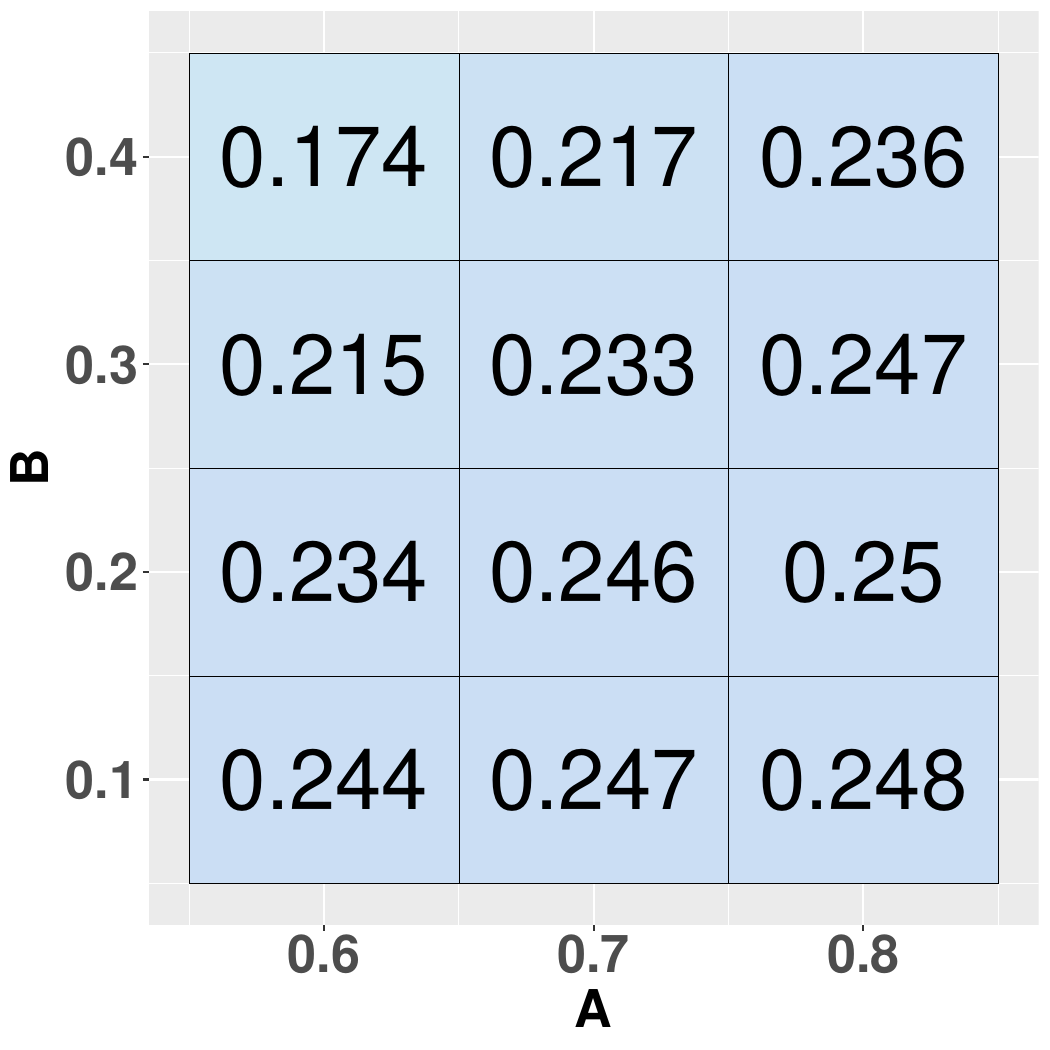} &
    \includegraphics[width=0.1\textwidth, valign = m]{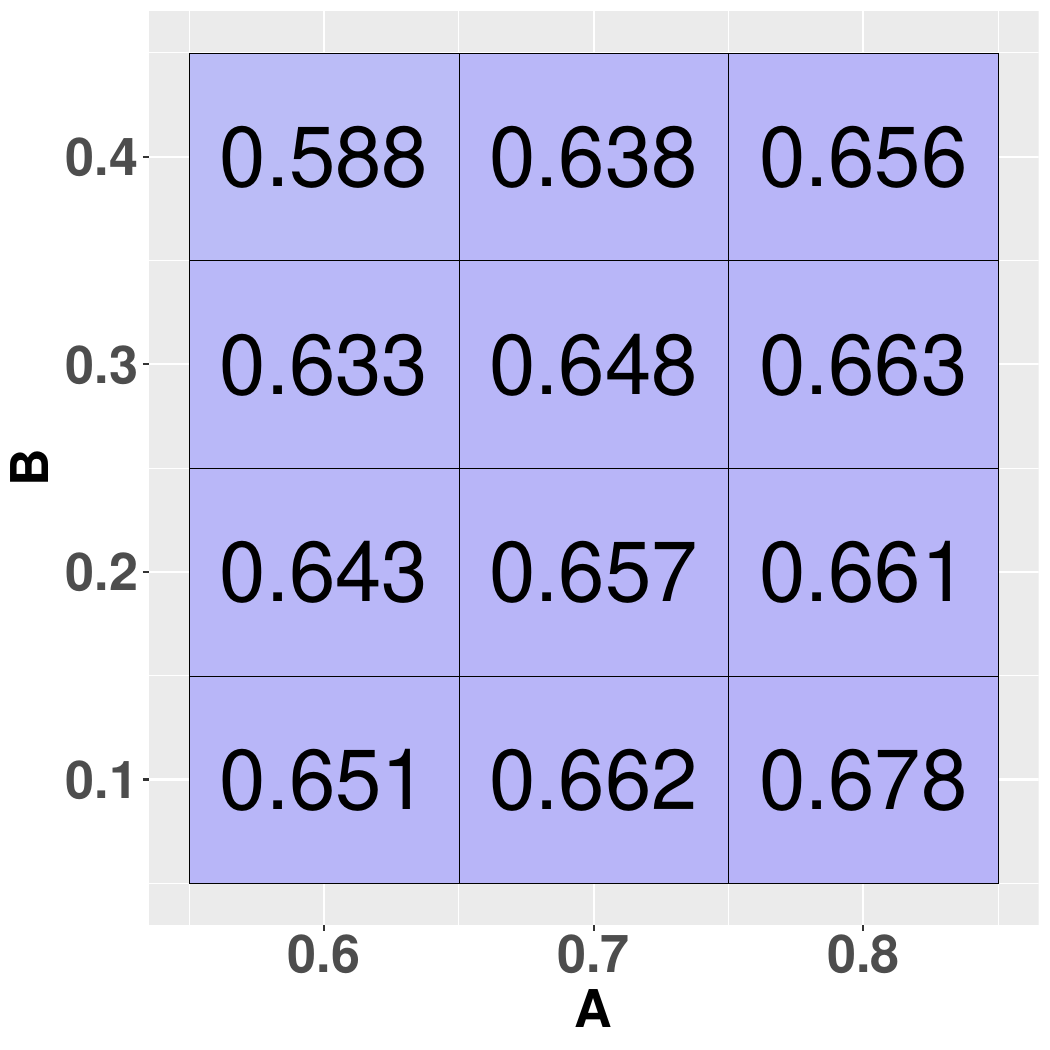} &
    \includegraphics[width=0.1\textwidth, valign = m]{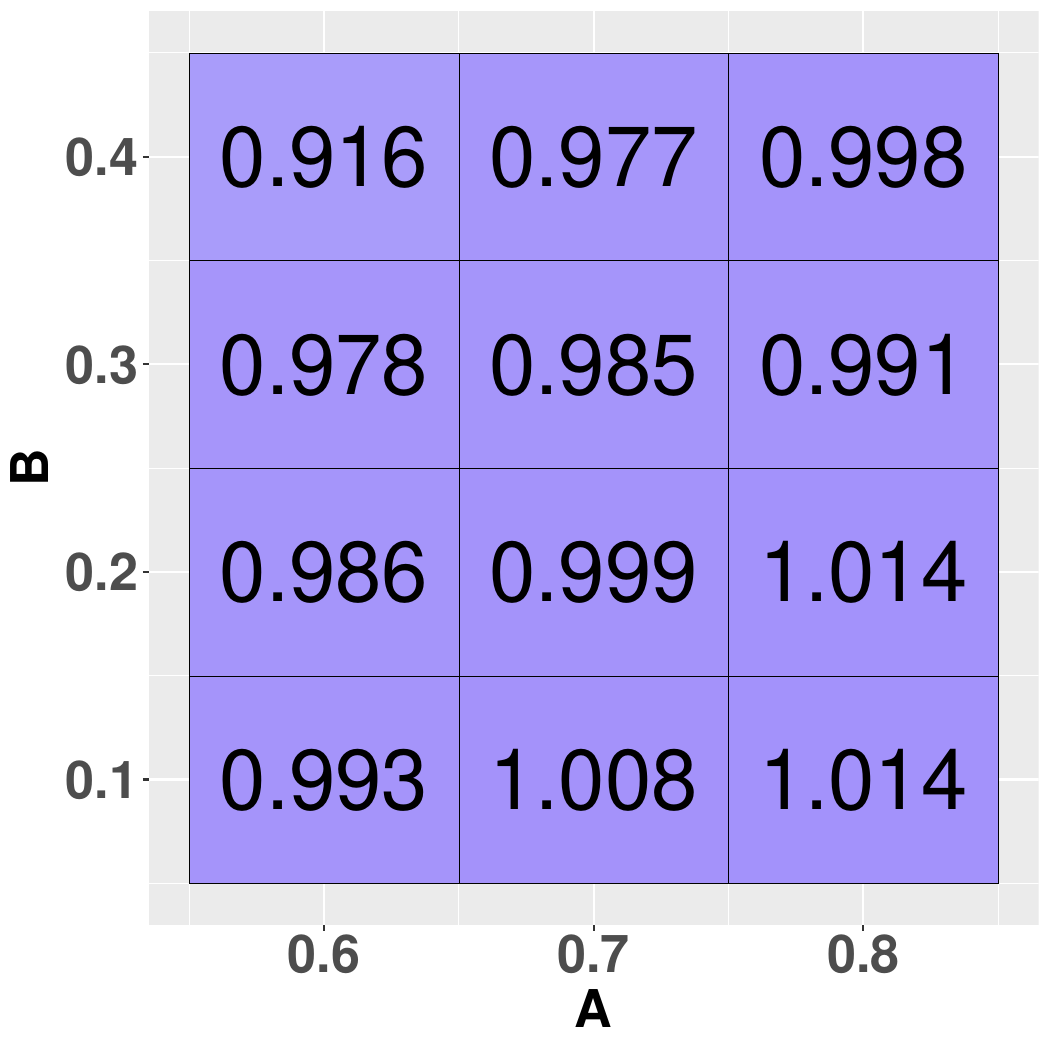} &
    \includegraphics[width=0.1\textwidth, valign = m]{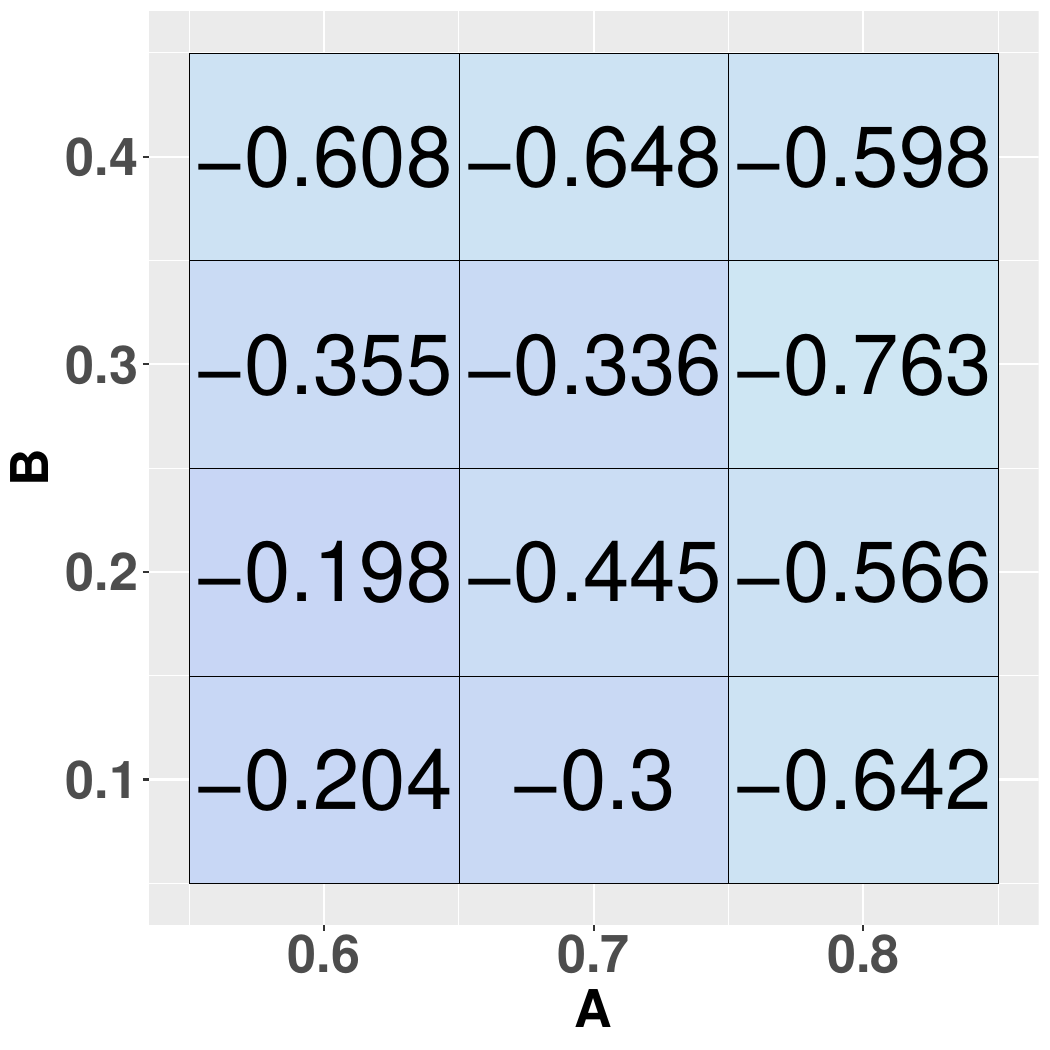} &
    \includegraphics[width=0.1\textwidth, valign = m]{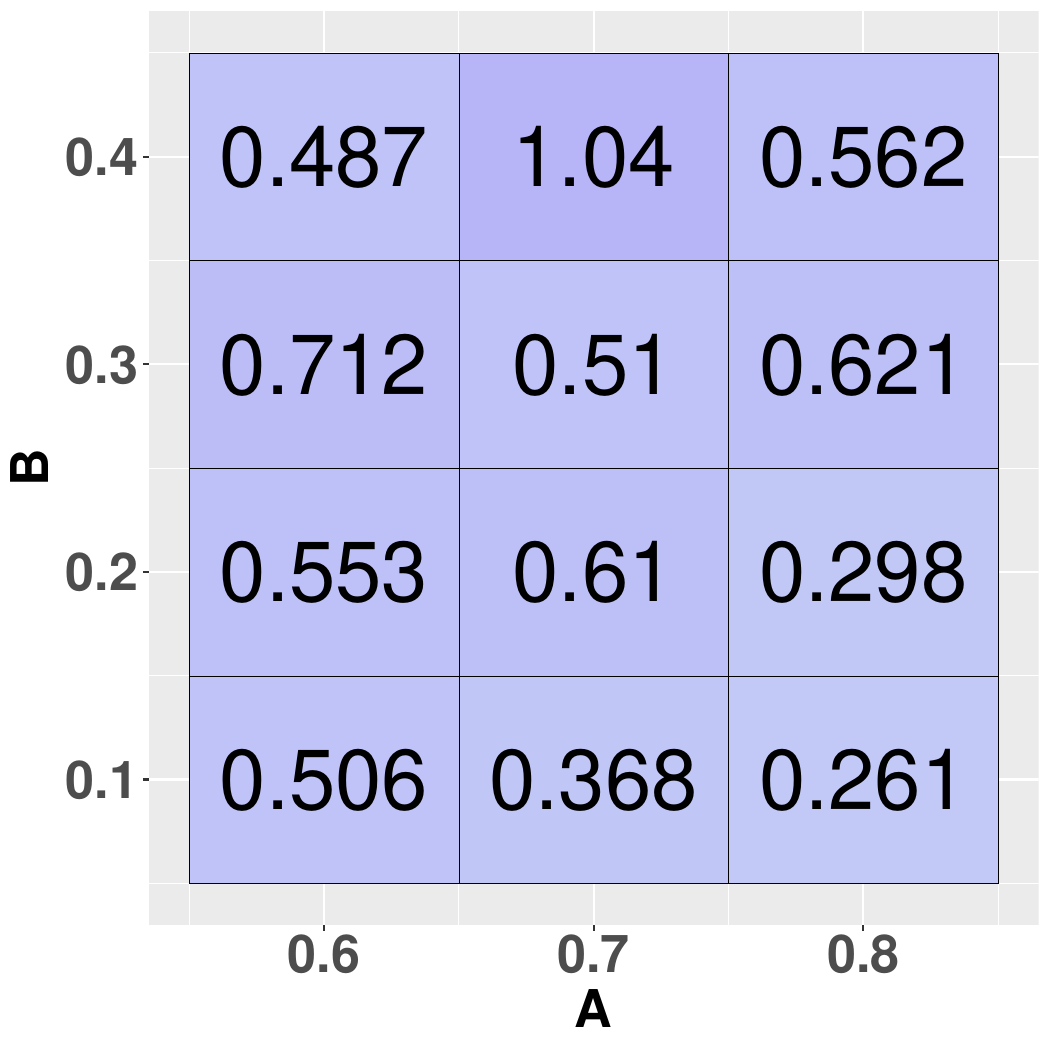} &
    \includegraphics[width=0.1\textwidth, valign = m]{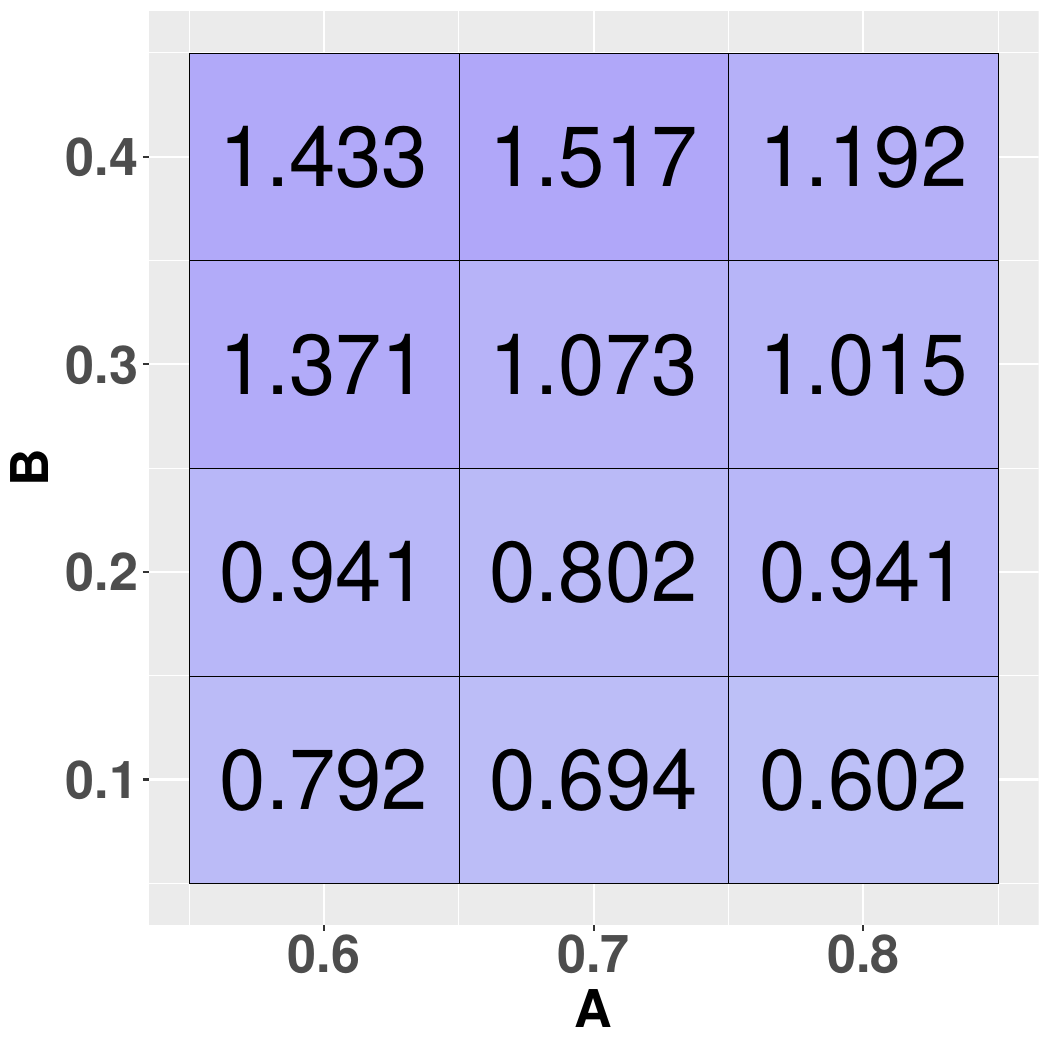} &
    \includegraphics[width=0.1\textwidth, valign = m]{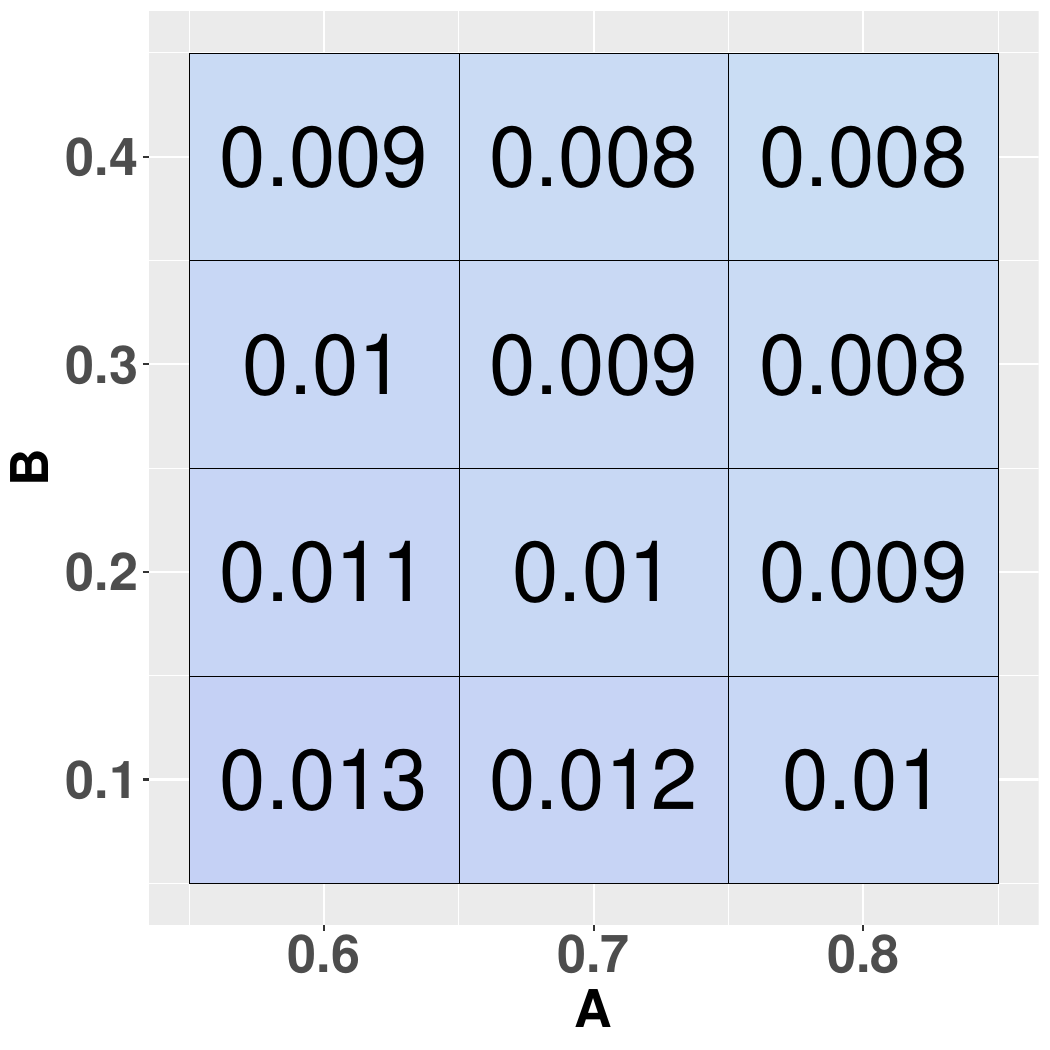} &
    \includegraphics[width=0.1\textwidth, valign = m]{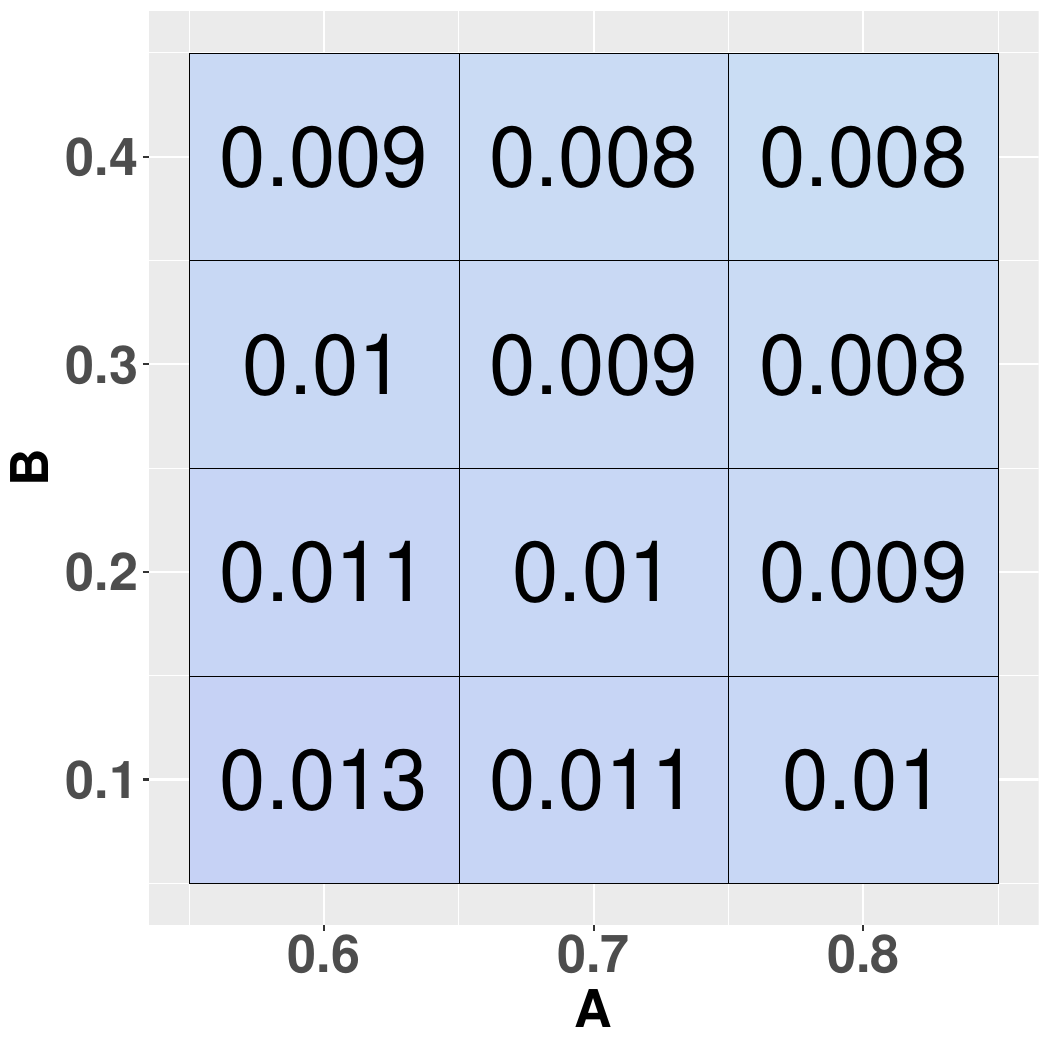} &
    \includegraphics[width=0.1\textwidth, valign = m]{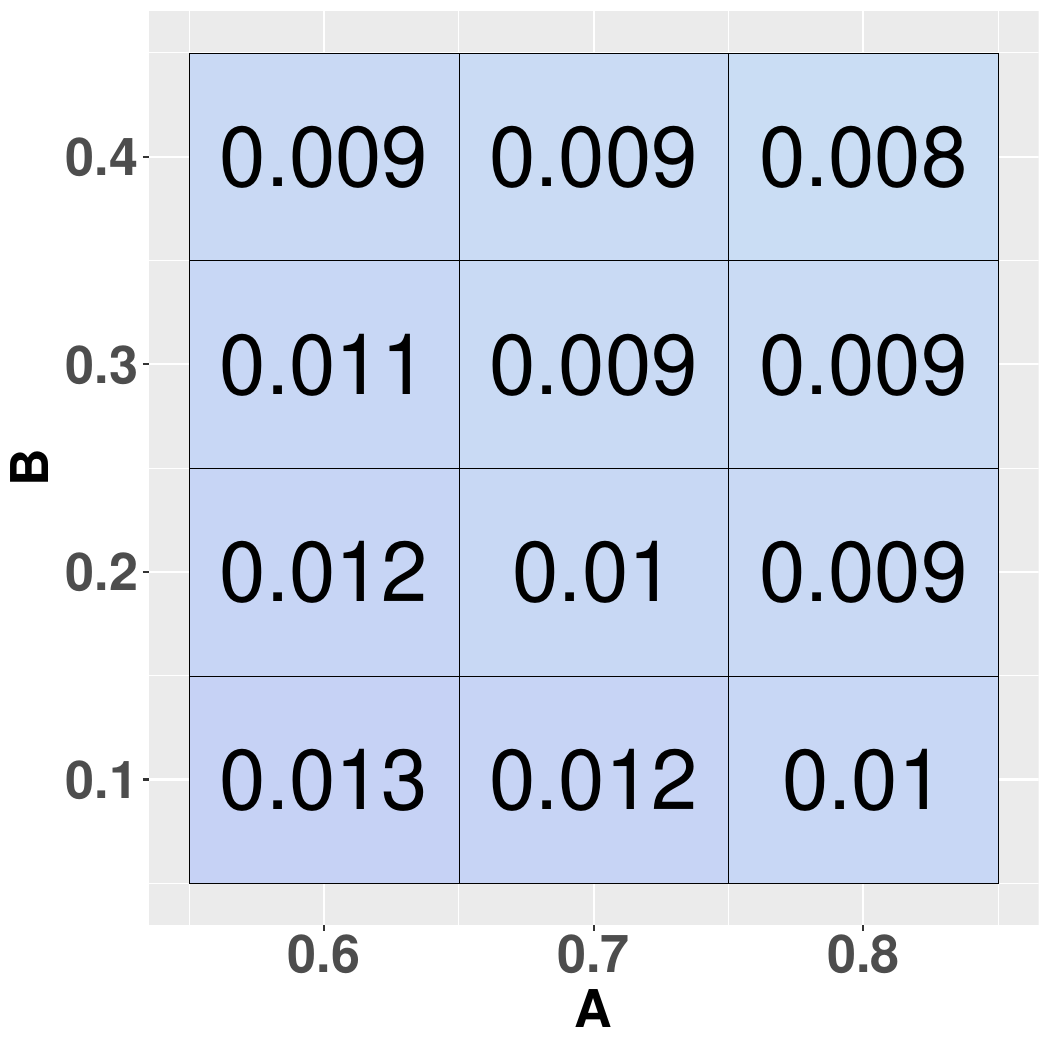} \\ 
    \end{tabular}
    }
    {
    \caption{\label{fig:heatmap}
     The heatmap of the posterior means of the social influence parameters estimated from the three models, when the group cohesion of network and item response data are weak, moderate, and strong in Scenarios 1.1 to 1.3, and 2.
    }
    }
\end{figure}
\end{landscape}

{
To compare the proposed approach with the two existing models, the network autocorrelation and linear-in-means models, we present the heatmap of the posterior means of the social influence parameters estimated from these three models (Figure \ref{fig:heatmap}). Comparisons are made for weak, moderate, and strong group cohesion of network and item response data, where the level of group cohesion is determined by the within-connection and between-connection probabilities in ${\bf P}_{\mbox{net}}$ and ${\bf P}_{\mbox{irt}}$. For strong group cohesion, the within-connection probability is high, while the between-connection probability is low. In contrast, weak-group cohesion is characterized by a low within-connection probability and a high between-connection probability. We select $a$ = 0.8 and $b$ = 0.1 for strong cohesion of the network, $c$ = 0.8, $d$ = 0.1, and $e$ = 0 for strong cohesion of item response, $a$ = 0.7 and $b$ = 0.2 for moderate cohesion of network, $c$ = 0.7, $d$ = 0.2, and $e$ = 0.05 for moderate cohesion of item response, $a$ = 0.6 and $b$ = 0.4 for weak cohesion of network, and $c$ = 0.6, $d$ = 0.4, and $e$ = 0.1 for weak cohesion of item response. Then, we draw in Figure \ref{fig:heatmap} the heatmaps for the 12 simulation settings that correspond to each level of network cohesion and item response cohesion. 
}

{
The results show that 
in the case of the proposed approach, when fixing the network cohesion level, the social influence parameter estimate increases as the within-connection probability of the item-response network rises and the between-connection probability of the item-response network drops. This is the case across all scenarios. Moreover, when the item-response network cohesion level is fixed, the same trend is observed with the proposed approach, although 
the increase of social influence tends to be somewhat weaker. Together, these results suggest that 1) the proposed model accurately reflects our simulation intentions: the stronger the social homophily, the greater the social influence; 2) the cohesion level of an item-response network has a greater impact on the social influence parameter than the cohesion level of a peer network.
}

{
Before discussing results from existing models, it is useful to repeat that the social influence parameter estimates from the models in the comparison may not be directly comparable due to the differences in how social influence is defined and how the behavioral measures are processed for data analysis. Thus, it is important to keep in mind that these differences are likely to contribute to variations in the estimated results.}

{
That said, it is worth noting that the linear-in-means model and the network autocorrelation model show pretty different patterns of social influence estimates than what is intended in the data simulation. For example, in the case of the linear-in-means model, when the level of cohesion for the item-response network is fixed as either moderate or strong, the estimates of the social influence parameter increase as the within-connection probability of the network drops and the between-connection probability of the network rises, across all scenarios. This trend is also observed in all heatmaps for Scenario 1.2. This result suggests that weaker social homophily leads to greater social influence, which is the opposite of the setting intended in this simulation. On the other hand, the network autocorrelation model shows little difference in the social influence across scenarios, not capturing the infused differences between the simulation conditions. These results imply that the linear-in-means model and the network autocorrelation model may not be able to capture social influence accurately when there is a distinctive dependence structure among item-level behavioral measures. 
}

\begin{table}[p]
\begin{center}
{
\small{
\begin{tabular}{c|c r r r r}
\hline
& & Scenario 1.1 & Scenario 1.2 & Scenario 1.3 & Scenario 2 \\ \hline
& Intercept & 0.160 (0.017) & 0.158 (0.011) & 0.173 ($<$0.001) & 0.199 ($<$0.001) \\
& a &  0.206 ($<$0.001) &  0.680 ($<$0.001) &  0.231 ($<$0.001) &  0.172 ($<$0.001) \\
Proposed
& b & -0.176 ($<$0.001) & -0.601 ($<$0.001) & -0.192 ($<$0.001) & -0.135 ($<$0.001) \\
Model
& c &  2.132 (0.002) &  2.073 ($<$0.001) &  2.082 ($<$0.001) &  1.068 ($<$0.001) \\
& d & -2.916 ($<$0.001) & -2.307 ($<$0.001) & -2.837 ($<$0.001) & -1.750 ($<$0.001) \\ 
& e & & -2.547 ($<$0.001) & -0.575 ($<$0.001) & \\
& Adj. R$^2$ & 0.972 & 0.917 & 0.972 & 0.978 \\ \hline
& Intercept & -0.264 ($<$0.001) & 2.030 ($<$0.001) & -0.428 (0.004) & -0.185 (0.139) \\
& a & -0.472 (0.068) & -3.644 ($<$0.001) & -0.276 (0.078) & -0.898 ($<$0.001) \\
Linear-in-means
& b &  3.218 ($<$0.001) & 12.308 ($<$0.001) & 2.855 ($<$0.001) & 0.661 ($<$0.001) \\
Model
& c &  3.276 ($<$0.001) &  0.956 ($<$0.001) &  3.398 ($<$0.001) &  2.634 ($<$0.001) \\
& d & -2.916 ($<$0.001) &  0.378 (0.007) & -4.118 ($<$0.001) & -3.252 ($<$0.001) \\ 
& e & & -0.965 (0.012) & -1.094 (0.001) & \\
& Adj. R$^2$ & 0.203 & 0.955 & 0.834 & 0.871 \\ \hline
& Intercept & 0.011 ($<$0.001) & 0.016 ($<$0.001) & 0.018 ($<$0.001) & 0.009 ($<$0.001) \\
& a & -0.002 (0.196) & -0.004 ($<$0.001) & -0.001 ($<$0.001) & -0.004 (0.001) \\
Network 
& b & -0.019 ($<$0.001) & -0.023 ($<$0.001) & -0.020 ($<$0.001) & -0.009 ($<$0.001) \\
Autocorrelation 
& c &  0.009 ($<$0.001) &  0.004 ($<$0.001) &  0.005 ($<$0.001) &  0.007 (0.123) \\
& d &  0.001 (0.273) &  0.001 (0.148) & 0.000 (0.378) & 0.001 (0.378) \\ 
& e & & -0.002 (0.083) &  0.002 (0.262) & \\ 
& Adj. R$^2$ & 0.970 & 0.952 & 0.954 & 0.965 \\ 
\hline
\end{tabular}
\caption{\label{tab:regression}
The regression coefficient of the estimated $\delta$ in the proposed model on $a$, $b$, $c$, $d$, and $e$ in Scenarios 1.1 to 1.3, and 2. As we described in Section \ref{sec:setting}, $a$ and $c$ range from 0.6 to 0.8 with an increment of 0.1 and $b$ and $d$ range from 0.1 to 0.4 with an increment of 0.1, and $e$ range from 0 to 0.1 with an increment of 0.05. 
}
}
}
\end{center}
\end{table}

{
To confirm the results shown in the heatmap, we conducted multiple linear regressions by taking the estimates of social influence as response variables and taking the five probabilities of connection as predictors ($a$, $b$, $c$, $d$, $e$) under the three models. The regression results with adjusted $R^2$ values are summarized in Table \ref{tab:regression}. The regression coefficients of $a$ and $c$ are expected to be positive, while those of $b$ and $d$ are expected to be negative because the social influence increases as the within-connection probabilities increase and the between-connection probabilities decrease. The result shown in Table \ref{tab:regression} for the proposed approach is consistent with this trend. Moreover, the regression coefficients for the connection probability of item responses have higher absolute values than those for the connection probability of the network, indicating that the cohesion level of the item responses has a greater impact on social influence. In the linear-in-means model, the coefficients of $a$ are negative while those of $b$ are positive, implying that weak social homophily has greater social influence (which is the opposite of the intended structure). For the network autocorrelation model, the result shows negative coefficients for all connection probabilities for the network and positive coefficients for all connection probabilities for the item response. In summary, the results of the linear regression analysis confirm that our proposed model may not be able to capture social influence accurately when there is an inherent dependence structure in the item response data, echoing the results shown in the heatmap analysis. 
}


\section{Discussion}

We have introduced a new approach for assessing the social influence of individuals' social networks at time $t$ on their behavior at time $t+1$. In our proposed model, evaluating social influence comes down to the task of integrating two latent space models, one for network data at time $t$ and the other for item-level behavior data at time $t+1$. In our framework, social influence is defined as the impact of social homophily on the respondents' behavior, where social homophily is reflected by the respondents' latent positions, and respondents' behavior is their responses to the test items. 
{
The adapted latent space, or interaction map, estimated from the proposed approach shows the structure and patterns of differential social influence across specific, item-level behavioral measures, while the magnitude of overall social influence can be evaluated by the parameters $\delta$ and $\omega$.} We developed a fully Bayesian approach to estimate the proposed model and illustrated the proposed approach with an empirical data example. We conducted simulation studies to understand the operating characteristics of the proposed model and evaluate the performance of our proposed approaches. 
{
To our knowledge, our proposed approach is the first attempt to analyze differential social influence effects across item-level behavioral measures in the literature.}

{
In closing, we discuss some additional points related to the proposed approach. First, our definition and measurement of social influence somewhat differ from how social influence is measured in the literature. However, as many variations of social influence definition are available, bringing a new perspective on social influence, such as ours, can be seen as a contribution to the field. Second, the current study presented the proposed approach in a setting with binary item response data and no covariates. The proposed model can be extended to accommodate more complex situations. For example, the model can accommodate polytomous item responses by adopting an appropriate link function, such as an adjacent logit or a baseline logit link (which means that the base model for LSIRM becomes a partial credit model or a nominal model, respectively). In addition, person covariates can be included in the model, similar to standard latent space models. Third, our proposed model is a regression-based approach; thus, the results should be treated as an association rather than causal effects. 
Nevertheless, it remains an interesting topic to describe how students’ social interactions are related to their behavior.  
Lastly, we estimate social influence in longitudinal data settings, where respondents' social network is measured at time $t$, while their behaviors (responses to behavioral items) are measured at time $t+1$. Although this approach may be seen as more reasonable than using network and behavioral measures obtained at the same time point $t$ \citep{Steglich2010, Sweet:2020, Frank2021}, 
our model may not fully disentangle confounding with social selection effects from social influence, unless specific, potentially stringent assumptions are additionally imposed. This is because respondents' latent positions estimated from LSM in our approach reflect social homophily, which can be seen as a social selection process that occurred prior to time $t$. The complete removal of the social selection confounding that happened before the measurement of a network from social influence is an important research topic, but it may require consideration of a rigorous causal inference setting, which falls outside the scope of our current study. Therefore, we leave such an investigation for future research. 
}

\section*{Acknowledgements}

This work was supported by the National Research Foundation of Korea [grant number NRF 2020R1A2C1A01009881; Basic Science Research Program awarded to IHJ] and the Yonsei University Research Grant of 2022 [grant number 2022-22-0439 awarded to JAP]. Correspondence should be addressed to Ick Hoon Jin, Department of Applied Statistics, Department of Statistics and Data Science, Yonsei University, Seoul, Republic of Korea. E-Mail: ijin@yonsei.ac.kr. 

\bibliographystyle{Chicago}
\bibliography{reference}
\end{document}